\documentclass{aa}

\usepackage{silence}
\usepackage{graphicx}

\usepackage{transparent}
\usepackage{xkeyval,xcolor}
\makeatletter
\newlength{\sfp@hseplen}\newlength{\sfp@vseplen}
\define@cmdkey{subfigpos}[sfp@]{pos}[ul]{}
\define@cmdkey{subfigpos}[sfp@]{font}[\small]{}
\define@cmdkey{subfigpos}[sfp@]{vsep}[0.75\baselineskip]{\setlength{\sfp@vseplen}{\sfp@vsep}}
\define@cmdkey{subfigpos}[sfp@]{hsep}[3.5pt]{\setlength{\sfp@hseplen}{\sfp@hsep}}
\newcommand{\subfigimg}[4][,]{%
	\setkeys{Gin,subfigpos}{pos,font,vsep,hsep,#1}
	\setbox1=\hbox{\includegraphics{#4}}
	\ifnum\pdfstrcmp{\sfp@pos}{ul}=0
		\leavevmode\rlap{\usebox1}
		\rlap{\hspace*{\sfp@hsep}\raisebox{\dimexpr\ht1-\sfp@vsep}{\transparent{#3}{\setlength{\fboxsep}{1pt}\colorbox{white}{%
\transparent{1}\sfp@font{#2}}}%
}}
		\phantom{\usebox1}
	\else\ifnum\pdfstrcmp{\sfp@pos}{ur}=0
		\leavevmode\usebox1
		\llap{\raisebox{\dimexpr\ht1-\sfp@vsep}{\sfp@font{#2}}\hspace*{\sfp@hsep}}
	\else\ifnum\pdfstrcmp{\sfp@pos}{lr}=0
		\leavevmode\usebox1
		\llap{\raisebox{\sfp@vsep}{\sfp@font{#2}}\hspace*{\sfp@hsep}}
	\else
		\leavevmode\rlap{\usebox1}
		\rlap{\hspace*{\sfp@hseplen}\raisebox{\sfp@vsep}{\sfp@font{#2}}}
		\phantom{\usebox1}
	\fi\fi\fi
}
\newcommand{\fontfig}[1]{\tiny$\!\!$\color{#1}\textbf}
\newcommand{\AspectRatio}[1]{\dimexpr 1pt * \wd#1 / \ht#1 \relax} 

\usepackage{mathtools}
\usepackage{mathrsfs}
\usepackage{nicefrac}

\usepackage{amsmath} 
\usepackage{array}
\usepackage{adjustbox}
\usepackage{multirow}
\newcolumntype{C}[1]{>{\centering\arraybackslash}m{#1}} 

\usepackage{algorithm}
\usepackage[noend]{algpseudocode}
\newcommand{\commentalgo}[1]{\Comment{{\tiny #1}}} 

\newcommand{\eq}[1]{Eq.~\eqref{#1}\xspace}
\newcommand{\eqs}[2]{Eqs.~(\ref{#1}) and (\ref{#2})\xspace}

\newcommand{\subfigref}[1]{#1} 
\newcommand{\fig}[1]{Fig.~\ref{#1}\xspace}
\newcommand{\figfull}[1]{Figure~\ref{#1}\xspace} 
\newcommand{\subfig}[2]{Fig.~\ref{#1}\subfigref{#2}\xspace}
\newcommand{\subfigpar}[2]{Fig.~\ref{#1}(\subfigref{#2})\xspace}
\newcommand{\subfigfull}[2]{Figure~\ref{#1}\subfigref{#2}\xspace} 
\newcommand{\subfigs}[2]{Figs.~\ref{#1}\subfigref{#2}\xspace}
\newcommand{\subfigspar}[2]{Fig.~\ref{#1}(\subfigref{#2})\xspace}
\newcommand{\subfigsfull}[2]{Figures~\ref{#1}\subfigref{#2}\xspace} 

\newcommand{\tab}[1]{Table~\ref{#1}\xspace}

\newcommand{\refsec}[1]{Sect.~\ref{#1}\xspace} 
\newcommand{\refsecs}[2]{Sects.~\ref{#1} and~\ref{#2}\xspace} 
\newcommand{\refapp}[1]{Appendix~\ref{#1}\xspace} 

\newcommand{\Tag}[1]{\text{#1}}	

\newcommand{\V}[1]{{\boldsymbol{#1}}}                 
\newcommand{\M}[1]{{\mathbf{#1}}}                     
\newcommand{\T}{^{\mathrm{T}}}                        
\newcommand{\Inv}{^{-1}}                              

\newcommand{\I}{\mathrm{i}}                           
\newcommand{\E}[1]{\mathrm{e}^{#1}}                   
\newcommand{\conj}[1]{#1^{\star}}                     

\newcommand{\D}{\mathrm{d}}                           

\newcommand{\FTnot}{\mathscr{F}}
\newcommand{\FTnotinv}{\mathscr{F}^{-1}}
\newcommand{\FTfull}[2]{\FTnot\left[#1\right]\left(#2\right)}
\newcommand{\FTfullinv}[2]{\FTnotinv\left[#1\right]\left(#2\right)}
\newcommand{\FT}[1]{\tilde{#1}}
\newcommand{\DFT}[1]{\Tag{DFT}\left[#1\right]}
\newcommand{\iDFT}[1]{\Tag{DFT}\Inv\left[#1\right]}

\newcommand{\ScaleAs}{\mathcal{O}}                    

\DeclarePairedDelimiterX{\paren}[1]{(}{)}{#1}
\newcommand{\Paren}[1]{\paren*{#1}}
\let\brace=\undefined 
\DeclarePairedDelimiterX{\brace}[1]{\{}{\}}{#1}
\newcommand{\Brace}[1]{\brace*{#1}}
\let\brack=\undefined 
\DeclarePairedDelimiterX{\brack}[1]{[}{]}{#1}
\newcommand{\Brack}[1]{\brack*{#1}}
\DeclarePairedDelimiterX{\abs}[1]{\rvert}{\lvert}{#1}     
\newcommand{\Abs}[1]{\abs*{#1}}
\DeclarePairedDelimiterX{\norm}[1]{\lVert}{\rVert}{#1}    
\newcommand{\Norm}[1]{\norm*{#1}}

\DeclarePairedDelimiterX{\avg}[1]{\langle}{\rangle}{#1}   

\DeclarePairedDelimiterX{\ceil}[1]{\lceil}{\rceil}{#1}     

\DeclarePairedDelimiterX{\floor}[1]{\lfloor}{\rfloor}{#1}  
\newcommand{\Floor}[1]{\floor*{#1}}

\newcommand{\VarD}[2]{\mathcal{V}_{#2}\paren*{#1}}

\newcommand{\Reals}{\mathbb{R}}
\newcommand{\PosIntegers}{\mathbb{N}}
\newcommand{\Integers}{\mathbb{Z}}
\newcommand{\Complexes}{\mathbb{C}}

\newcommand{\real}[1]{\mathscr{R}\left(#1\right)}

\newcommand{\Va}{{\V{a}}}                   
\newcommand{\Vc}{{\V{c}}}                   
\newcommand{\Vap}{{\V{a}^{\prime}}}         
\newcommand{\ap}{{a^{\prime}}}              
\newcommand{\Vx}{{\V{x}}}                   
\newcommand{\Vxp}{{\V{x^{\prime}}}}         
\newcommand{\Vu}{{\V{u}}}                   
\newcommand{\Vup}{{\V{u^{\prime}}}}         
\newcommand{\Vk}{{\V{k}}}                   
\newcommand{\Vkp}{{\V{k^{\prime}}}}         

\newcommand{\Dx}{{\D \Vx}}            
\newcommand{\Dxp}{{\D \Vxp}}          
\newcommand{\Du}{{\D \Vu}}            
\newcommand{\Dup}{{\D \Vup}}          
\newcommand{\Dk}{{\D \Vk}}            
\newcommand{\Dkp}{{\D \Vkp}}          

\newcommand{\Pupil}{\mathcal{P}}             
\newcommand{\pupil}{P}                       
\newcommand{\PSF}{h}                         
\newcommand{\FTPSF}{\FT{h}}
\newcommand{\PSFturb}{h_{\pha}}              
\newcommand{\FTPSFturb}{\FT{h}_{\pha}} 
\newcommand{\PSFtel}{h_{\Tag{tel}}}          
\newcommand{\FTPSFtel}{\FT{h}_{\Tag{tel}}} 
\newcommand{\PSFres}{h_{\epsilon}}           

\newcommand{\strehl}{\gamma_{\Tag{Strehl}}}  
\newcommand{\convprod}[2]{#1\star#2}         

\newcommand{\proja}{\pi_{a}}                    
\newcommand{\projap}{\pi_{a^{\prime}}}          
\DeclareMathOperator*{\argmin}{arg\,min}

\newcommand{\Daap}{\Brack{\M{D}}_{a,\ap}}
\newcommand{\Dapa}{\Brack{\M{D}}_{\ap,a}}
\newcommand{\PDa}{\Pi_{a}}

\newcommand{\pha}{\varphi}                  
\newcommand{\phac}{\pha_\Tag{cor}}          

\newcommand{\sigAO}{\sigma_{\epsilon}}      
\newcommand{\phaAO}{\pha_{\epsilon}}        
\newcommand{\FTphaAO}{\FT{\pha}_{\epsilon}} 

\newcommand{\IFref}{\pha_{0}}               
\newcommand{\FTIFref}{\FT{\pha}_{0}}        
\newcommand{\IFa}{\pha_{a}}                 
\newcommand{\IFap}{\pha_{\ap}}              

\newcommand{\IFperp}{\psi_{\perp}}          
\newcommand{\FTIFperp}{\FT{\psi}_{\perp}}   
\newcommand{\IFperpa}{\psi_{a}}             
\newcommand{\FTIFperpa}{\FT{\psi}_{a}}      
\newcommand{\IFperpap}{\psi_{\ap}}          
\newcommand{\FTIFperpap}{\FT{\psi}_{\ap}}   

\newcommand{\AvgT}[1]{\avg*{#1}_{t}}          
\newcommand{\PSD}[1]{\Phi_{#1}}               
\newcommand{\PSDres}{\Phi_{\epsilon}}          
\newcommand{\PSDperp}{\Phi_\perp}             

\newcommand{\SF}[1]{D_{#1}}         
\newcommand{\SFc}{\SF{\Tag{cor}}}   
\newcommand{\SFpha}{\SF{\pha}}      
\newcommand{\SFAO}{\SF{\epsilon}}   
\newcommand{\FTSF}[1]{\FT{D}_{#1}}  
\newcommand{\FTSFpha}{\FTSF{\pha}}      
\newcommand{\FTSFAO}{\FTSF{\epsilon}}   
\newcommand{\SFd}[1]{D^{\Tag{dir}}_{#1}} 
\newcommand{\SFf}[1]{D^{\Tag{freq}}_{#1}} 

\newcommand{\nPiston}{n_{\circ}}              
\newcommand{\mPiston}{m_{\circ}}              
\newcommand{\PSDpiston}{\Phi_{\circ}}         
\newcommand{\nTip}{n_{\leftrightarrow}}       
\newcommand{\mTip}{m_{\leftrightarrow}}       
\newcommand{\PSDtip}{\Phi_{\leftrightarrow}}  
\newcommand{\nTilt}{n_{\updownarrow}}         
\newcommand{\mTilt}{m_{\updownarrow}}         
\newcommand{\PSDtilt}{\Phi_{\updownarrow}}    

\newcommand{\ca}{c_{a}}               
\newcommand{\setactu}{\mathcal{A}}    
\newcommand{\pitch}{\Delta}           
\newcommand{\Vpitch}{\V{\pitch}}      
\newcommand{\Pa}{\V{p}_{a}}           
\newcommand{\Pap}{\V{p}_{\ap}}        
\newcommand{\nactu}{n_{\Tag{act}}}    
\newcommand{\ntotactu}{m_{\Tag{act}}} 

\newcommand{\EWprod}{\otimes}         

\newcommand{\Vspacenot}{\mathscr{V}}          
\newcommand{\Vspace}[1]{\Vspacenot\Paren{#1}} 

\newcommand{\weight}{w}                  
\newcommand{\wa}{\weight_{a}}            
\newcommand{\wap}{\weight_{\ap}}         
\newcommand{\wqOne}{\weight_{\pairOne}}  
\newcommand{\wqTwo}{\weight_{\pairTwo}}  
\newcommand{\pair}{{\V{q}}}              
\newcommand{\pairOne}{{\pair_{1}}}       
\newcommand{\pairTwo}{{\pair_{2}}}       
\newcommand{\nw}{n_{\Tag{w}}}            
\newcommand{\nwpair}{\nw\Paren{\pair}}   

\newcommand{\rFried}{r_{0}}             
\newcommand{\nPha}{n_{\pha}}            

\newcommand{\pix}{\theta}                 
\newcommand{\Vpix}{\V{\pix}}              
\newcommand{\Vpixp}{\V{\pix}^{\prime}}    
\newcommand{\pixk}{\nu}                   
\newcommand{\Vpixk}{\V{\pixk}}            
\newcommand{\Dsim}{D}                     
\newcommand{\nPad}{n^{\Tag{pad}}}         
\newcommand{\nPix}{n_{\Tag{pix}}}         

\newcommand{\shift}{{\V{s}}}                    
\newcommand{\IFit}[1]{{\IFperp^{#1}}}   
\newcommand{\IFiti}{{\IFit{i}}}                 
\newcommand{\IFitip}{{\IFit{i+1}}}              
\newcommand{\IFits}{{\psi_{\shift}^{i}}}
\newcommand{\Vj}{{\V{j}}}                               
\newcommand{\jOne}{{j_{1}}}                             
\newcommand{\jTwo}{{j_{2}}}                             
\newcommand{\RMS}{{\epsilon_{\Tag{RMS}}}} 
\newcommand{\IFrefproj}{\bar{\pha}_{0}} 

\newcommand{\Vpitchper}{\tilde{\Vpitch}}\newcommand{\Sup}{\mathcal{D}}                      
\newcommand{\Supa}{\Sup_{a}}                        
\newcommand{\Supref}{\Sup_{0}}                      

\newcommand{\fLF}{f_{\Tag{LF}}}           
\newcommand{\sinc}{\Tag{sinc}}            

\newcommand{\eg}{e.g.\xspace}          
\newcommand{\ie}{i.e.\xspace}          
\newcommand{\st}{\text{ s.t. }\xspace}          
\newcommand{\vs}{\emph{vs.}\xspace}          

\usepackage[squaren,thickspace,thickqspace]{SIunits}
\addunit{\adu}{adu}             
\addunit{\au}{arbitrary~unit}   
\newcommand{\percent}[1]{#1\,\%}                

\usepackage{flushend}

\usepackage{txfonts}
\usepackage{hyperref}
\begin{document}

\title{Analytical modelling of adaptive optics systems: Role of the influence function}

\author{Anthony Berdeu\inst{1,2,4}
	\and
	Michel Tallon\inst{3}
	\and
	\'{E}ric Thiébaut\inst{3}
	\and
	Maud Langlois\inst{3}
}

\institute{
	National Astronomical Research Institute of Thailand, Center for Optics and Photonics, 260 Moo 4, T. Donkaew, A. Maerim, Chiang Mai 50180, Thailand
	\and
	Department of Physics, Faculty of Science, Chulalongkorn University,
254 Phayathai Road, Pathumwan, Bangkok 10330, Thailand
	\and
	Université de Lyon, Université Lyon1, ENS de Lyon, CNRS, Centre de Recherche Astrophysique de Lyon UMR 5574,
F-69230, Saint-Genis-Laval, France
	\and
	LESIA, Observatoire de Paris, Université PSL, Sorbonne Université, Université Paris Cité, CNRS, 5 place Jules Janssen, 92195 Meudon, France
	\\
	\email{anthony.berdeu@obspm.fr}
}

\date{Received 29 October 2022 / Accepted 26 February 2023}

 
\abstract
{
Adaptive optics (AO) is now a tool commonly deployed in astronomy. The real time correction of the atmospheric turbulence that AO enables allows telescopes to perform close to the diffraction limit at the core of their point spread function (PSF). Among other factors, AO-corrected PSFs depend on the ability of the wavefront corrector (WFC), generally a deformable mirror, to fit the incident wavefront corrugations.
}
{
In this work, we focus on this error introduced by the WFC, the so-called fitting error. To date, analytical models only depend on the WFC cut-off frequency, and Monte Carlo simulations are the only solution for studying the impact of the WFC influence function shape on the AO-corrected PSF. We aim to develop an analytical model accounting for the influence function shape.
}
{
We first obtain a general analytical model of the fitting error structure function. With additional hypotheses, we then derive an analytical model of the AO-corrected power spectral density (PSD). These two analytical solutions are compared with Monte Carlo simulations on different ideal profiles (piston, pyramid, Gaussian) as well as with real hardware (DM192 from ALPAO).
}
{Our analytical predictions show a very good agreement with the Monte Carlo simulations. We show that in the image plane, the depth of the correction as well as the transition profile between the AO-corrected area and the remaining turbulent halo depend on the influence functions of the WFC. We also show that the generally assumed hypothesis of stationarity of the AO correction is actually not met.
}
{
As the fitting error is the intrinsic optimal limit of an AO system, our analytical model allows for the assessment of the theoretical limits of extreme AO systems limited by the WFC  in high-contrast imaging through a context where other errors become comparable.
}


\maketitle
%

\let\oldpageref\pageref
\renewcommand{\pageref}{\oldpageref*}

\section{Introduction}

\newcommand{\EmphPar}[1]{}

\EmphPar{Adaptive optics systems} Since first being introduced into astronomy in the 1990s, adaptive optics (AO) systems have become a standard tool of big telescopes in ground-based observatories~\citep{Tyson:15_principles_of_AO}. Via real time compensation of atmospheric turbulence, the core of the point spread function (PSF) can be corrected close to the diffraction limit of an instrument. During the design phase of an instrument (to predict its performance) or when in operation (to improve the data reduction algorithms and push  the instrument's limits further), it becomes increasingly critical to accurately predict~\citep{Fusco:06_SPHERE_design}, model~\citep{Veran:97_PSF_AO_telemetry}, and reconstruct~\citep{Beltramo:20_PSF_reconstruction_Review} the PSF after correction by an AO system (\eg deconvolution for imaging, astrometry for stellar population, contrast analysis for high-contrast imaging). If the shape of the corrected area as well as the halo of the turbulence residuals beyond it naturally depend on the turbulence and the performances of the AO control loop, they also depend on the intrinsic limits given by the AO system's optical design and the properties of its optical elements, such as the profile shape of the wavefront corrector's (WFC) influence functions, which is in charge of correcting the wavefront~\citep{Hudgin:77}.

\EmphPar{Fitting error} In the following, we focus on the influence function shapes. They are directly linked with the fitting error by way of the fitting residuals after the optimal correction by the WFC of the AO system. This optimal correction is defined as the closest shape to the incident wavefront aberrations that the WFC can apply according to the least squares norm, and it is equivalent to minimising the residuals' variance in the instrument pupil~\citep{Hudgin:77, Ellerbroek:05_PSD_AO_model} and maximising the Strehl ratio~\citep{Herrmann_1992}. Thus, by definition, the fitting error is the intrinsic optimal limit of an AO system and the best correction that the AO system can apply on the incident wavefront. For extreme AO systems (XAOs) dedicated to high-contrast imaging, having an accurate model of the fitting error is of prime interest since all the other error terms \citep[\eg aliasing, WFC noise, servo-lag, ][]{Rigaut:98_SH_error} become comparable, as the performances of coronagraph instruments under good observing conditions are then limited by the residual non-common path aberration speckle and photon noises.

\EmphPar{Temporal statistic} In the photon-starved context of astronomy, observations imply long exposures that average the temporal evolution of the turbulence. Thus, one needs to model the long-exposure PSF or, equivalently in the frequency domain, the long-exposure optical transfer function (OTF) of the telescope from the knowledge of the atmospheric turbulence statistic~\citep{Roddier:81}. In the spatial domain, this statistic is given by the structure function of the wavefront. When the statistic is stationary, that is to say, when it does not depend on the position in the instrument's pupil, the statistic can be equivalently described in the frequency domain via its power spectrum density (PSD). Different solutions exist to model the statistic of the fitting residuals after the AO correction as well as its impact on the long-exposure PSF.

\EmphPar{Modal correction} One approach to model such data is to consider that the WFC perfectly compensates some given modes of the incident wavefront on the full pupil, generally through Zernike or Karhunen-Loève polynomials. This can either be done on the structure function~\citep{Conan:94_PhD,Dai:95_Modal_Zernike_Karhunen-Loeve} or on the PSD~\citep{Noll:76_Zernike, Sasiela:94, Stone:94_Anisoplanatism}. These methods are convenient since they carry a lot of physics information and most of the aberrations in optical systems are generally described on such modal bases (including piston, tip and tilt, defocus, and astigmatism). But such bases do not accurately model most of the real WFCs. Indeed, in astronomy most of the  WFCs are deformable mirrors (DM) controlled with actuators on a square or hexagonal grid~\citep{Tyson:15_principles_of_AO}. By construction, the WFCs are zonal correctors and cannot perfectly compensate a given global mode.

\EmphPar{Binary PSD filtering} Another approach uses the fact that the discrete layout of the actuators of 
a WFC is associated with a spatial cut-off frequency.

Some authors have already introduced a full analytical model that depends on the spatial properties of the WFC, such as~\cite{Veran:97_PSF_AO_telemetry} and~\cite{Tokovinin:00_MCAO_isoplanatism}. But when it comes to the numerical applications, the impact of the WFC on the PSD is reduced to a binary filter where the WFC corrects all the frequencies below its cut-off limit. This assumption is common to all the analytical models~\citep{Ellerbroek:05_PSD_AO_model, Jolissaint:06_AO_analytical, Correia:14_anti-aliasing}.

These models based on modal correction or binary PSD filtering can be used for design and performance predictions \citep{Fusco:06_SPHERE_design, Dohlen:16_SPHERE_performance_vs_prediction}. They can also be part of the instrumental modelling~\citep{Flicker:08_Keck_AO, Sauvage:10_AO_long_exp_perfect_coro, Herscovici:17_coronagraph_PSF_analytical} or non-common path aberration correction between the wavefront sensor channel and the scientific channel~\citep{Paul:14_COFFEE_SPHERE, Herscovici:19_NCPA_correction_coronagraph_analytic}. Nonetheless, a common feature of these analytical models is that they all assume the stationarity of the incident wavefront statistics after an AO system. This assumption has been mentioned by \cite{Conan:94_PhD} and \cite{Dai:95_Modal_Zernike_Karhunen-Loeve} in the case of modal corrections (Zernike or Karhunen-Loève polynomials) mainly impacting the high spatial frequencies, as shown by \cite{Veran:97_PSF_AO_telemetry}. But for most of the applications that imply zonal correctors, the impact on the wavefront statistics stationarity requires more thorough analyses.

\EmphPar{Monte Carlo simulations} To date, the only means to account for the shape of the influence functions in the fitting error (and potentially the non-stationarity of the fitting residuals) is to run Monte Carlo simulations. Many codes are available for this purpose, such as YAO~\citep[Yorick Adaptive Optics;][]{Rigaut:13_YAO}, OOMAO~\citep[Object-Oriented Matlab Adaptive
Optics;][]{Conan:14_OOMAO}, COMPASS~\citep[COMputing Platform for Adaptive opticS Systems, C++/GPU with Python interface;][]{Gratadour:14_COMPASS}, HCIPy~\citep[High Contrast Imaging for Python;][]{Por:18_HCIPy}, AOtools~\citep[Python;][]{Townson:19_AOtools}, and SOAPY~\citep[Python;][]{Reeves_16:Soapy}. In such end-to-end codes, it is possible to easily change the influence function shapes, as was done by~\cite{Flicker:08_Keck_AO} for the Keck observatory for example. The PSD is numerically estimated for a given set of influence functions (implying the stationarity) and scaled as necessary according to   the turbulence strength. If the Monte Carlo simulations are extremely versatile, the computational burden can be very high, as many short exposure frames must then be simulated close to the turbulence temporal evolution to be stacked in order to mimic a long exposure.

\EmphPar{A new analytical model} In this work, our aim is to develop an analytical model accounting for the influence function shapes in the fitting residuals in order to compute the AO-corrected PSF without the need of Monte Carlo simulations. We note here that the notations used throughout this paper are introduced in~\refapp{app:notations}. To develop our analytical model, we introduce the basic equations of turbulence correction through a perfect AO system in \refsec{sec:LE_PSF}. Focusing on the fitting error, no other error term is added (\eg correction delay, anisoplanetism, noise in the wavefront sensor measurements, noise propagation to the command estimation). In \refsec{sec:SF_analytical}, we present a general analytical model to predict the non-stationary structure function of the fitting residuals for a given set of influence functions. Enforcing the stationarity and with additional hypotheses, we also present an analytical model of the AO-corrected PSD in \refsec{sec:PSD_analytical}. The mathematical details on how to obtain these analytical models are given in Appendices~\ref{app:proj_opti} to~\ref{app:ortho_IF}. In \refsec{sec:num_implementation}, we describe how to numerically implement the analytical PSD model and how the computational burden can be reduced. We also detail how to account for the correction of the tip and tilt modes by use of a dedicated mirror in addition to the WFC. \refapp{app:SF} gives further details on how to numerically estimate the structure function and the PSD from a given set of simulated random wavefronts. In \refsec{sec:results}, we compare our analytical structure function and PSD models to Monte Carlo simulations performed on different ideal profiles often used in the literature (local piston, pyramid, Gaussian) as well as on a realistic case (DM192 from ALPAO). The structure function's inhomogeneities and their periodicity are further discussed in \refapp{app:SF_periodic}. \refapp{app:add_materials} presents additional materials to further expand the results of our simulations.

\section{Long-exposure point spread function and structure function}
\label{sec:LE_PSF}

This section summarises the main equations already gathered by~\cite{Roddier:81}. In this work, we use the common assumption that the turbulence and the AO correction only induce a pure temporally dependent phase~$\pha\Paren{\Vx,t}$ in the pupil~$\pupil\Paren{\Vx}$ of the instrument (there is no scintillation). For non-dimensional reduced variables, as defined by~\cite{Roddier:81}, the instantaneous PSF~$\PSF\Paren{\Vk,t}$ in the focal plane of the instrument is given, except for a normalisation factor, by the square modulus of its Fourier transform
\begin{equation}
	\label{eq:PSF_short}
	\PSF\Paren{\Vk,t} \triangleq \Abs{\FTfull{\pupil\Paren{\Vx}\E{\I\pha\Paren{\Vx,t}}}{\Vk}}^{2}
	\,,
\end{equation}
whose Fourier transform is classically given by the autocorrelation of the complex amplitude in the pupil plane
\begin{equation}
	\label{eq:FTPSF_short}
	\FTPSF\Paren{\Vx,t} = \int\pupil\Paren{\Vxp}\conj{\pupil}\Paren{\Vxp+\Vx}\E{\I\Paren{\pha\Paren{\Vxp,t}-\pha\Paren{\Vxp+\Vx,t}}}\Dxp
	\,.
\end{equation}The long-exposure PSF~$\PSF\Paren{\Vk}$ is given by the temporal expectation of \eq{eq:FTPSF_short}
\begin{align}
	\FTPSF\Paren{\Vx} {}={} & \AvgT{\FTPSF\Paren{\Vx,t}}
	\\
	\label{eq:FTlongPSF_pha}
	{}={} & \int\pupil\Paren{\Vxp}\conj{\pupil}\Paren{\Vxp+\Vx}\AvgT{\E{\I\Paren{\pha\Paren{\Vxp,t}-\pha\Paren{\Vxp+\Vx,t}}}}\Dxp
	\\
	\label{eq:FTlongPSF_SF}
	{}={} & \int\pupil\Paren{\Vxp}\conj{\pupil}\Paren{\Vxp+\Vx}\E{-\frac{1}{2}\SF{\pha}\Paren{\Vxp,\Vxp+\Vx}}\Dxp
	\,,
\end{align}
where
\begin{equation} 
	\label{eq:SF_def}
	\SF{\pha}\Paren{\Vx,\Vxp} \triangleq \AvgT{\Paren{\pha\Paren{\Vx,t} - \pha\Paren{\Vxp,t}}^{2}}
	\,,
\end{equation}
is the structure function of~$\pha\Paren{\Vx}$. The relation between \eq{eq:FTlongPSF_pha} and \eq{eq:FTlongPSF_SF} is detailed by \cite{Roddier:81}.

We note that if the structure function is stationary, such as for a turbulence with a Kolomogorov statistic~\citep{Roddier:81}, we will obtain the classical results that
\begin{equation}
	\label{eq:SF_stat}
	\SF{\pha}\Paren{\Vx,\Vxp}
	= \SF{\pha}\Paren{\Vxp-\Vx}
	= \SF{\pha}\Paren{\Vx-\Vxp}
	\,,
\end{equation}
and
\begin{equation}
	\label{eq:FTlongPSF_SF_stat}
	\FTPSF\Paren{\Vx} =\FTPSFturb\Paren{\Vx} \FTPSFtel\Paren{\Vx}
	\text{ with }
	\begin{cases}
		\FTPSFturb\Paren{\Vx} \triangleq \E{-\frac{1}{2}\SF{\pha}\Paren{\Vx}}
		\\
		\FTPSFtel\Paren{\Vx} \triangleq  \int\pupil\Paren{\Vxp}\conj{\pupil}\Paren{\Vxp+\Vx}\Dxp
	\end{cases}
	\,.
\end{equation}
This expresses the fact that the long-exposure PSF is the convolution of the diffraction-limited PSF of the telescope~$\PSFtel$ with the equivalent PSF of the incident turbulence wavefront~$\PSFturb$.

\section{Analytical structure function of the fitting residuals}
\label{sec:SF_analytical}

In this work, we only focus on the fitting error of the WFC, which is generally a DM. No additional term, such as the wavefront sensor aliasing term, the noise of the measurements, the delay of the command, or the wavefront reconstruction accuracy are added~\citep{Jolissaint:06_AO_analytical}.

For a given phase screen~$\pha\Paren{\Vx,t}$ at time~$t$, we note
\begin{equation} 
	\label{eq:phac}
	\phac\Paren{\Vx,t} \triangleq \sum_{a\in\setactu} \ca\Paren{t}\IFa\Paren{\Vx}
	\,,
\end{equation}
the correction applied with the WFC where~$\setactu$ is the set of actuators and~$\IFa\Paren{\Vx}$ is the influence function of the actuator~$a$. As the PSF is insensitive to the phase piston on the support~$\Pupil$ of the pupil, all the influence functions were assumed to be piston-free:
\begin{equation} 
	\label{eq:piston_free}
	\forall a\in\setactu, \int \IFa\Paren{\Vx}\Dx = 0 
	\,.
\end{equation}
the optimal command~$\V{c}\Paren{t}$ that minimises the variance of the fitting residuals
\begin{equation} 
	\label{eq:pha_AO}
	\phaAO\Paren{\Vx,t} = \pha\Paren{\Vx,t}-\phac\Paren{\Vx,t}
	\,,
\end{equation}
is given by
\begin{equation} 
	\label{eq:c_opti}
	\forall a \in \setactu, \ca\Paren{t} \triangleq \int \proja\Paren{\Vx}\pha\Paren{\Vx,t}\Dx
	\,,
\end{equation}
where~$\proja\Paren{\Vx}$ is the optimal projector of the wavefront~$\pha\Paren{\Vx,t}$ on the actuator~$a$, as defined in \refapp{app:proj_opti}. We show in \refapp{app:SF_analytical} that the analytical expression of the structure function~$\SFAO\Paren{\Vx,\Vxp}$ of these fitting residuals can then be written
as\begin{equation} 
	\label{eq:SF_analytic}
	\begin{aligned}
		\SFAO\Paren{\Vx,\Vxp} {}={} & \SFpha\Paren{\Vx,\Vxp}
		\\
		&
		+ \sum_{a\in\setactu} \Paren{\IFa\Paren{\Vx}-\IFa\Paren{\Vxp}}
		\\
		& \quad \times 
		\int\proja\Paren{\Vu}\Paren{\SFpha\Paren{\Vu,\Vx}-\SFpha\Paren{\Vu,\Vxp}}\Du
		\\
		& -\frac{1}{2}\sum_{a \in \setactu}\sum_{\ap \in \setactu}
		\Daap
		\Paren{\IFa\Paren{\Vx}-\IFa\Paren{\Vxp}}
		\\
		& \quad \times 
		\Paren{\IFap\Paren{\Vx}-\IFap\Paren{\Vxp}}
		\,,
	\end{aligned}
\end{equation}
with
\begin{equation}
	\label{eq:Daap}
	\Daap \triangleq \iint \SFpha\Paren{\Vu,\Vup}\proja\Paren{\Vu}\projap\Paren{\Vup}\Du\Dup
	\,.
\end{equation}
We note here that for a turbulence with a stationary structure function (see \eq{eq:SF_stat}), the projection of~$\SFpha$ via~$\proja$ is a convolution
\begin{equation} 
	\label{eq:D_conv}
	\int\proja\Paren{\Vu}\SFpha\Paren{\Vu,\Vx}\Du
	= \int\proja\Paren{\Vu}\SFpha\Paren{\Vx-\Vu}\Du
	\,,
\end{equation}
which can be numerically computed in the frequency domain using discrete Fourier transform (DFT).
 
\section{Analytical power spectrum density of the fitting residuals}
\label{sec:PSD_analytical}

From \eq{eq:SF_analytic}, it is clear that the structure function of the fitting residuals is not stationary
\begin{equation}
	\SFAO\Paren{\Vx,\Vxp} \neq \SFAO\Paren{\Vx-\Vxp}
	\,.
\end{equation}
Nonetheless, making this assumption is sometimes necessary in order to push the analytical formulations forwards, such as when obtaining the PSD of the AO-corrected turbulence~\citep{Jolissaint:06_AO_analytical} or the long-exposure PSF of a coronagraph system~\citep{Herscovici:17_coronagraph_PSF_analytical}. Imposing stationarity implies, at least, that (1) the position~$\Pa$ of the actuators is located on a regular Cartesian grid of pitch~$\Vpitch\in\Reals^{2}$
and (2) all the influence functions are identical to a translated reference influence function, noted~$\IFref\Paren{\Vx}$ in the following such that $\IFa\paren{\Vx} = \IFref\Paren{\Vx - \Pa}$.
In this work, we study~$\PSDres\Paren{\Vk} = \AvgT{\Abs{\FTphaAO\Paren{\Vk,t}}^{2}}$, the PSD of the AO fitting residuals~$\phaAO\Paren{\Vx,t}$, knowing~$\PSD{\pha}\Paren{\Vk}=\AvgT{\Abs{\FT{\pha}\Paren{\Vk,t}}^{2}}$, and the PSD of the incident wavefront phase~$\pha\Paren{\Vx,t}$.

In the literature~\citep[see][]{Veran:97_PSF_AO_telemetry, Tokovinin:00_MCAO_isoplanatism, Ellerbroek:05_PSD_AO_model, Jolissaint:06_AO_analytical, Correia:14_anti-aliasing}, a binary high-pass filter is assumed
\begin{equation}
	\PSDres\Paren{\Vk} = \Paren{1-\fLF\Paren{\Vk}}\PSD{\pha}\Paren{\Vk}
	\,,
\end{equation}
with
\begin{equation} 
	\label{eq:binary_mask}
	\fLF\Paren{\Vk} \triangleq 
	\begin{cases}
		1 \text{ if } \forall i\in\Brace{1,2}, \,
		\Abs{k_{i}}<\Paren{2\pitch_{i}}^{-1}
		\\
		0 \text{ otherwise}     
		\end{cases}
	\hspace{-0.5cm}
	\,,
\end{equation}
which implies a perfect correction below the cut-off frequency of the WFC and no correction beyond. This approximation is based on the fact that the WFC cannot compensate for spatial scales smaller than the pitch of its actuators. The aim of this paper is to obtain and study a finer analytical expression of $\PSDres\Paren{\Vk}$ that depends on the reference influence function shape~$\IFref\Paren{\Vx}$.

We show in \refapp{app:PSD_analytical} that this expression can be written
as
\begin{equation} 
	\label{eq:PSDres}
	\begin{aligned}
	\PSDres\Paren{\Vk} {}={} &
	\Paren{1 - 2 \ntotactu\PSDperp\Paren{\Vk}}\PSD{\pha}\Paren{\Vk} 
	+
	\sum_{a\in\setactu} \sum_{\ap\in\setactu} \PSDperp\Paren{\Vk}\E{-2\I\pi\Paren{\Pa-\Pap}\T\Vk}
	\\
	& \quad \times \int{\PSD{\pha}\Paren{\Vkp}\PSDperp\Paren{\Vkp}\E{-2\I\pi\Paren{\Pa-\Pap}\T\Vkp}\Dkp}
	\,,
	\end{aligned}
\end{equation}
where~$\ntotactu$ is the total number of actuators and~$\PSDperp\Paren{\Vk}$ is the PSD of~$\IFperp\Paren{\Vx}$, which is the orthonormalised actuator shape obtained from~$\IFref\Paren{\Vx}$ as described in \refapp{app:ortho_IF}. This 2D expression is further studied in the following sections.

One can also discuss the integral term of \eq{eq:PSDres} that gives the expected variance of the fitting residuals, that is, the `fitting error'. Indeed, using the Parseval–Plancherel identity and the linearity of the integral to commute the temporal average,
\begin{equation}
	\label{eq:var_PSD}
	\AvgT{\sigAO^{2}}
	{}={}
	\AvgT{\int{\phaAO^{2}\Paren{\Vx,t}\Dx}}
	{}={}
	\AvgT{\int{\Abs{\FTphaAO\Paren{\Vk,t}}^{2}\Dk}}
	{}={}
	\int{\PSDres\Paren{\Vk}\Dk}
	\,.
\end{equation}
With the notations of \refapp{app:PSD_analytical} and as~$\IFperpap\Paren{\Vx}\in\Reals$,
\begin{align}
	\int{\PSDperp\Paren{\Vk}\E{-2\I\pi\Paren{\Pa-\Pap}\T\Vk}\Dk}
	& {}={}
	\int{\FTIFperpa\Paren{\Vk}\conj{\FTIFperpap}\Paren{\Vk}\Dk}
	\\
	& {}={}
	\int{\IFperpa\Paren{\Vx}\conj{\IFperpap}\Paren{\Vx}\Dx}
	{}={} \delta_{\Va,\Vap}
	\,.
\end{align}
Then, from \eq{eq:PSDres} comes
\begin{equation} 
	\label{eq:sigAO}
	\AvgT{\sigAO^{2}}
	{}={}
	\int{\Paren{1 - \ntotactu\PSDperp\Paren{\Vk}}\PSD{\pha}\Paren{\Vk}\Dk}
	\,.
\end{equation}
This equation is similar to the classical subtraction of the Zernike or Karhunen-Loève modes, such as piston and tip-tilt~\citep[see][]{Noll:76_Zernike, Sasiela:94, Stone:94_Anisoplanatism, Dai:95_Modal_Zernike_Karhunen-Loeve}, used to filter the wavefront PSD, but here they are applied to the WFC modes that are driven by~$\IFperp\Paren{\Vx}$. Comparing \eq{eq:binary_mask} and \eq{eq:sigAO}, we can conclude that an influence function shape that filters out all the energy carried by the frequencies~$\forall i\in\Brace{1,2} ,\, \Abs{k_{i}}<\Paren{2\pitch_{i}}^{-1}$ but leaves the other frequencies untouched must satisfy
\begin{equation} 
	\label{eq:prof_SINC}
	\ntotactu\PSDperp\Paren{\Vk} = \fLF\Paren{\Vk}
	\,,
\end{equation}
which is obtained, for example, with the 2D~$\sinc$ function
\begin{equation}
	\IFperp\Paren{\Vx} \propto \sinc\Paren{\pi\frac{x_{1}}{\Delta_{1}}}\times\sinc\Paren{\pi\frac{x_{2}}{\Delta_{2}}}
	\,.
\end{equation}
We note here that this conclusion has already been mentioned by~\cite{Ellerbroek:05_PSD_AO_model}.

\section{Numerical implementation}
\label{sec:num_implementation}

In this section, we discuss the numerical implementation of \eq{eq:PSDres}. This process implies additional assumptions.

\subsection{Domain periodisation}
\label{sec:Domain_periodisation}

\begin{figure}[t!] 
	\centering
	
	\newcommand{\LineRatio}{1}
	
	\newcommand{\FigOne}{figures_actu_pairing_map_W.pdf}
	\newcommand{\FigTwo}{figures_actu_pairing_map_n_pair.pdf}       
	\newcommand{\subfigColor}{white}	
	
	\sbox1{\includegraphics{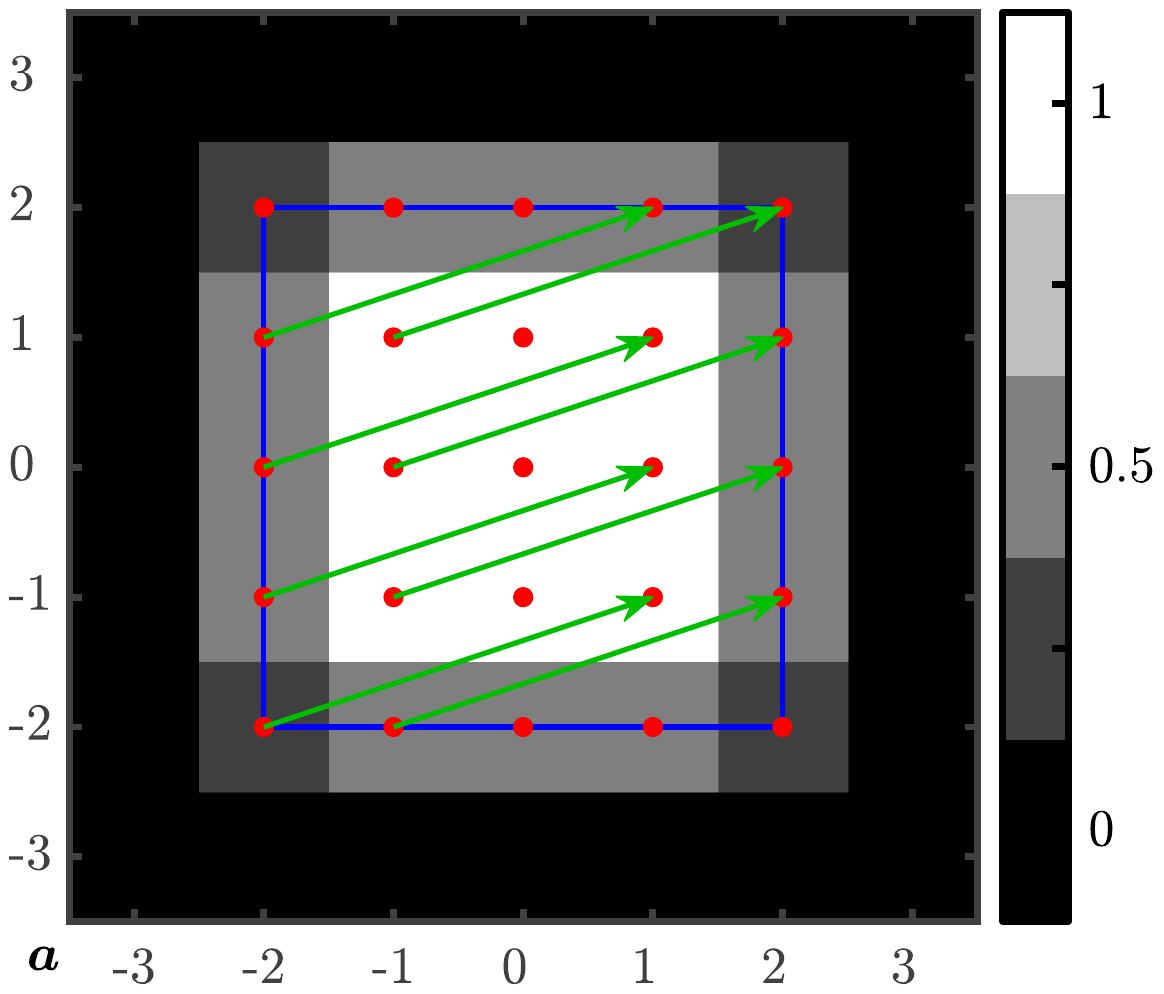}}	       
	\sbox2{\includegraphics{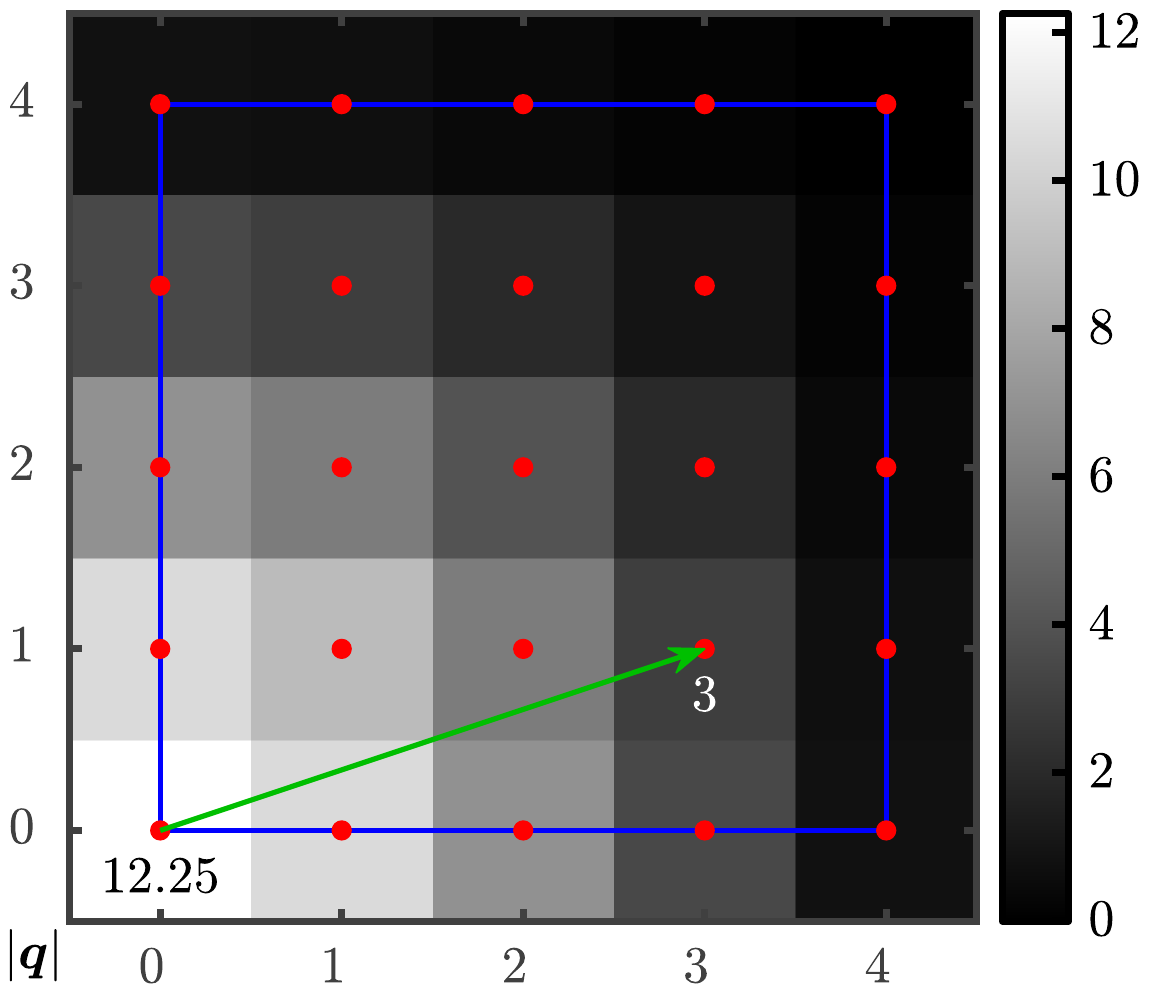}}       
	\newcommand{\ColumnWidth}[1]
		{\dimexpr \LineRatio \linewidth * \AspectRatio{#1} / (\AspectRatio{1} + \AspectRatio{2}) \relax
		}
	\newcommand{\ColumnGap}{\hspace {\dimexpr \linewidth /3 - \LineRatio\linewidth /3 }}

	\begin{tabular}{
		@{\ColumnGap}
		m{\ColumnWidth{1}}
		@{\ColumnGap}
		m{\ColumnWidth{2}}
		@{\ColumnGap}
		}
		\subfigimg[width=\linewidth,pos=ul,font=\fontfig{\subfigColor}]{$\quad$(a)}{0.0}{\FigOne} &
		\subfigimg[width=\linewidth,pos=ul,font=\fontfig{\subfigColor}]{$\quad$(b)}{0.0}{\FigTwo}
	\end{tabular}
	\caption{\label{fig:W_map} Pairing the actuators on a periodic mesh for~$\nactu = 5$. Panel~(a):~Map of the weight~$\wa$ given to each actuator. The blue frame is the simulated domain. The red dots are the actuator positions. The green arrows emphasise the different possible pairings for the spacing~$\pair=\Paren{3,1}$. Panel~(b):~Map of the weighted number of pairs~$\nwpair$ according to the interactuator spacing~$\pair$. The blue frame encompasses all the possible positive pairings~$\Abs{\pair}$ on the domain. The green arrow points towards the weighted number of pairs~$\nw\Paren{\pair = \Paren{3,1}}$.}
\end{figure}

Simulations were restrained to a pixelated domain with a finite size and number of actuators. In addition, the use of DFT algorithms implied periodic boundary conditions. To respect this periodicity and avoid any boundary issue, the number of actuators across the simulated domain diameter must be an integer~$\nactu \in \PosIntegers$, as mentioned by~\cite{Flicker:08_Keck_AO}. In this work, we use the convention presented in \subfig{fig:W_map}{a}, where an actuator lies exactly on the domain edge. As a consequence, the actuators had to be weighted according to their contribution to the domain, as emphasised in the figure: $\wa=1$ if the actuator is inside the domain, $\wa=\nicefrac{1}{2}$ if the actuator is on an edge, $\wa=\nicefrac{1}{4}$ if the actuator is on a corner, and $\wa=0$ if the actuator is outside the domain.

Looking back at \eq{eq:PSDres}, it consequently follows that the total number of actuators in the domain is
\begin{equation}
	\ntotactu = \sum_{a\in\setactu} \wa = \Paren{\nactu-1}^{2}
	\,.
\end{equation}
In addition, the double sum in \eq{eq:PSDres} had to be weighted accordingly
\begin{equation}
	\begin{aligned}
	\Sigma {}\triangleq{} \sum_{a\in\setactu} \sum_{\ap\in\setactu} & \wa \wap \PSDperp\Paren{\Vk}\E{-2\I\pi\Paren{\Pa-\Pap}\T\Vk}
	\\
	& \times \Paren{\int{\PSD{\pha}\Paren{\Vkp}\PSDperp\Paren{\Vkp}\E{-2\I\pi\Paren{\Pa-\Pap}\T\Vkp}\Dkp}}
	\,,
	\end{aligned}
\end{equation}
which scales as~$\ScaleAs\Paren{\nactu^{4}}$. Since the summed terms only depend on the actuator spacing~$\Pap-\Pa = \Vpitch\EWprod\pair$ with~$\pair\in\Integers^{2}$, the computational burden can be decreased by counting the weighted number of actuator pairs~$\nwpair$ in the domain. Noting~$\weight_{\pair}$ the zero-padded 2D map of the weights according to the position on the Cartesian grid~$\pair$, shown in \subfig{fig:W_map}{a}, $\nwpair$ can be computed using DFT, as shown in \subfig{fig:W_map}{b},
\begin{align} 
	\label{eq:n_DFT}
	\nwpair {}={} & \sum_{
		\left\lbrace
		\substack{\pairOne,\pairTwo
		\\ \pair = \pairTwo-\pairOne
		}
		\right.
	} \wqOne \wqTwo
	= \sum_{\pairOne} \wqOne \weight_{\pair+\pairOne}
	\\
	{}={} & \iDFT{\conj{\DFT{\wqOne}}\DFT{\wqOne}}\Paren{\pair}
	\\
	{}={} & \iDFT{\Abs{\DFT{\wqOne}}^{2}}\Paren{\pair}
	\,,
\end{align}
leading to
\begin{equation}
	\begin{aligned}
	\Sigma {}={} \sum_{\pair\in\Integers^{2}} & \nwpair 
	\PSDperp\Paren{\Vk}\E{-2\I\pi\Paren{\Vpitch\EWprod\pair}\T\Vk}
	\\
	& \times \Paren{\int{\PSD{\pha}\Paren{\Vkp}\PSDperp\Paren{\Vkp}\E{-2\I\pi\Paren{\Vpitch\EWprod\pair}\T\Vkp}\Dkp}}
	\,,
	\end{aligned}
\end{equation}
which scales as~$\ScaleAs\Paren{\nactu^{2}}$. We note here that the number of needed computations could be further reduced by playing with the domain and function symmetries, such as~$\nwpair = \nw\Paren{\Abs{\pair}}$.

We emphasise that the actuator weighting is essential. For example, with~$\nactu = 5$ as presented in \fig{fig:W_map}, it seems that the domain encompasses eight pairs~$\pair = \Paren{3,1}$ in \subfig{fig:W_map}{a}. But accounting for the weighting, there are only three pairs in total, as emphasised in \subfig{fig:W_map}{b}. Similarly for the actuators paired with themselves, the total weighted count gives~$12.25 < \nactu^{2} = 25$.

\subsection{Piston and tip-tilt modes removal}
\label{sec:piston_TT}

The piston mode must be removed from the incident wavefront $\pha\Paren{\Vx,t}$ since the final image is insensitive to this mode. Tip and tilt modes also need to be removed if they are independently corrected by a fast steering mirror. This section explains how the PSD $\PSD{\pha}\Paren{\Vk}$ is modified by these operations.

As before, we state that~$\pupil$ is the pupil mask of the telescope aperture. In the following, this pupil is circular. Working on a grid for the simulation discretised on pixel~$\Vpix=\Paren{\pix_{x}, \pix_{y}}$, the pupil is apodised as described by \cite{Berdeu:22_SPIE_EvWaCo_AO_loop}:
\begin{equation} 
	\label{eq:pupil}
	\pupil\Paren{\Vpix} = \text{min}\Paren{1, \text{max}\Paren{0, \nicefrac{1}{2} - d_{\Tag{P}}\Paren{\Vpix}/\D{\pix}}}
	\,,
\end{equation}
where~$\Vpix$ is the pixel position in the aperture plane, $\D{\pix}$ is the pixel pitch of the simulation, and~$d_{\Tag{P}}$ is the relative distance of~$\Vpix$ compared to the pupil edge, which is positive if outside the aperture and negative if inside. In this work, all the numerical integrals on the pupil are weighted by this mask.

As mentioned by~\cite{Jolissaint:06_AO_analytical}, the system is not sensitive to the piston on the aperture, and its contribution should be removed from the PSD. This mode is written as follows:
\begin{equation} 
	\label{eq:mPiston}
	\mPiston\Paren{\Vpix} = \nPiston\pupil\Paren{\Vpix}
	\,,
\end{equation}
where~$\nPiston$ is a normalisation factor that ensures that the modal basis is normalised $\sum_{\Vpix} \mPiston^{2}\Paren{\Vpix} = 1$. Thus,
\begin{equation} 
	\label{eq:PSDpiston}
	\PSDpiston\Paren{\Vpixk} = \Abs{\DFT{\mPiston\Paren{\Vpix}}}^{2}
	\,,
\end{equation}
where~$\Vpixk$ is the position in the pixelated frequency domain. The PSD of the incident wavefront $\PSD{\pha}$ is thus filtered as follows:
\begin{equation} 
	\label{eq:PSDpiston_rmv}
	\PSD{\pha}\Paren{\Vpixk} \leftarrow \PSD{\pha}\Paren{\Vpixk}\times\Paren{1-\PSDpiston\Paren{\Vpixk}}
	\,.
\end{equation}

As mentioned by~\cite{Sasiela:94}, the tip and tilt modes may also have to be removed, for example, if a fast steering mirror is used to correct the tip-tilt and leaves higher modes to the WFC. These modes are written as
\begin{equation} 
	\label{eq:mTipTilt}
	\mTip\Paren{\Vpix} = \nTip \pix_{x} \pupil\Paren{\Vpix}
	\text{ and }
	\mTilt\Paren{\Vpix} = \nTilt \pix_{y} \pupil\Paren{\Vpix}
	\,,
\end{equation}
where~$\nTip$ and~$\nTilt$ are normalisation factors that ensure the modal basis is normalised $\sum_{\Vpix} \mTip^{2}\Paren{\Vpix} = \sum_{\Vpix} \mTilt^{2}\Paren{\Vpix} = 1$. And
\begin{equation} 
	\label{eq:PSDtt}
	\PSDtip\Paren{\Vpixk} = \Abs{\DFT{\mTip\Paren{\Vpix}}}^{2}
	\text{ and }
	\PSDtilt\Paren{\Vpixk} = \Abs{\DFT{\mTilt\Paren{\Vpix}}}^{2}
	\,.
\end{equation}
We note here that the frame origin is assumed to be at the centroid of the aperture so that these modes do not possess any piston contribution $\sum_{\Vpix} \mTip\Paren{\Vpix} = \sum_{\Vpix} \mTilt\Paren{\Vpix} = 0$, which ensures that the basis is orthogonal. The PSD of the incoming wavefront $\PSD{\pha}\Paren{\Vpixk}$ is then filtered as follows
\begin{equation} 
	\label{eq:PSDtt_rmv}
	\PSD{\pha}\Paren{\Vpixk} \leftarrow \PSD{\pha}\Paren{\Vpixk}\times\Paren{1-\PSDpiston\Paren{\Vpixk} - \PSDtip\Paren{\Vpixk} - \PSDtilt\Paren{\Vpixk}}
	\,.
\end{equation}

We note that obtaining \eq{eq:PSDres} relies on the assumption that the incident wavefront structure function is stationary, as described in \refapp{app:PSD_analytical} via \eq{eq:SF_var}. But strictly speaking, removing the piston and tip-tilt breaks this assumption by respectively breaking the stationarity of the expected wavefront variance $\AvgT{\pha^{2}\Paren{\V{x},t}}$ and of the structure function. Thus to be mathematically correct, the PSD of the incoming wavefront $\PSD{\pha}\Paren{\Vpixk}$ must first be filtered via \eq{eq:PSDres} before applying \eq{eq:PSDpiston_rmv} or  \eq{eq:PSDtt_rmv} on $\PSDres\Paren{\Vpixk}$. Since the imaging system and its PSF are insensitive to the wavefront piston, applying the piston filter after the AO system filter makes sense. This filtering sequence is less pertinent for the tip-tilt removal since it is equivalent to placing the fast steering mirror after the WFC, which greatly reduces its impact on the wavefront correction and limits the offloading of these modes from the WFC, as discussed in \refsec{sec:res_ALPAO}.

\section{Results}
\label{sec:results}

\subsection{Simulation of the turbulent phase screens}
\label{sec:res_Screen}

We tested our model on a classical Kolomogorov statistic for the turbulence~\citep{Jolissaint:06_AO_analytical}
\begin{equation}
	\PSD{\pha}\Paren{\Vpixk} \triangleq 0.023\rFried^{-5/3}\Norm{\Vk}^{-11/3}
	\Leftrightarrow
	\SF{\pha}\Paren{\Vx} \triangleq 6.88\Paren{\Norm{\Vx}/\rFried}^{5/3}
	\,.
\end{equation}
As described by~\cite{McGlamery:76_Turb_PSD}, $\nPha = 10000$ random phase screens were simulated in the frequency domain on a square grid with a side of~$\nPix = 129$ pixels. The low frequencies of the Kolomogorov spectrum were injected via the sub-harmonics method proposed by~\cite{Lane:92_Subharmonics}. Sixteen sub-harmonics were added, and as suggested by~\cite{Lane:92_Subharmonics}, the screens were simulated on a grid twice as big and only the central~$129 \times 129$ pixels were kept to avoid aliasing.

$\Paren{\Vx, \Vk}$ denote the values obtained according to the theoretical profiles, while $\Paren{\Vpix, \Vpixk}$ denote the values obtained through the simulation on a discrete grid. We emphasise here that the frequencies are given in terms of cycle per simulation diameter~$\Dsim^{-1}$.

\begin{figure}[t!] 
	\centering
	
	\newcommand{\LineRatio}{1}
	
	\newcommand{\FigOne}{figures_SF_no_AO_SF_prof.pdf}
	\newcommand{\FigTwo}{figures_SF_no_AO_PSD_map.pdf}
	
	\sbox1{\includegraphics{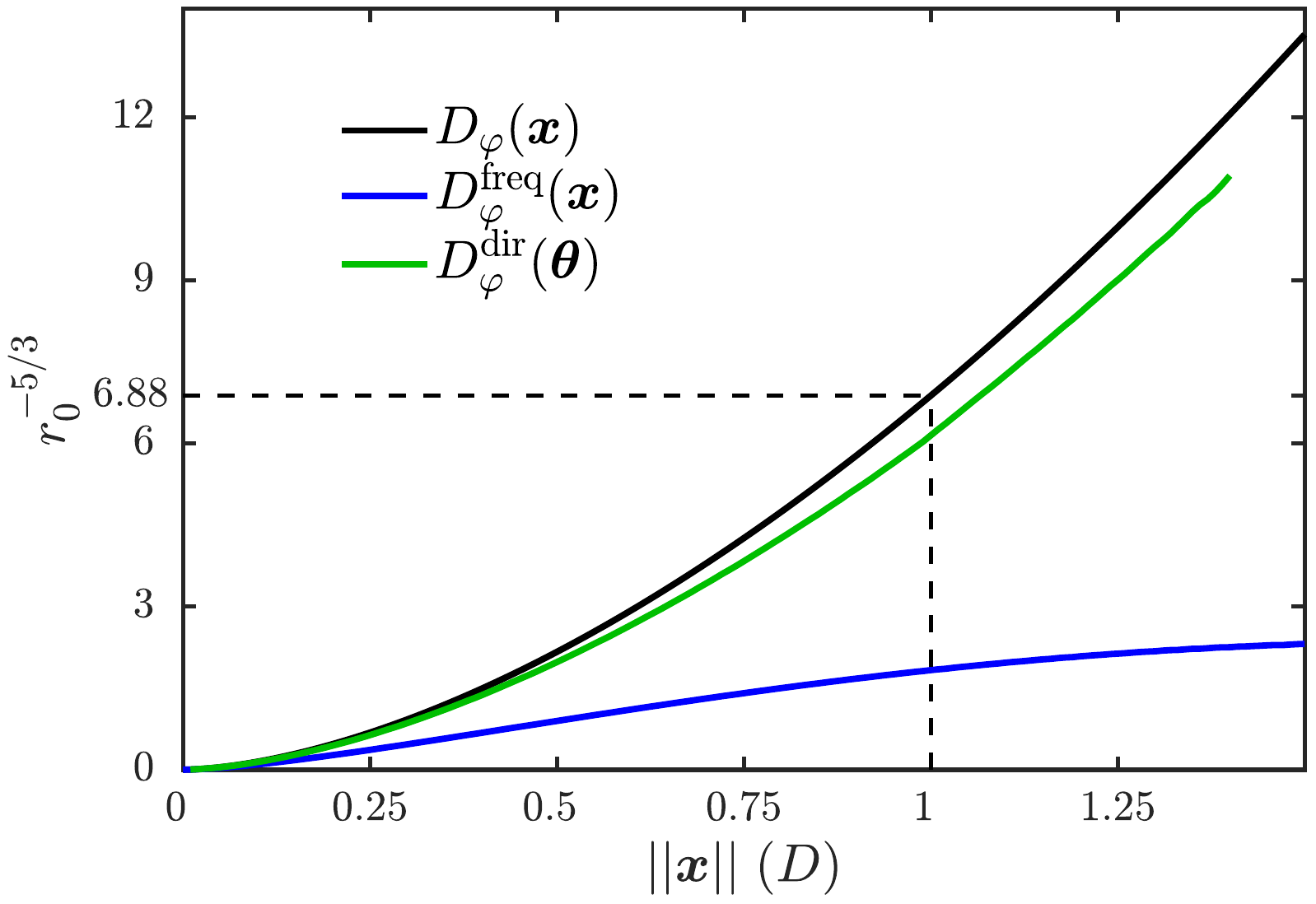}}	       
	\sbox2{\includegraphics{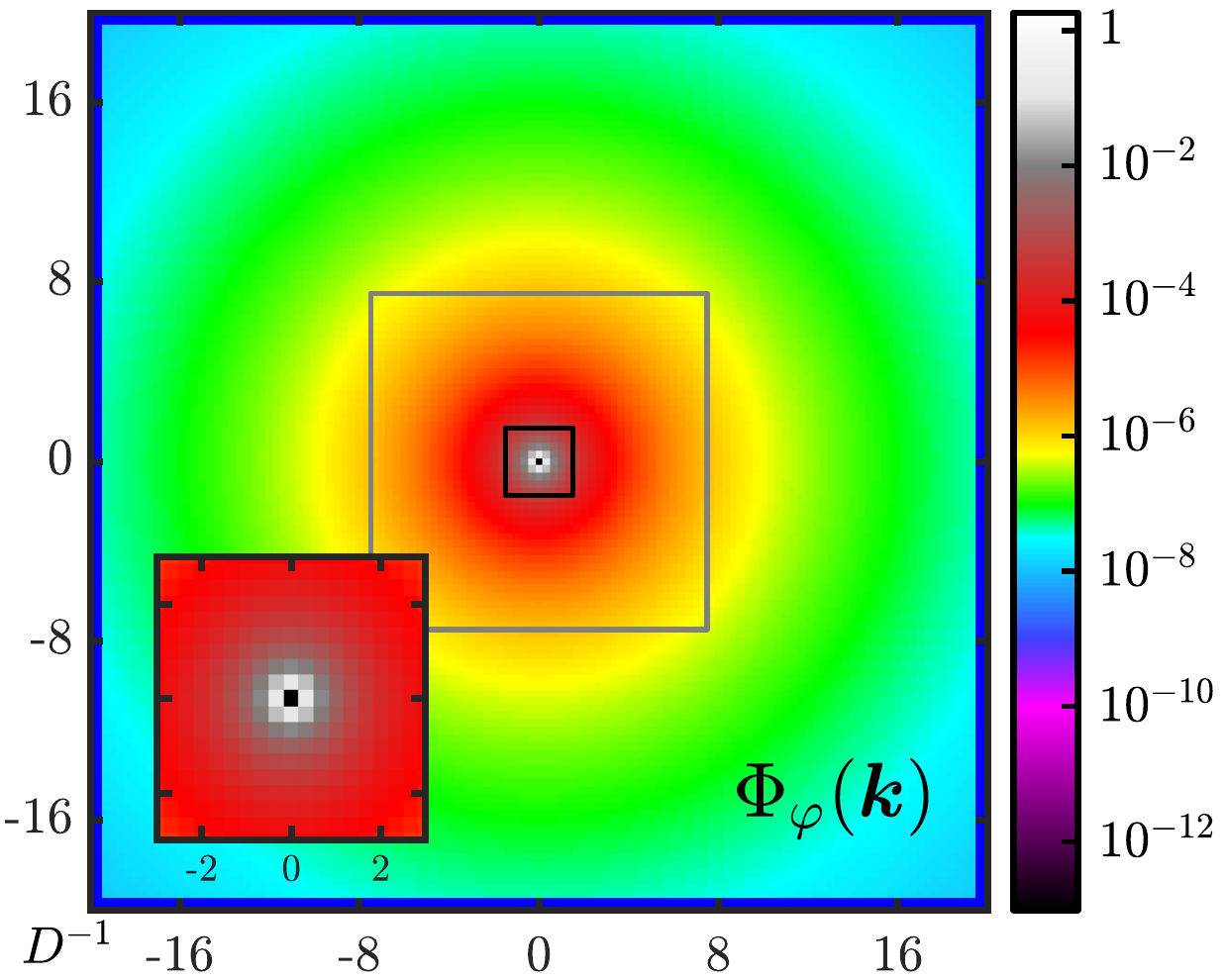}}       
	\newcommand{\ColumnWidth}[1]
		{\dimexpr \LineRatio \linewidth * \AspectRatio{#1} / (\AspectRatio{1} + \AspectRatio{2}) \relax
		}
	\newcommand{\ColumnGap}{\hspace {\dimexpr \linewidth /3 - \LineRatio\linewidth /3 }}

	\begin{tabular}{
		@{\ColumnGap}
		m{\ColumnWidth{1}}
		@{\ColumnGap}
		m{\ColumnWidth{2}}
		@{\ColumnGap}
		}
		\subfigimg[width=\linewidth,pos=ul,font=\fontfig{black}]{$\;\;$(a)}{0.0}{\FigOne} &
		\subfigimg[width=\linewidth,pos=ul,font=\fontfig{black}]{$\quad\;$(b)}{0.0}{\FigTwo}
	\end{tabular}
	
	\caption{\label{fig:SF_no_AO} Structure function and PSD of the $\nPha = 10000$ simulated turbulent phase screens normalised by~$\rFried^{-5/3}$. Panel~(a):~Average radial profiles of the 2D structure function: Theoretical profile ($\SF{\pha}\Paren{\Vx}$; black), profile of the simulated set computed in the direct space ($\SFd{\pha}\Paren{\Vpix}$; green) and theoretical profile computed in the frequency domain ($\SFf{\pha}\Paren{\Vx}$; blue). Panel~(b):~Theoretical 2D map of the PSD $\PSD{\pha}\Paren{\Vk}$. The grey square represents the cut-off frequency~$\Abs{k_{i\in\Brace{1,2}}}<\Paren{2\pitch}^{-1}$.}
\end{figure}

\subfigfull{fig:SF_no_AO}{a} presents the theoretical structure function profile as well as its numerical estimate on the generated set of phase screens. The results were normalised by~$\rFried^{-5/3}$ to remove the dependency on that parameter. The structure functions were estimated via the methods described in \refapp{app:SF}. The figure shows that the sub-harmonics method proposed by~\cite{Lane:92_Subharmonics} succeeds in generating screens whose structure function (green) is close to the theory (black). We note here that the structure function cannot be estimated beyond~$\sqrt{2\Dsim}$ since there is no pair in the pupil beyond that scale.

We highlight that simulating the screens with the method proposed by~\cite{McGlamery:76_Turb_PSD} via the frequency domain would lead to the theoretical PSD of \subfig{fig:SF_no_AO}{b}. But as presented in the inset of the figure, the lack of value at the origin (because of the divergence of the Kolomogorov spectrum but null value at~$\Vk=\V{0}$ for piston-free screens) leads to an underestimation of the large structures in the phase screens and a structure function profile that is below the theoretical one (in blue in \subfig{fig:SF_no_AO}{a}).

\subsection{Modelled influence function profiles}

In this work, we placed $\nactu=16$ actuators across the diameter~$\Dsim$ of the simulation. The pitch~$\Vpitch$ of the actuators is regular
\begin{equation}
	\pitch_{1} = \pitch_{2} = \pitch = \Dsim/\Paren{\nactu-1}
	\,.
\end{equation}
Different influence functions~$\IFref$ were tested~\citep[see][]{Hudgin:77}, including
a 2D local piston
\begin{equation}
	\IFref\Paren{\Vx} = 
	\begin{cases}
		1 & \text{if } \max\Paren{\Abs{x_{1}}, \Abs{x_{2}}}\leq \pitch
		\\
		0 & \text{otherwise}
	\end{cases}
	\,,
\end{equation}
a 2D pyramid,
\begin{equation}
	\IFref\Paren{\Vx} = 
	\begin{cases}
		\Paren{1-\nicefrac{\Abs{x_{1}}}{\pitch}}
		\Paren{1-\nicefrac{\Abs{x_{2}}}{\pitch}} & \text{if } \max\Paren{\Abs{x_{1}}, \Abs{x_{1}}}\leq \pitch
		\\
		0 & \text{otherwise}
	\end{cases}
	\,,
\end{equation}
a 2D axisymmetric Gaussian profile
\begin{equation}
	\IFref\Paren{\Vx} = \E{-0.5\Norm{\Vx}^{2}/\pitch^{2}}
	\,,
\end{equation}
a 2D axisymmetric profile experimentally fitted on a deformable from ALPAO (DM192) with~$\nactu = 16$ actuators across the pupil and a 2D~$\sinc$ function as defined by \eq{eq:prof_SINC}.
Beyond the above mentioned profiles, the results obtained with the binary mask analytical model of \eq{eq:binary_mask} are presented in the following sections for comparison purposes.

\begin{figure}[t!] 
	\centering
		
	
	\newcommand{\LineRatio}{1}
	
	\newcommand{\FigOne}  {figures_IF_profile_Piston.pdf}
	\newcommand{\FigTwo}  {figures_IF_profile_Pyramid.pdf}  
	\newcommand{\FigThree}{figures_IF_profile_Gaussian.pdf}
	\newcommand{\subfigColor}{white}
	
	\sbox1{\includegraphics{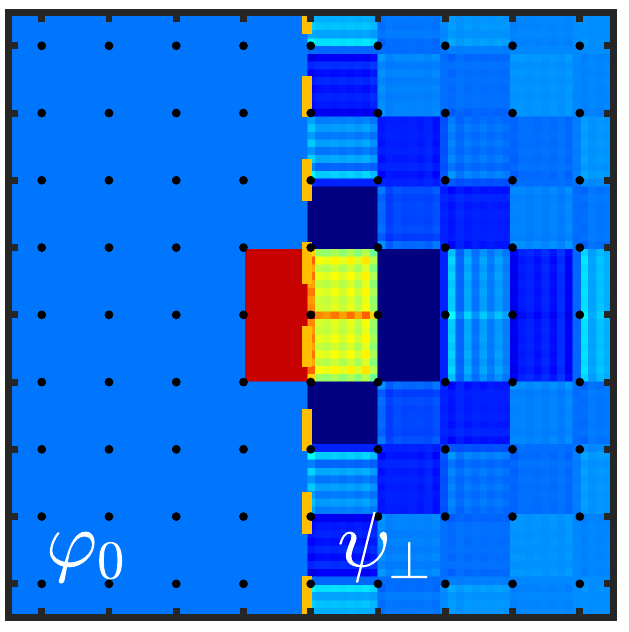}}	       
	\sbox2{\includegraphics{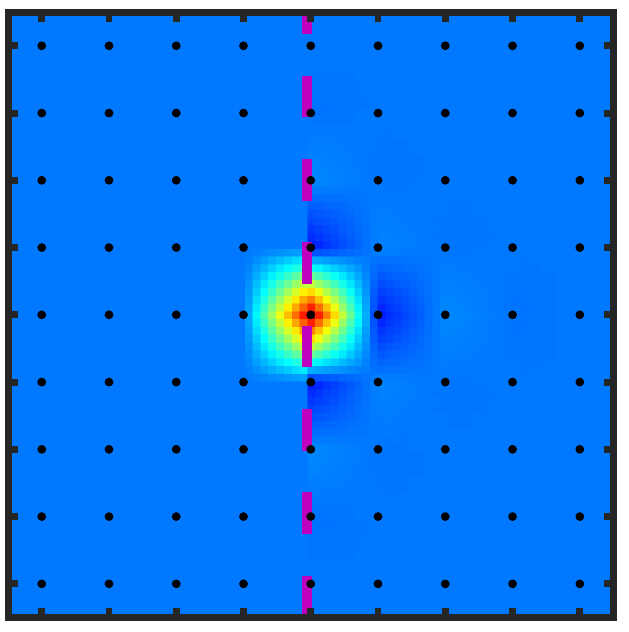}}	   
	\sbox3{\includegraphics{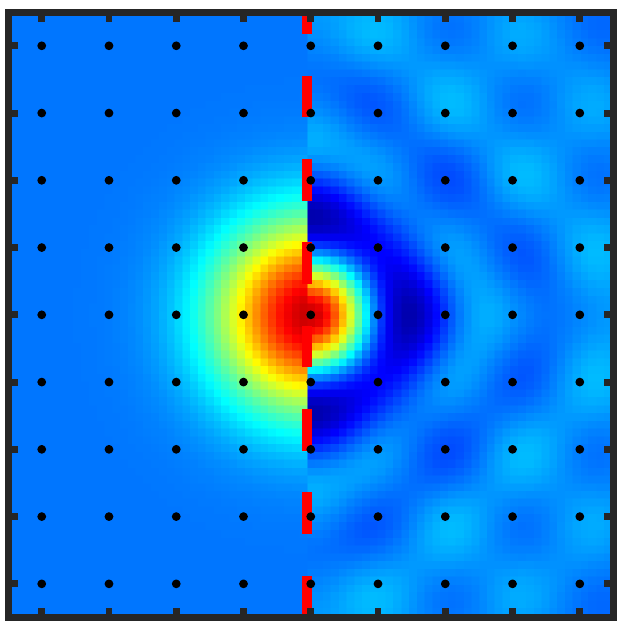}}     
	\newcommand{\ColumnWidth}[1]
		{\dimexpr \LineRatio \linewidth * \AspectRatio{#1} / (\AspectRatio{1} + \AspectRatio{2} + \AspectRatio{3}) \relax
		}
	\newcommand{\ColumnGap}{\hspace {\dimexpr \linewidth /4 - \LineRatio\linewidth /4 }}

	\begin{tabular}{
		@{\ColumnGap}
		m{\ColumnWidth{1}}
		@{\ColumnGap}
		m{\ColumnWidth{2}}
		@{\ColumnGap}
		m{\ColumnWidth{3}}
		@{\ColumnGap}
		}
		\subfigimg[width=\linewidth,pos=ul,font=\fontfig{\subfigColor}]{$\,$(a)}{0.0}{\FigOne} &
		\subfigimg[width=\linewidth,pos=ul,font=\fontfig{\subfigColor}]{$\,$(b)}{0.0}{\FigTwo} &
		\subfigimg[width=\linewidth,pos=ul,font=\fontfig{\subfigColor}]{$\,$(c)}{0.0}{\FigThree}
	\end{tabular}

	
	\renewcommand{\LineRatio}{0.75}
	
	\renewcommand{\FigOne}  {figures_IF_profile_ALPAO_repair.pdf}
	\renewcommand{\FigTwo}  {figures_IF_profile_SINC.pdf}   
	\renewcommand{\FigThree}{figures_IF_profile_bar.pdf}
	
	\sbox1{\includegraphics{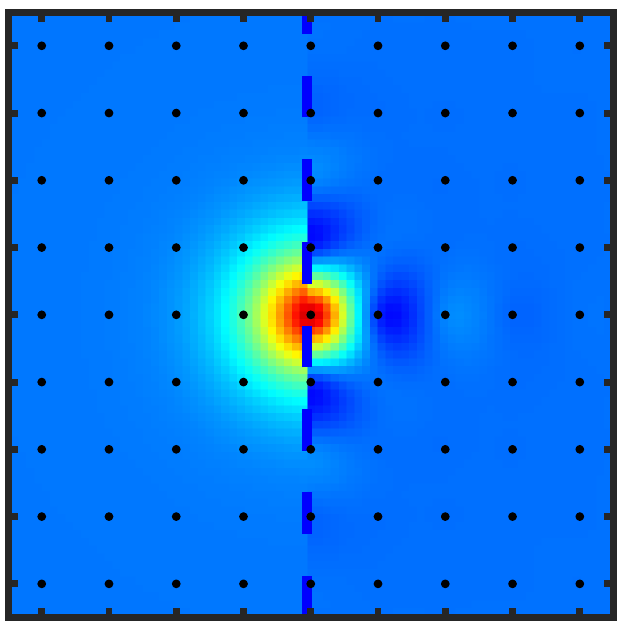}}	       
	\sbox2{\includegraphics{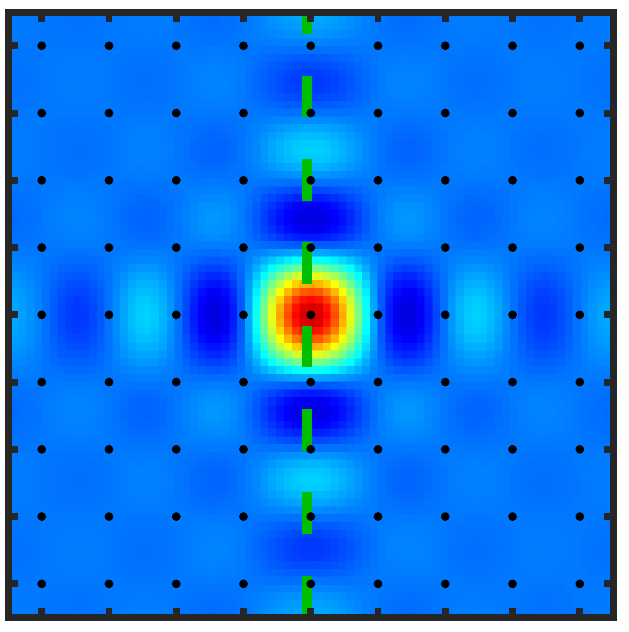}}	   
	\sbox3{\includegraphics{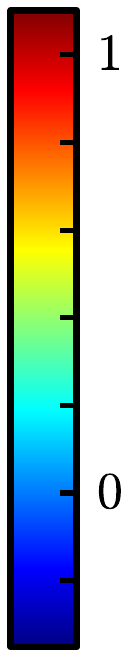}}     
	\renewcommand{\ColumnWidth}[1]
		{\dimexpr \LineRatio \linewidth * \AspectRatio{#1} / (\AspectRatio{1} + \AspectRatio{2} + \AspectRatio{3}) \relax
		}
	\newcommand{\ColumnGapTwo}{\hspace {\dimexpr \linewidth /2 - \LineRatio\linewidth /2 }}
	
	\vspace{-5.5pt}
	
	\begin{tabular}{
		@{\ColumnGapTwo}
		m{\ColumnWidth{1}}
		@{}
		m{\ColumnWidth{2}}
		@{}
		m{\ColumnWidth{3}}
		@{\ColumnGapTwo}
		}
		\subfigimg[width=\linewidth,pos=ul,font=\fontfig{\subfigColor}]{$\,$(d)}{0.0}{\FigOne} &
		\subfigimg[width=\linewidth,pos=ul,font=\fontfig{\subfigColor}]{$\,$(e)}{0.0}{\FigTwo} &
		\subfigimg[width=\linewidth,pos=ul,font=\fontfig{\subfigColor}]{$\,$}{0.0}{\FigThree}
	\end{tabular}

	\vspace{-4pt}

	\centering
	\subfigimg[width=0.95\linewidth,pos=ul,font=\fontfig{black}]{$\!\!\!\!\!\!\!\!$(f)}{0.0}{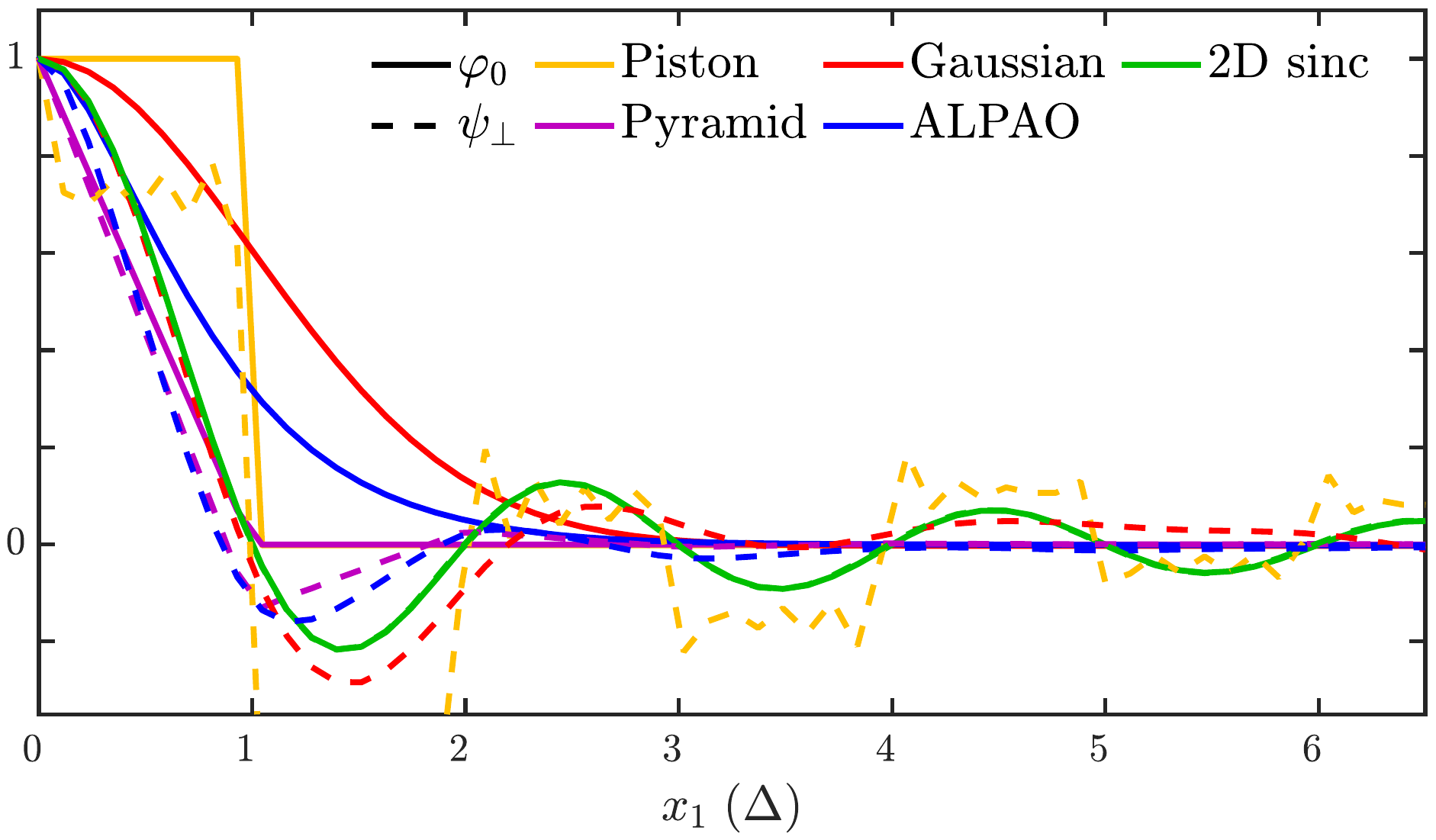}

	\caption{\label{fig:IF_profile} Visualisation of the different influence functions before ($\IFref$, left side of the panels) and after ($\IFperp$, right side of the panels) orthogonalisation for the different profiles. The 2D maps are normalised to their maximal value. Panel  (a):~2D local piston. Panel~(b):~2D pyramid. Panel~(c):~2D axisymmetric Gaussian profile. Panel~(d):~2D axisymmetric ALPAO profile. Panel~(e):~2D~$\sinc$ function. Panel~(f):~$x_{1}$-profiles of the different influence functions before (solid lines) and after (dashed lines) orthogonalisation.}
\end{figure}

The DM192 was procured by the National Astronomical Research Institute of Thailand (NARIT) in the framework of the Evanescent Wave Coronagraph project~\citep[EvWaCo, ][]{Buisset:18_EvWaCo_spec, Alagao:21_EvWaCo_exp_contrast}. In this work, the simulations are based on the influence functions measured and kindly provided by ALPAO and fitted as described by \cite{Berdeu:22_SPIE_EvWaCo_AO_bench}.

To avoid aliasing and to increase the resolution in the frequency domain, the simulated diameter was zero padded three times,~$\nPad = 3$. The influence function~$\IFref$ was orthogonalised on this extended domain of side~$\nPad\times\Dsim$. As a consequence, the number of actuators was adapted accordingly. In addition, the piston was removed, and~$\IFref$ was normalised on this extended square domain.

\figfull{fig:IF_profile} introduces the 2D maps and the $x_{1}$-profiles of the different influence functions~$\IFref$ as well as their orthogonalised counterparts~$\IFperp$. Interestingly, it appears that after orthogonalisation, the different profiles show oscillations similar to the 2D~$\sinc$ case. \subfigsfull{fig:IF_profile}{a,f} show that due to the sharp edges of the local piston, the orthogonalisation leads to aliasing artefacts because of shifts that do not correspond to an integer number of pixels.

From \subfigs{fig:IF_profile}{c,f}, it appears that the 2D Gaussian profile is the one that, after orthogonalisation, gives the closest result to the 2D~$\sinc$ function. The two first oscillations are visible despite both being slightly too large and the first also being slightly too deep. In contrast, the core of the profiles match perfectly with the full widths at half-maximum.

Finally, from \subfigs{fig:IF_profile}{e,f}, there is no difference between the 2D~$\sinc$ shape before and after orthogonalisation. This means that the function discussed in \refsec{sec:PSD_analytical} is already orthogonalised.

\begin{figure}[t!] 
	\centering
		
	
	\newcommand{\LineRatio}{1}
	
	\newcommand{\FigOne}  {figures_IF_profile_Piston_f.pdf}
	\newcommand{\FigTwo}  {figures_IF_profile_Pyramid_f.pdf}	
	\newcommand{\FigThree}{figures_IF_profile_Gaussian_f.pdf}
	\newcommand{\subfigColor}{white}
	
	\sbox1{\includegraphics{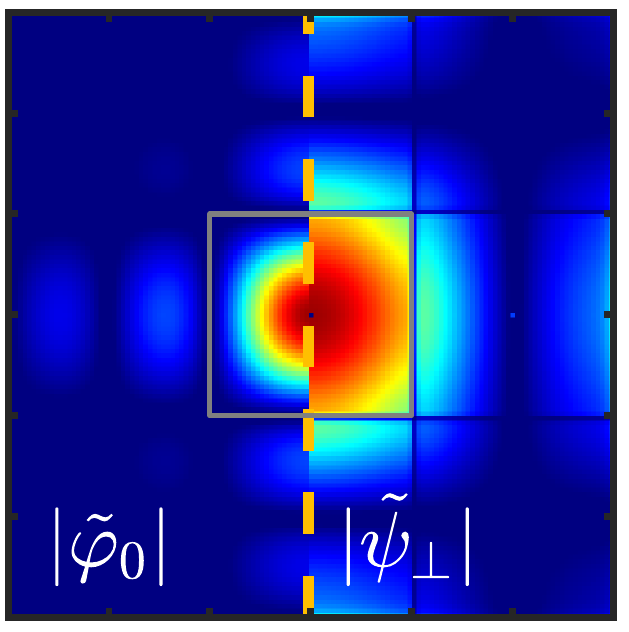}}	       
	\sbox2{\includegraphics{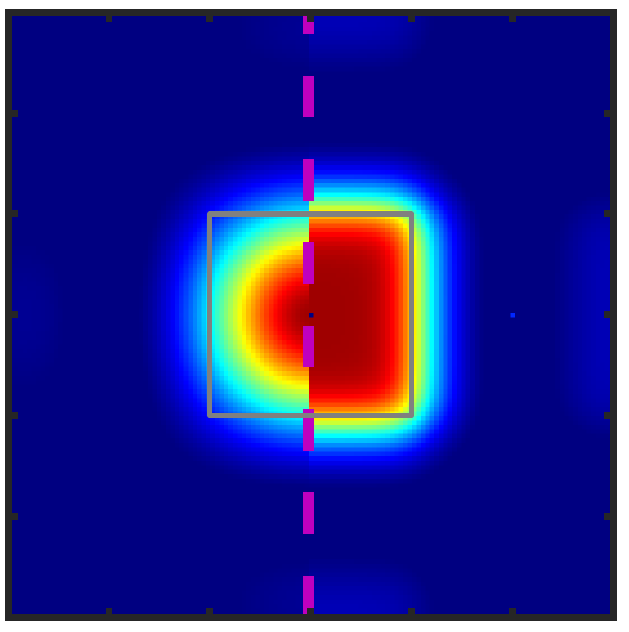}}	   
	\sbox3{\includegraphics{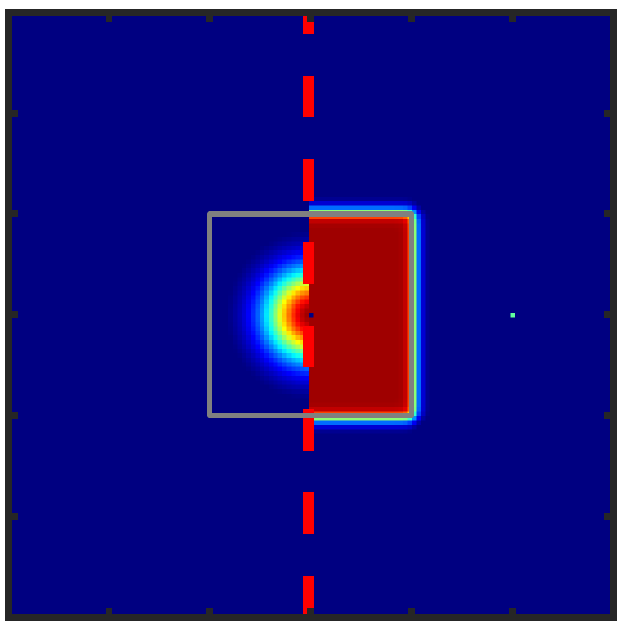}}     
	\newcommand{\ColumnWidth}[1]
		{\dimexpr \LineRatio \linewidth * \AspectRatio{#1} / (\AspectRatio{1} + \AspectRatio{2} + \AspectRatio{3}) \relax
		}
	\newcommand{\ColumnGap}{\hspace {\dimexpr \linewidth /4 - \LineRatio\linewidth /4 }}

	\begin{tabular}{
		@{\ColumnGap}
		m{\ColumnWidth{1}}
		@{\ColumnGap}
		m{\ColumnWidth{2}}
		@{\ColumnGap}
		m{\ColumnWidth{3}}
		@{\ColumnGap}
		}
		\subfigimg[width=\linewidth,pos=ul,font=\fontfig{\subfigColor}]{$\,$(a)}{0.0}{\FigOne} &
		\subfigimg[width=\linewidth,pos=ul,font=\fontfig{\subfigColor}]{$\,$(b)}{0.0}{\FigTwo} &
		\subfigimg[width=\linewidth,pos=ul,font=\fontfig{\subfigColor}]{$\,$(c)}{0.0}{\FigThree}
	\end{tabular}

	
	\renewcommand{\LineRatio}{0.725}
	
	\renewcommand{\FigOne}  {figures_IF_profile_ALPAO_repair_f.pdf}
	\renewcommand{\FigTwo}  {figures_IF_profile_SINC_f.pdf} 
	\renewcommand{\FigThree}{figures_IF_profile_bar_f.pdf}
	
	\sbox1{\includegraphics{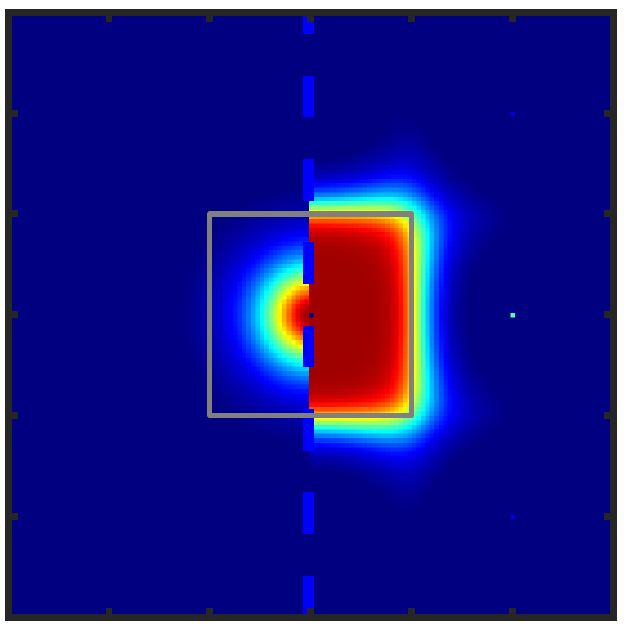}}	       
	\sbox2{\includegraphics{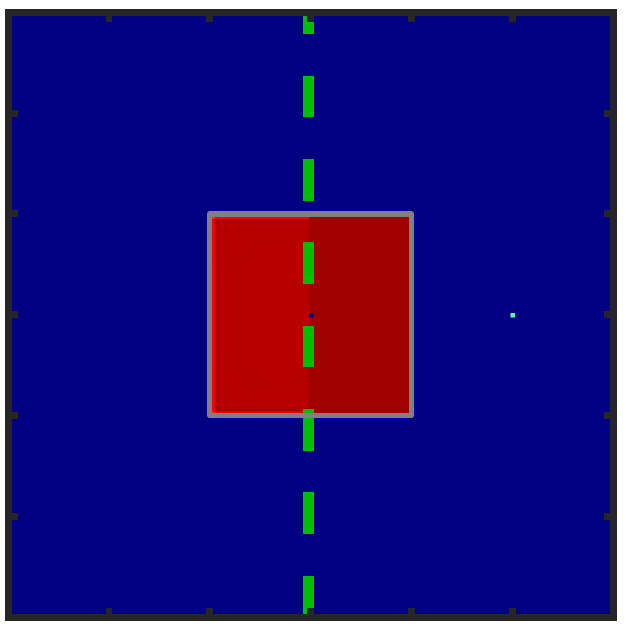}}	   
	\sbox3{\includegraphics{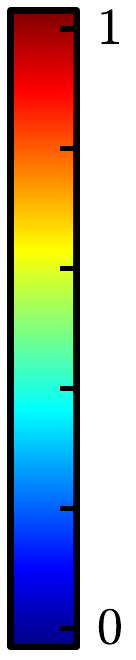}}     
	\renewcommand{\ColumnWidth}[1]
		{\dimexpr \LineRatio \linewidth * \AspectRatio{#1} / (\AspectRatio{1} + \AspectRatio{2} + \AspectRatio{3}) \relax
		}
	\newcommand{\ColumnGapTwo}{\hspace {\dimexpr \linewidth /2 - \LineRatio\linewidth /2 }}
	
	\vspace{-5.5pt}
	
	\begin{tabular}{
		@{\ColumnGapTwo}
		m{\ColumnWidth{1}}
		@{}
		m{\ColumnWidth{2}}
		@{}
		m{\ColumnWidth{3}}
		@{\ColumnGapTwo}
		}
		\subfigimg[width=\linewidth,pos=ul,font=\fontfig{\subfigColor}]{$\,$(d)}{0.0}{\FigOne} &
		\subfigimg[width=\linewidth,pos=ul,font=\fontfig{\subfigColor}]{$\,$(e)}{0.0}{\FigTwo} &
		\subfigimg[width=\linewidth,pos=ul,font=\fontfig{\subfigColor}]{$\,$}{0.0}{\FigThree}
	\end{tabular}
	
	\caption{\label{fig:IF_profile_f} Visualisation of the modulus of the Fourier transform of the different influence functions before ($\Abs{\FTIFref}$, left side of the panels) and after ($\Abs{\FTIFperp}$, right side of the panels) orthogonalisation for the different profiles. The 2D maps were normalised to their maximal value. The grey squares represent the cut-off frequency~$\Abs{k_{i\in\Brace{1,2}}}<\Paren{2\pitch}^{-1}$. Panel~(a):~2D local piston. Panel~(b):~2D pyramid. Panel~(c):~2D axisymmetric Gaussian profile. Panel~(d):~2D axisymmetric ALPAO profile. Panel~(e):~2D~$\sinc$ function.}
\end{figure}

For additional information, \fig{fig:IF_profile_f} presents the 2D Fourier transforms of the different influence functions~$\Abs{\FTIFref}$ as well as their orthogonalised counterparts~$\Abs{\FTIFperp}$. It gives an idea of the correction efficiency of a given orthogonalised profile in the frequency domain. Once again, from \subfigs{fig:IF_profile}{c,e}, it appears that after orthogonalisation, the 2D Gaussian profile behaves in a manner very similar to the 2D~$\sinc$ pattern, with a homogeneous amplitude on the domain~$\forall i\in\Brace{1,2} , \, \Abs{k_{i}}<\Paren{2\pitch_{i}}^{-1}$.

\subsection{Simulated apertures and actuator grids}

\begin{figure}[t!] 
	\centering
	
	\newcommand{\LineRatio}{0.9}
	
	\newcommand{\FigOne}{figures_aperture_square.pdf}
	\newcommand{\FigTwo}{figures_aperture_disk.pdf} 
	\newcommand{\FigThree}{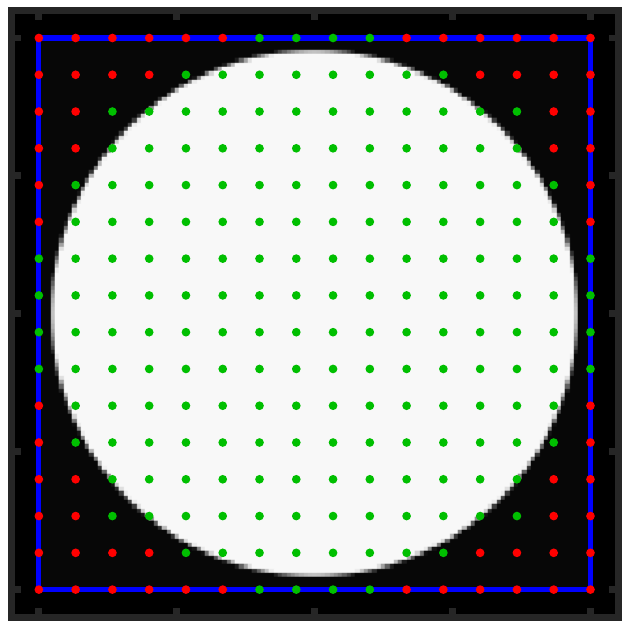}    
	\newcommand{\FigFour}{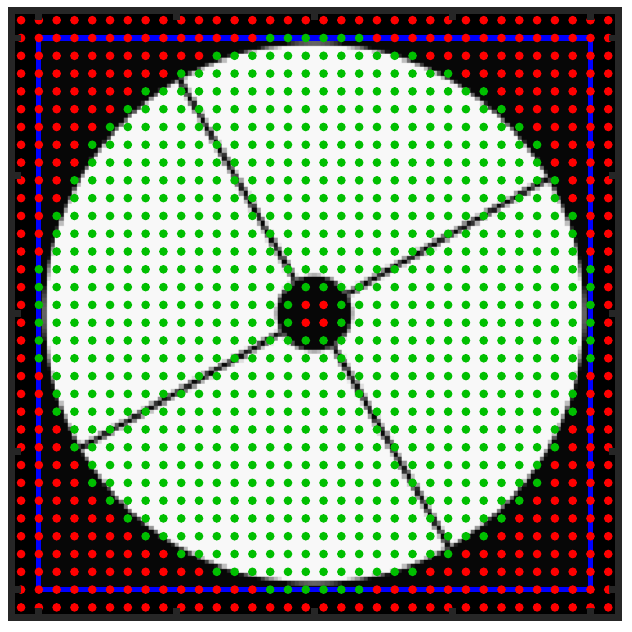} 
	\newcommand{\FigFive}{figures_aperture_bar.pdf} 
	\newcommand{\subfigColor}{white}
	
	\sbox1{\includegraphics{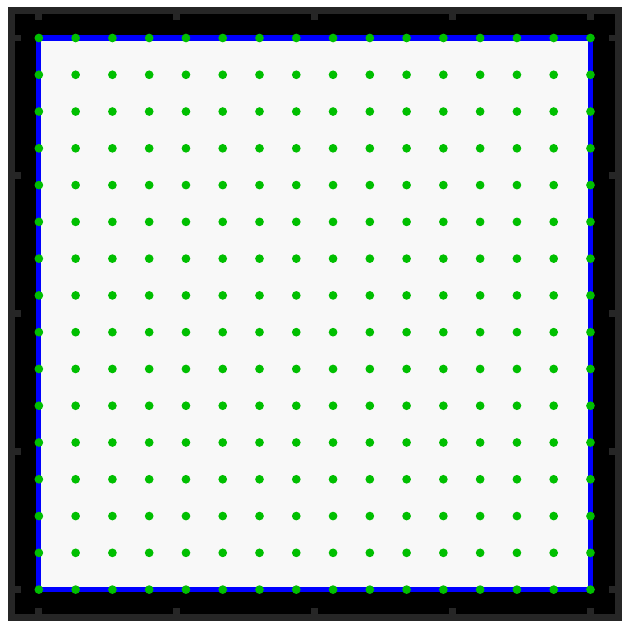}}	       
	\sbox2{\includegraphics{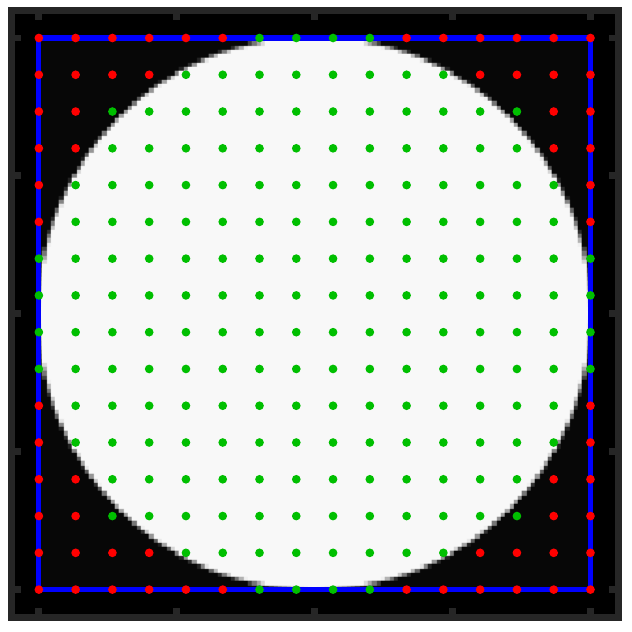}}       
	\sbox3{\includegraphics{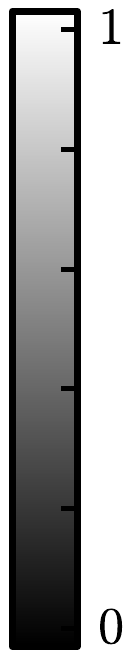}}      
	\newcommand{\ColumnWidth}[1]
		{\dimexpr \LineRatio \linewidth * \AspectRatio{#1} / (\AspectRatio{1} + \AspectRatio{2} + \AspectRatio{3}) \relax
		}
	\newcommand{\ColumnGap}{\hspace {\dimexpr \linewidth /3 - \LineRatio\linewidth /3 }}

	\begin{tabular}{
		@{\ColumnGap}
		m{\ColumnWidth{1}}
		@{\ColumnGap}
		m{\ColumnWidth{2}}
		@{}
		m{\ColumnWidth{3}}
		@{\ColumnGap}
		}
		\subfigimg[width=\linewidth,pos=ul,font=\fontfig{\subfigColor}]{$\;$(a)}{0.0}{\FigOne} &
		\subfigimg[width=\linewidth,pos=ul,font=\fontfig{\subfigColor}]{$\;$(b)}{0.0}{\FigTwo} &

\subfigimg[width=\linewidth,pos=ul,font=\fontfig{\subfigColor}]{}{0.0}{\FigFive}
		\\      \subfigimg[width=\linewidth,pos=ul,font=\fontfig{\subfigColor}]{$\;$(c)}{0.0}{\FigThree} &
		\subfigimg[width=\linewidth,pos=ul,font=\fontfig{\subfigColor}]{$\;$(d)}{0.0}{\FigFour}
	\end{tabular}   

	\caption{\label{fig:aperture} Weighted apertures used for the simulations. The blue frames represent the simulation domain~$\Dsim$. The red dots represent the regular grid of the actuators. The green dots represent the active actuators for a given aperture. Panel~(a):~Square aperture on the full simulated domain. Panel~(b):~Circular aperture of diameter~$\Dsim$. Panel~(c):~Circular aperture of diameter~$\percent{95}\Dsim$. Panel~(d): VLT-like aperture with a central obscuration of~$\percent{14}\Dsim$ and vane width of~$\percent{1}\Dsim$ relative to the aperture diameter of~$\percent{97.5}\Dsim$.}
\end{figure}

\figfull{fig:aperture} shows the two apertures used in the simulations: a square and a disc. As mentioned above, the pixels lying on the aperture edges were weighted according to \eq{eq:pupil}.

For the circular apertures in \subfigs{fig:aperture}{b,c}, the 192 active positions correspond to the 192 actuators of the DM192 from ALPAO. In \subfig{fig:aperture}{b}, the actuators lie exactly on the edge of the pupil, and some peripheral parts of the pupil are not inside the active actuator grid. Thus, the turbulence compensation would not be optimal and definitely not stationary on the pupil edge. The third simulated aperture, \subfig{fig:aperture}{c}, was aimed at solving this issue. The outer active positions allowed for a correction beyond the aperture in order to get better compensation on the edges of the pupil and more stationary fitting residuals. Finally, to test more complex pupils, such as those encountered in XAO systems, a pupil with a central obscuration and a four-vane spider similar to the Very Large Telescope (VLT) was studied (see \subfig{fig:aperture}{c}). In this case, the number of actuators was doubled across the diameter, from $\nactu=16$ to $\nactu=32$.

The results corresponding only to the third case are presented in the following sections in order to compare them with the simulations obtained with the real ALPAO DM192, \refsec{sec:res_ALPAO}. The three other cases are presented in \refapp{app:add_materials}.

\subsection{Predicted point spread function}
\label{sec:res_PSF}

In this section, we present the results obtained with a Fried parameter~$\rFried=\pitch$, the interactuator pitch. We used three different methods to obtain the long-exposure PSF: Monte Carlo simulations, the structure function method, and the power spectrum density method.

For the Monte Carlo simulations, the $\nPha = 10000$ random phase screens were optimally corrected as described in \refapp{app:proj_opti} and then propagated through the aperture via \eq{eq:FTPSF_short}, where~$\pupil\Paren{\Vx}$ is given by \eq{eq:pupil}. The long exposure is the average of the $\nPha$ results.

For the structure function method, the analytic structure function after AO correction was computed via \eq{eq:SF_analytic}. The integrals were estimated by summing the discrete variables. Then, the long exposure was obtained via its Fourier transform using \eq{eq:FTlongPSF_SF}.

For the power spectrum density method, the analytic PSD after AO correction was computed via \eq{eq:PSDres}. Then the stationary structure function was obtained via \eq{eq:SF_FT}, allowing convolution of the diffraction-limited PSF~$\PSFtel$ with~$\PSFres$ using \eq{eq:FTlongPSF_SF_stat} in order to produce the long-exposure Fourier transform.

In addition, for each method, the tip-tilt can be removed. For the two first methods, this was done by adding the tip and the tilt modes to the list of the influence functions. For the third method, it was done as described in \refsec{sec:piston_TT}, via \eq{eq:PSDtt_rmv}.

\begin{figure}[t!]
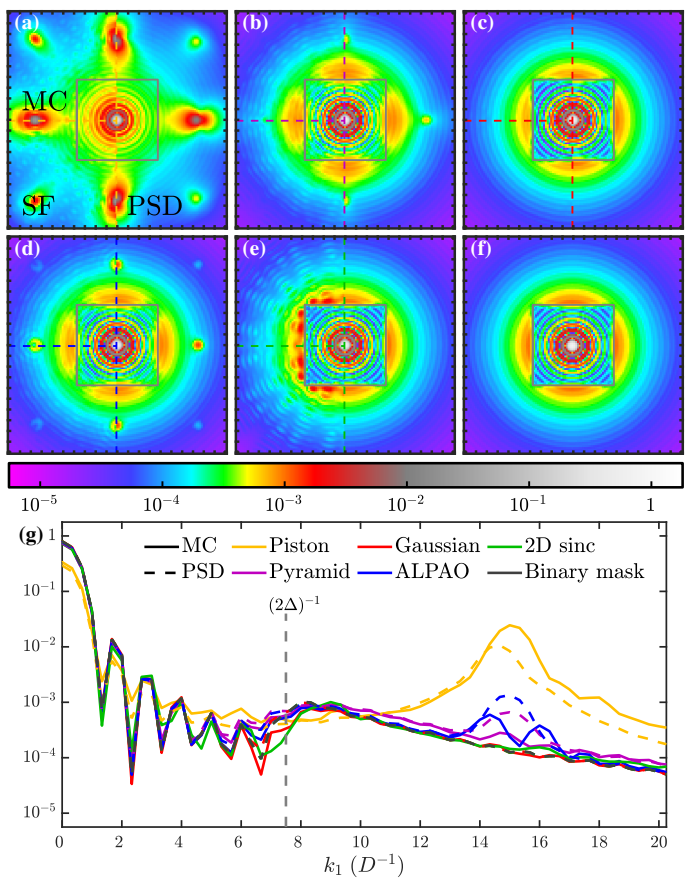
 
	\centering
	
	\newcommand{\PathFig}{figures_PSF_}
	\newcommand{\FlagAperture}{disk_95}
	\newcommand{\subfigColor}{white}
	
	\newcommand{\LineRatio}{1}
	
	\newcommand{\FigOne}{\PathFig \FlagAperture _Piston.pdf}
	\newcommand{\FigTwo}{\PathFig \FlagAperture _Pyramid.pdf}       
	\newcommand{\FigThree}{\PathFig \FlagAperture _Gaussian.pdf}
	
	\sbox1{\includegraphics{\FigOne}}	       
	\sbox2{\includegraphics{\FigTwo}}       
	\sbox3{\includegraphics{\FigThree}}     
	\newcommand{\ColumnWidth}[1]
		{\dimexpr \LineRatio \linewidth * \AspectRatio{#1} / (\AspectRatio{1} + \AspectRatio{2} + \AspectRatio{3}) \relax
		}
	\newcommand{\ColumnGap}{\hspace {\dimexpr \linewidth /2 - \LineRatio\linewidth /2 }}

	\begin{tabular}{
		@{\ColumnGap}
		m{\ColumnWidth{1}}
		@{}
		m{\ColumnWidth{2}}
		@{}
		m{\ColumnWidth{3}}
		@{\ColumnGap}
		}
		\subfigimg[width=\linewidth,pos=ul,font=\fontfig{\subfigColor}]{$\;$(a)}{0.0}{\FigOne} &
		\subfigimg[width=\linewidth,pos=ul,font=\fontfig{\subfigColor}]{$\;$(b)}{0.0}{\FigTwo} &
		\subfigimg[width=\linewidth,pos=ul,font=\fontfig{\subfigColor}]{$\;$(c)}{0.0}{\FigThree}
	\end{tabular}
	
	
	
	\renewcommand{\FigOne}  {\PathFig \FlagAperture _ALPAO_repair.pdf}
	\renewcommand{\FigTwo}  {\PathFig \FlagAperture _SINC.pdf}      
	\renewcommand{\FigThree}{\PathFig \FlagAperture _Binary_mask.pdf}
	
	\sbox1{\includegraphics{\FigOne}}	       
	\sbox2{\includegraphics{\FigTwo}}	   
	\sbox3{\includegraphics{\FigThree}}     
	\renewcommand{\ColumnWidth}[1]
		{\dimexpr \LineRatio \linewidth * \AspectRatio{#1} / (\AspectRatio{1} + \AspectRatio{2} + \AspectRatio{3}) \relax
		}
	\newcommand{\ColumnGapTwo}{\hspace {\dimexpr \linewidth /2 - \LineRatio\linewidth /2 }}
	
	\vspace{-5.5pt}
	
	\begin{tabular}{
		@{\ColumnGapTwo}
		m{\ColumnWidth{1}}
		@{}
		m{\ColumnWidth{2}}
		@{}
		m{\ColumnWidth{3}}
		@{\ColumnGapTwo}
		}
		\subfigimg[width=\linewidth,pos=ul,font=\fontfig{\subfigColor}]{$\;$(d)}{0.0}{\FigOne} &
		\subfigimg[width=\linewidth,pos=ul,font=\fontfig{\subfigColor}]{$\;$(e)}{0.0}{\FigTwo} &
		\subfigimg[width=\linewidth,pos=ul,font=\fontfig{\subfigColor}]{$\;$(f)}{0.0}{\FigThree}
	\end{tabular}
	
	\vspace{-5.5pt}

	\centering
	\subfigimg[width=\linewidth,pos=ul,font=\fontfig{black}]{}{0.0}{\PathFig bar.pdf}

	\centering
	\subfigimg[width=0.95\linewidth,pos=ul,font=\fontfig{black}]{$\!\!$(g)}{0.0}{\PathFig \FlagAperture _profile.pdf}

	\caption{\label{fig:PSF_disk_95} Simulated long-exposure PSF~$\PSF$ for the different profiles. Each panel shows the long exposure simulated via Monte Carlo simulations (MC, upper-left corner), via the analytical structure function (SF,  lower-left corner), and via the analytical PSD (right). The grey squares represent the cut-off frequency~$\Abs{k_{i\in\Brace{1,2}}}<\Paren{2\pitch}^{-1}$. Panel~(a):~2D local piston. Panel~(b):~2D pyramid. Panel~(c):~2D axisymmetric Gaussian profile. Panel~(d):~2D axisymmetric ALPAO profile. Panel~(e):~2D~$\sinc$ function. Panel~(f):~Binary mask analytical model. Panel~(g):~$k_{1}$-profiles of Panels~(a-f) of the Monte Carlo simulations (solid lines) and the PSD analytical models (dashed lines). -- Aperture: Disc of diameter~$\percent{95}\Dsim$.}
\end{figure}

\figfull{fig:PSF_disk_95} compares the different methods on the different influence function profiles for the aperture disc of diameter~$\percent{95}\Dsim$. The tip-tilt has not been not removed. Additional situations are presented in \refapp{app:add_materials} for all the apertures presented in \fig{fig:aperture}, from which similar conclusions can be derived.

Regarding global features of the PSF, all but the Gaussian and 2D~$\sinc$ shapes present secondary spots. As further discussed in \refsec{sec:res_ALPAO}, these spots can be interpreted by the capacity of the influence functions to provide a correct piston and tip-tilt. If not, the incident light is scattered, producing the secondary spots. These replicates, which directly depend on the influence function shapes, cannot be simulated via the standard PSD techniques assuming \eq{eq:binary_mask}. Finally, the secondary spots are reduced if a steering mirror is used to compensate for the tip-tilt, as shown in \subfigspar{fig:PSF_appen}{2,3}.

All the PSFs have a correction in the expected zone (grey square in \fig{fig:PSF_disk_95}) whose quality depends on the influence function profile. The Gaussian and the 2D~$\sinc$ function give results close to the diffraction limit ($10^{-4.5}$, dark blue diffraction rings in \subfig{fig:PSF_disk_95}{c}), while the local piston influence function leads to the worst quality ($10^{-3.5}$). The pyramid and ALPAO influence functions provide intermediate results.

When comparing the Monte Carlo and structure function methods, it appears that they are identical, showing the validity of \eq{eq:SF_analytic}. Overall, the PSD method gives results in good agreement with the predictions of the Monte Carlo simulations and of the analytical structure function. This supports the method described in \refsecs{sec:PSD_analytical}{sec:num_implementation}.

Nonetheless, some differences can be noticed. Firstly, the size and shape of the secondary spots are slightly different, being a bit brighter in the PSD model. Secondly, the halo beyond the corrected zone is slightly overestimated (orange versus yellow in the images), as shown by the profiles in \subfig{fig:PSF_disk_95}{g}.
On top of these minor discrepancies, the PSD method fails at explaining the speckle-like patterns mainly present in the piston and 2D~$\sinc$ shapes, which are even more visible on the full aperture disc (see \subfigpar{fig:PSF_appen}{2}). The patterns come from a breaking of the assumption of stationarity implied by the PSD technique, as discussed below in \refsec{sec:res_SF}. As for the secondary spots, the artefacts of the PSD method are strongly reduced if a steering mirror is used to compensate for the tip-tilt in the model (compare \subfigpar{fig:PSF_appen}{2} and \subfigpar{fig:PSF_appen}{3}). Indeed, part of the energy of the low orders not fitted by the WFC on the aperture edges is filtered out by the steering mirror.

It is worth mentioning that the analytical PSD of the Gaussian and the 2D~$\sinc$ shapes provide results extremely similar to the optimal case of the binary mask model (\subfig{fig:PSF_disk_95}{f}). Similar conclusions can be obtained on the full square and VLT-like apertures presented in \subfigspar{fig:PSF_appen}{1,4}. We emphasise here that in the case of the VLT-like aperture, the resolution of the simulation is not sufficient to properly sample the sharp piston shape due to the increased number of actuators. As a consequence, the analytical model fails at reproducing the Monte Carlo simulations for this case. To avoid this problem for the 2D~$\sinc$ shape in the Fourier domain in this specific case, we increased $\nPad$ to four.

\subsection{Predicted power spectrum density and pupil variance}
\label{sec:res_PSD}

\begin{figure}[t!]
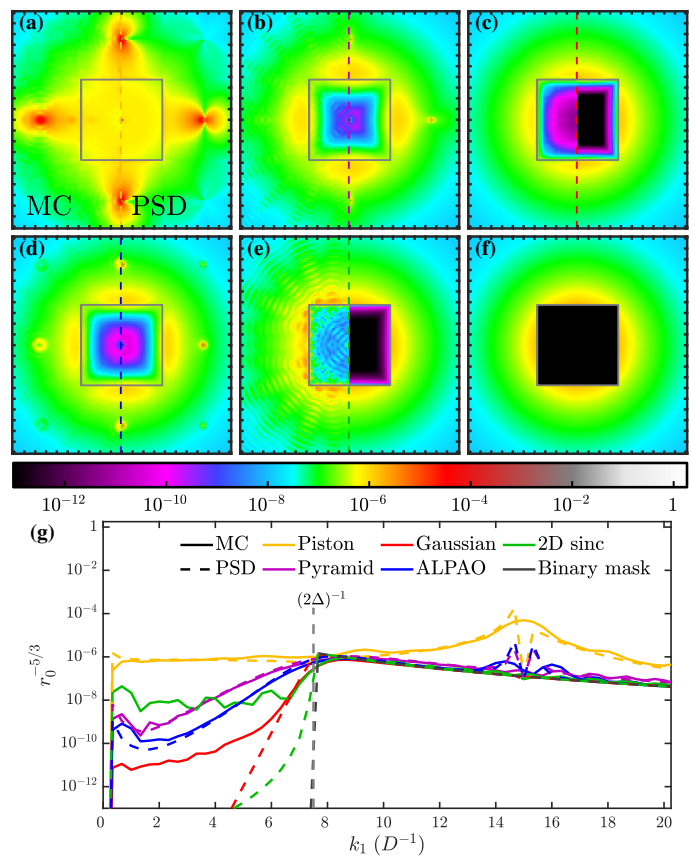
 
	\centering
	
	\newcommand{\PathFig}{figures_PSD_}
	\newcommand{\FlagAperture}{disk_95}
	\newcommand{\subfigColor}{black}
	
	\newcommand{\LineRatio}{1}
	
	\newcommand{\FigOne}{\PathFig \FlagAperture _Piston.pdf}
	\newcommand{\FigTwo}{\PathFig \FlagAperture _Pyramid.pdf}       
	\newcommand{\FigThree}{\PathFig \FlagAperture _Gaussian.pdf}
	
	\sbox1{\includegraphics{\FigOne}}	       
	\sbox2{\includegraphics{\FigTwo}}       
	\sbox3{\includegraphics{\FigThree}}     
	\newcommand{\ColumnWidth}[1]
		{\dimexpr \LineRatio \linewidth * \AspectRatio{#1} / (\AspectRatio{1} + \AspectRatio{2} + \AspectRatio{3}) \relax
		}
	\newcommand{\ColumnGap}{\hspace {\dimexpr \linewidth /2 - \LineRatio\linewidth /2 }}

	\begin{tabular}{
		@{\ColumnGap}
		m{\ColumnWidth{1}}
		@{}
		m{\ColumnWidth{2}}
		@{}
		m{\ColumnWidth{3}}
		@{\ColumnGap}
		}
		\subfigimg[width=\linewidth,pos=ul,font=\fontfig{\subfigColor}]{$\;$(a)}{0.0}{\FigOne} &
		\subfigimg[width=\linewidth,pos=ul,font=\fontfig{\subfigColor}]{$\;$(b)}{0.0}{\FigTwo} &
		\subfigimg[width=\linewidth,pos=ul,font=\fontfig{\subfigColor}]{$\;$(c)}{0.0}{\FigThree}
	\end{tabular}
	
	
	
	\renewcommand{\FigOne}  {\PathFig \FlagAperture _ALPAO_repair.pdf}
	\renewcommand{\FigTwo}  {\PathFig \FlagAperture _SINC.pdf}      
	\renewcommand{\FigThree}{\PathFig \FlagAperture _Binary_mask.pdf}
	
	\sbox1{\includegraphics{\FigOne}}	       
	\sbox2{\includegraphics{\FigTwo}}	   
	\sbox3{\includegraphics{\FigThree}}     
	\renewcommand{\ColumnWidth}[1]
		{\dimexpr \LineRatio \linewidth * \AspectRatio{#1} / (\AspectRatio{1} + \AspectRatio{2} + \AspectRatio{3}) \relax
		}
	\newcommand{\ColumnGapTwo}{\hspace {\dimexpr \linewidth /2 - \LineRatio\linewidth /2 }}
	
	\vspace{-5.5pt}
	
	\begin{tabular}{
		@{\ColumnGapTwo}
		m{\ColumnWidth{1}}
		@{}
		m{\ColumnWidth{2}}
		@{}
		m{\ColumnWidth{3}}
		@{\ColumnGapTwo}
		}
		\subfigimg[width=\linewidth,pos=ul,font=\fontfig{\subfigColor}]{$\;$(d)}{0.0}{\FigOne} &
		\subfigimg[width=\linewidth,pos=ul,font=\fontfig{\subfigColor}]{$\;$(e)}{0.0}{\FigTwo} &
		\subfigimg[width=\linewidth,pos=ul,font=\fontfig{\subfigColor}]{$\;$(f)}{0.0}{\FigThree}
	\end{tabular}
	
	\vspace{-5.5pt}

	\centering
	\subfigimg[width=\linewidth,pos=ul,font=\fontfig{black}]{}{0.0}{\PathFig bar.pdf}

	\centering      \subfigimg[width=0.95\linewidth,pos=ul,font=\fontfig{black}]{(g)}{0.0}{\PathFig \FlagAperture _profile.pdf}
	
	\caption{\label{fig:PSD_disk_95} Simulated PSD of the fitting residuals~$\PSDres$ normalised by~$\rFried^{-5/3}$ for the different profiles. Each panel shows the PSD simulated via Monte Carlo simulations (MC, upper-left corner), via the analytical structure function (SF, lower-left corner), and via the analytical PSD (right). The grey squares represent the cut-off frequency~$\Abs{k_{i\in\Brace{1,2}}}<\Paren{2\pitch}^{-1}$. Panel~(a):~2D local piston. Panel~(b):~2D pyramid. Panel~(c):~2D axisymmetric Gaussian profile. Panel~(d):~2D axisymmetric ALPAO profile. Panel~(e):~2D~$\sinc$ function. Panel~(f):~Binary mask analytical model. Panel~(g):~$k_{1}$-profiles of Panels~(a-f) of the Monte Carlo simulations (solid lines) and the PSD analytical models (dashed lines). -- Aperture: Disc of diameter~$\percent{95}\Dsim$.}
\end{figure}

In this section, we compare the analytical PSD obtained via \eq{eq:PSDres}, which assumes stationarity, with the PSD of the fitting residuals of the Monte Carlo simulations. The Monte Carlo PSD was estimated with the method described in \refapp{app:SFf}, padding the pupil by a factor of~$\nPad = 3$. The analytical PSD was computed on the same grid. The results are given in \fig{fig:PSD_disk_95}. Additional situations are presented in \refapp{app:add_materials}, which need to be compared with the uncorrected turbulent PSD in \subfig{fig:SF_no_AO}{b} (same logarithmic colour bar).

In terms of morphology comparison, similar conclusions to those in \refsec{sec:res_PSF} can be deduced, as the 2D maps are extremely similar, except for the 2D~$\sinc$ case where several orders of magnitude of discrepancy can be observed in the corrected zone as well as a speckle-like pattern beyond it. In other words, as purposely designed, the 2D~$\sinc$ function gives a perfect correction with the analytical computation via \eq{eq:PSDres}, that is, a correction equal to the binary mask ideal situation (\subfig{fig:PSD_disk_95}{f}). The discrepancy between the two, on the order of~$10^{-10}$, can be imputable to numerical effects, as the analytical 2D~$\sinc$ (not spatially bounded) is not exactly a square after a discrete Fourier transform on a finite grid. Nonetheless, from the Monte Carlo simulations, we can conclude otherwise, as the normalised correction is not below~$10^{-8}$. This result implies, once again, a problem with the stationary assumption, as further discussed in \refsec{sec:res_SF}. This also means that the assumption of \eq{eq:binary_mask} made in the different analytical models of the literature does not hold.

\figfull{fig:PSD_disk_95} shows that the PSD depends on the influence function shape. Discrepancies in the different PSD are expected since it is already known that the average variance of the residual phase depends on this profile~\citep{Hudgin:77, Tyson:15_principles_of_AO}. But our work allows for more information beyond this averaged value by computing the 2D map without needing Monte Carlo simulations. Thus, we show that the depth of the correction and the transition between the corrected and the uncorrected zones depend on~$\IFref$.

\tab{tab:var_disk} and \tab{tab:var_sup} compare the fitting error, that is, the theoretical expected variance on the pupil, obtained empirically via the Monte Carlo simulations and via the analytical model using \eq{eq:var_PSD} for the different influence function shapes. As expected, the empirical variance differs from the theoretical values because local sub- or over-optimal correction can be applied on the circular pupil edge depending on where its edge is locally situated on the actuator grid. Nonetheless, the standard deviations of this parameter (indicated by the~$\pm$ sign in the tables) show that their values are compatible with the theory.

\begin{table}[t!] 
	\caption{\label{tab:var_disk} Fitting error~$\AvgT{\sigAO^{2}}$ on the pupil normalised by~$\Paren{\pitch/\rFried}^{5/3}$.}
	\centering
	\begin{tabular}{cccc}
	\hline
	\hline
	\multirow{2}{*}{$\IFref$} & \multicolumn{3}{c}{$\AvgT{\sigAO^{2}} / \Paren{\pitch/\rFried}^{5/3}$}
	\\ \cline{2-4} 
			 & \multicolumn{1}{c}{\cite{Tyson:15_principles_of_AO}} & \multicolumn{1}{c}{Monte Carlo} & \multicolumn{1}{c}{$\;\;\quad$PSD$\;\;\quad$}
    \\
	\hline
	Piston & 1.26 & 1.22$\pm 0.44$ & 1.23
	\\
	Pyramid & 0.28 & 0.29$\pm 0.04$ & 0.30
	\\
	Gaussian & 0.23 & 0.22$\pm 0.01$ & 0.23
	\\
	ALPAO & -- & 0.26$\pm 0.02$ & 0.26
	\\
	2D~$\sinc$ & -- & 0.36$\pm 0.13$ & 0.23
	\\
	Binary mask & -- & -- & 0.23
	\\
	\hline
	\end{tabular}
	\tablefoot{Comparison between the Monte Carlo simulations (empirical average), the analytical PSD prediction, and the literature~\citep{Tyson:15_principles_of_AO}. Aperture: Disc of diameter~$\percent{95}\Dsim$.}
\end{table}

In contrast, the dependence of the analytical PSD estimate on the pupil, only via \eqs{eq:PSDpiston_rmv}{eq:PSDtt_rmv}, is negligible (see \tab{tab:var_sup}). Thus, the expected variance only depends on the influence function profile~$\IFref$ and the actuator pitch~$\pitch$. The estimated values match the theoretical values. For the VLT-like aperture, the noticeable discrepancies with the theoretical values can be attributed to the resolution of the simulation, as it is not sufficient to correctly sample the piston and pyramid sharp profiles for such a high actuator density. This limitation also impacts the centre of \subfigspar{fig:PSD_appen}{4a,4b}, where the PSD should not have a central peak.

Interestingly, for the expected variance in \tab{tab:var_disk} and \tab{tab:var_sup}, the Gaussian profile gives equivalent results to those of the ideal cases of the 2D~$\sinc$ and the binary mask model. This suggests that this profile is also optimal when it comes to minimising the variance, making it similar to the binary mask of \eq{eq:binary_mask}.

The profiles given in \subfig{fig:PSD_disk_95}{g} confirm the analysis performed on the 2D maps. As mentioned above, the 2D~$\sinc$ correction saturates around~$10^{-8}$, diverging from the PSD model prediction. For the other profiles, the PSD analytical model provides a very good estimate of the transition between the corrected zone and the halo of the PSD for~$\Abs{k_{1}}$ in the vicinity of the cut-off frequency~$\Paren{2\pitch}^{-1}$. The depth of the corrected PSD is also very nicely matched, except for the 2D Gaussian profile. This divergence can come from the approximation of the structure function stationarity, which becomes the main source of errors, as discussed in \refsec{sec:res_SF}. Indeed, as shown by the dashed red curve in \subfig{fig:PSD_disk_95}{f}, the 2D Gaussian profile provides a very effective correction, quickly dropping below~$10^{-9}$ according to the analytical model. As a consequence, any error on the corrected frequencies becomes a non-negligible source of error.

Similar conclusions can be obtained from \fig{fig:PSD_appen}. All the transitions between the corrected zone and the halo of the PSD were properly retrieved. But for the very deep correction of the Gaussian profile, small errors dominate the central region of the PSD.

\subsection{Predicted turbulent PSF and achievable contrast}
\label{sec:res_contrast}

\begin{figure}[t!]
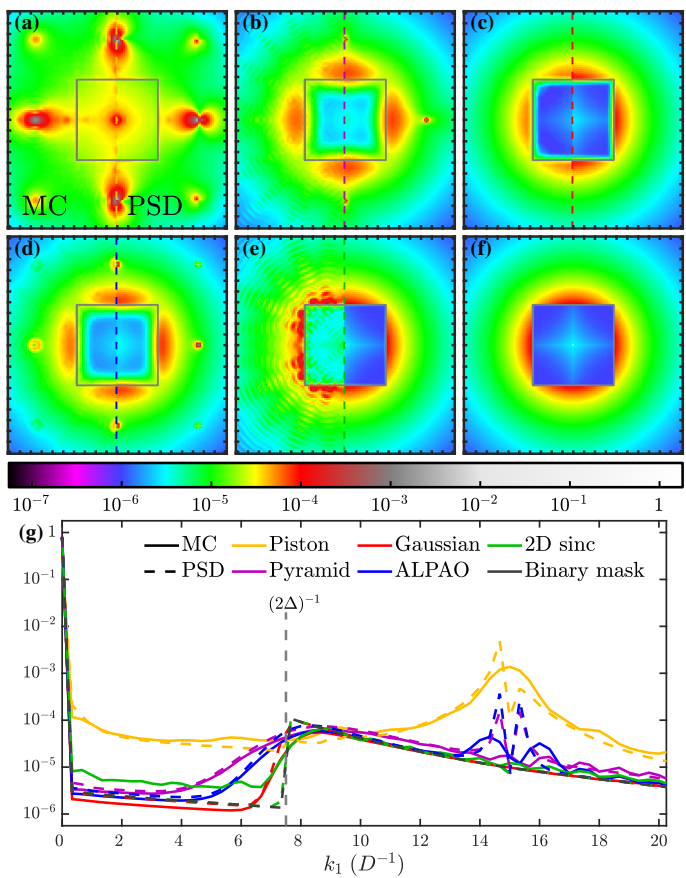
 
	\centering
	
	\newcommand{\PathFig}{figures_PSF_PSD_}
	\newcommand{\FlagAperture}{disk_95}
	\newcommand{\subfigColor}{black}	
	
	\newcommand{\LineRatio}{1}
	
	\newcommand{\FigOne}{\PathFig \FlagAperture _Piston.pdf}
	\newcommand{\FigTwo}{\PathFig \FlagAperture _Pyramid.pdf}       
	\newcommand{\FigThree}{\PathFig \FlagAperture _Gaussian.pdf}
	
	\sbox1{\includegraphics{\FigOne}}	       
	\sbox2{\includegraphics{\FigTwo}}       
	\sbox3{\includegraphics{\FigThree}}     
	\newcommand{\ColumnWidth}[1]
		{\dimexpr \LineRatio \linewidth * \AspectRatio{#1} / (\AspectRatio{1} + \AspectRatio{2} + \AspectRatio{3}) \relax
		}
	\newcommand{\ColumnGap}{\hspace {\dimexpr \linewidth /2 - \LineRatio\linewidth /2 }}

	\begin{tabular}{
		@{\ColumnGap}
		m{\ColumnWidth{1}}
		@{}
		m{\ColumnWidth{2}}
		@{}
		m{\ColumnWidth{3}}
		@{\ColumnGap}
		}
		\subfigimg[width=\linewidth,pos=ul,font=\fontfig{\subfigColor}]{$\;$(a)}{0.0}{\FigOne} &
		\subfigimg[width=\linewidth,pos=ul,font=\fontfig{\subfigColor}]{$\;$(b)}{0.0}{\FigTwo} &
		\subfigimg[width=\linewidth,pos=ul,font=\fontfig{\subfigColor}]{$\;$(c)}{0.0}{\FigThree}
	\end{tabular}
	
	

	\renewcommand{\FigOne}  {\PathFig \FlagAperture _ALPAO_repair.pdf}
	\renewcommand{\FigTwo}  {\PathFig \FlagAperture _SINC.pdf}      
	\renewcommand{\FigThree}{\PathFig \FlagAperture _Binary_mask.pdf}
	
	\sbox1{\includegraphics{\FigOne}}	       
	\sbox2{\includegraphics{\FigTwo}}	   
	\sbox3{\includegraphics{\FigThree}}     
	\renewcommand{\ColumnWidth}[1]
		{\dimexpr \LineRatio \linewidth * \AspectRatio{#1} / (\AspectRatio{1} + \AspectRatio{2} + \AspectRatio{3}) \relax
		}
	\newcommand{\ColumnGapTwo}{\hspace {\dimexpr \linewidth /2 - \LineRatio\linewidth /2 }}
	
	\vspace{-5.5pt}
	
	\begin{tabular}{
		@{\ColumnGapTwo}
		m{\ColumnWidth{1}}
		@{}
		m{\ColumnWidth{2}}
		@{}
		m{\ColumnWidth{3}}
		@{\ColumnGapTwo}
		}
		\subfigimg[width=\linewidth,pos=ul,font=\fontfig{\subfigColor}]{$\;$(d)}{0.0}{\FigOne} &
		\subfigimg[width=\linewidth,pos=ul,font=\fontfig{\subfigColor}]{$\;$(e)}{0.0}{\FigTwo} &
		\subfigimg[width=\linewidth,pos=ul,font=\fontfig{\subfigColor}]{$\;$(f)}{0.0}{\FigThree}
	\end{tabular}
	
	\vspace{-5.5pt}

	\centering
	\subfigimg[width=\linewidth,pos=ul,font=\fontfig{black}]{}{0.0}{\PathFig bar.pdf}

	\centering      
	\subfigimg[width=0.95\linewidth,pos=ul,font=\fontfig{black}]{$\!\!$(g)}{0.0}{\PathFig \FlagAperture _profile.pdf}
	
	\caption{\label{fig:PSF_pha_disk_95} Simulated PSF of the fitting residuals~$\PSFres$ for the different profiles. Each panel shows the PSF simulated via Monte Carlo simulations (MC, left) and via the analytical PSD (right). The grey squares represent the cut-off frequency~$\Abs{k_{i\in\Brace{1,2}}}<\Paren{2\pitch}^{-1}$. Panel~(a):~2D local piston. Panel~(b):~2D pyramid. Panel~(c):~2D axisymmetric Gaussian profile. Panel~(d):~2D axisymmetric ALPAO profile. Panel~(e):~2D~$\sinc$ function. Panel~(f):~Binary mask analytical model. Panel~(g):~$k_{1}$-profiles of Panels~(a-f) of the Monte Carlo simulations (solid lines) and the PSD analytical models (dashed lines). -- Aperture: Disc of diameter~$\percent{95}\Dsim$.}
\end{figure}

From the PSD, one can get the turbulence residual PSF, $\PSFres$, via \eqs{eq:SF_FT}{eq:FTlongPSF_SF_stat}. The results are shown in \fig{fig:PSF_pha_disk_95}. Additional situations are presented in \refapp{app:add_materials}.

Once again, the 2D maps are highly similar for the different profiles, except for the 2D~$\sinc$ influence function. The Gaussian profile also gives a slightly sharper transition via the analytical PSD than the prediction from the Monte Carlo simulations.

Another element can be extracted from the 2D PSF maps. The Strehl ratio, which generally quantifies the efficiency of an AO system \citep{Tyson:15_principles_of_AO}, is defined as the ratio of the peak intensity of the AO-corrected PSF with the diffraction-limited PSF peak
\begin{equation} 
	\label{eq:Strehl}
	\strehl \triangleq \frac{\PSF\Paren{\V{0}}}{\PSFtel\Paren{\V{0}}}
	\,.
\end{equation}
With the 2D~$\PSFres$ map being highly peaked in~$\Vk=\V{0}$ in the middle of the AO-corrected zone, it is a safe approximation to consider that
\begin{equation} 
	\label{eq:Strehl_res}
	\strehl \simeq \PSFres\Paren{\V{0}}
	\,.
\end{equation}

\tab{tab:strehl_disk} and \tab{tab:strehl_sup} compare the Strehl ratios obtained from the Monte Carlo simulations via \eq{eq:Strehl} and from the analytical model via \eq{eq:Strehl_res} for the different influence function shapes. Once again, except for the 2D~$\sinc$ function, the Strehl ratios are in good agreement. As in \refsec{sec:res_PSD}, the Gaussian profile gives similar results to the prediction of the 2D~$\sinc$ perfect case, with~$\strehl \simeq \percent{80}$. The ALPAO DM also provides a similar performance of ~$\simeq\percent{77.5}$. Though not shown in this paper, we highlight that these values compare well with the expected variance on the pupil from \refsec{sec:res_PSD} via the extended Maréchal's approximation~\citep{Tyson:15_principles_of_AO}
\begin{equation} 
	\label{eq:Strehl_Marechal}
	\strehl \simeq \E{-\AvgT{\sigAO^{2}}}
	\,.
\end{equation}

\begin{table}[t!] 
	\caption{\label{tab:strehl_disk} Strehl ratio~$\strehl$.}
	\centering
	\begin{tabular}{ccc}
	\hline
	\hline
	\multirow{2}{*}{$\IFref$} & \multicolumn{2}{c}{$\strehl$}								 \\ \cline{2-3} 
			 & \multicolumn{1}{c}{Monte Carlo} & \multicolumn{1}{c}{Anaytical PSD}
    \\
	\hline
	Piston & $\percent{33.7}$ & $\percent{29.7}$
	\\
	Pyramid & $\percent{75.4}$ & $\percent{74.1}$
	\\
	Gaussian & $\percent{80.4}$ & $\percent{79.7}$
	\\
	ALPAO & $\percent{77.5}$ & $\percent{77.4}$
	\\
	2D~$\sinc$ & $\percent{72.4}$ & $\percent{79.8}$
	\\
	Binary mask & -- & $\percent{79.8}$
	\\ \hline
	\end{tabular}
	\tablefoot{Comparison of the Monte Carlo simulations and the analytical PSD prediction. -- Aperture: Disc of diameter~$\percent{95}\Dsim$.}
\end{table}

As described by \cite{Cavarroc:06_perfect_coro} and \cite{Sauvage:10_AO_long_exp_perfect_coro}, obtaining the on-axis instantaneous PSF through a perfect coronagraph consists of removing the diffraction-limited PSF~$\PSFtel$ with an energy of~$\E{-\sigAO^{2}\Paren{t}}$. This leads to
\begin{equation}
	\AvgT{\E{-\sigAO^{2}}}
	= \E{-\AvgT{\sigAO^{2}}}\E{\VarD{\sigAO^{2}}{t}/2}
	\simeq \E{-\AvgT{\sigAO^{2}}}\Paren{1+\VarD{\sigAO^{2}}{t}/2}
	\,,
\end{equation}
where $\VarD{\sigAO^{2}}{t}$ is the variance of~$\sigAO^{2}\Paren{t}$. Except for the piston case, the Monte Carlo simulations show that the variance on~$\sigAO^{2}\Paren{t}$ is only a few~$10^{-4}$ and thus
\begin{equation}
	\AvgT{\E{-\sigAO^{2}}} \simeq \E{-\AvgT{\sigAO^{2}}}
	\,.
\end{equation}
Combined with \eqs{eq:Strehl_res}{eq:Strehl_Marechal}, this means that the long-exposure coronagraph PSF is
\begin{equation} 
	\label{eq:FTlongPSF_coro}
	\PSF^{\Tag{coro}} = \convprod{\PSFres^{\Tag{coro}}}{\PSFtel}
	\text{ with }
	\PSFres^{\Tag{coro}}\Paren{\Vk} = 
	\begin{cases}
		0 & \text{ for } \Vk = \V{0}
		\\
		\PSFres\Paren{\Vk} & \text{ otherwise}
	\end{cases}
	\,.
\end{equation}
As a consequence, the 2D map of~$\PSFres$ represents the limit of the achievable contrast of a perfect coronagraph due to the fitting error, that is to say, the intrinsic limit of an AO system due to its WFC optical properties.

\figfull{fig:PSF_pha_disk_95} also shows that the transition between the AO-corrected zone and the turbulent halo highly depends on the influence function shape~$\IFref$. Thus, though the contrast is similar for the ALPAO and Gaussian profiles ($10^{-5.6}$ \vs $10^{-5.7}$), the area in which this contrast is achieved shrinks with the ALPAO profile by~$\sim 1.5 \Dsim^{-1}$ on each side of the PSF. This needs to be compared with the theoretical radius of the corrected area of~$1/\Paren{2\pitch} = 7.5\Dsim^{-1}$. These different features are well visible in the profiles shown in \subfig{fig:PSF_pha_disk_95}{g} where the PSD analytical model profiles match almost perfectly with the Monte Carlo simulations for all the profiles but the 2D~$\sinc$ shape. From its PSD model, and as purposely designed, the 2D~$\sinc$ shape gives a sharp and homogeneous corrected zone that is equal to the prediction of the binary mask model. But similar to the PSD profile (\refsec{sec:res_PSD}), this model based on the stationarity of the structure function failed at reproducing the Monte Carlo simulations.

We note here that the analytical PSD is a convenient tool to quickly assess the performance of the system for different Fried parameters~$\rFried$. Indeed, as seen in \eq{eq:PSDres}, the dependence in terms of~$\rFried^{-5/3}$ can be factorised. Thus, a normalised PSD for a given influence function profile can be computed once and for all and re-scaled according to the need for different~$\rFried$, avoiding a time-consuming rerunning of the Monte Carlo simulations for each scenario.

\begin{figure}[t!]
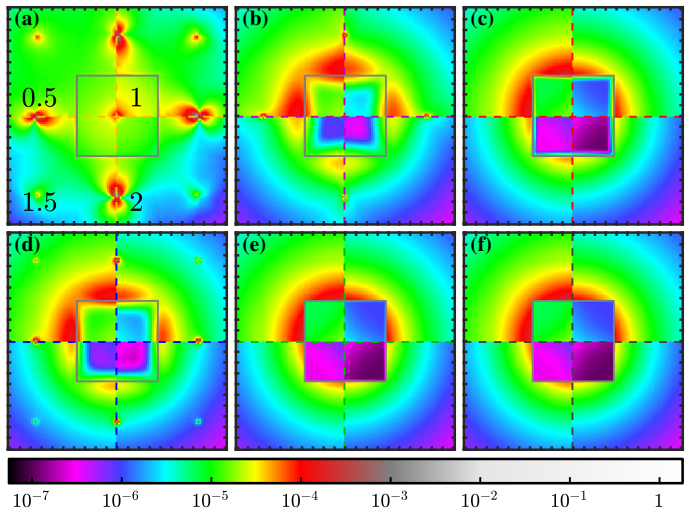
 
	\centering
	
	\newcommand{\PathFig}{figures_PSF_PSD_r_0_}
	\newcommand{\FlagAperture}{disk_95}
	\newcommand{\subfigColor}{black}	
	
	\newcommand{\LineRatio}{1}
	
	\newcommand{\FigOne}{\PathFig \FlagAperture _Piston.pdf}
	\newcommand{\FigTwo}{\PathFig \FlagAperture _Pyramid.pdf}       
	\newcommand{\FigThree}{\PathFig \FlagAperture _Gaussian.pdf}
	
	\sbox1{\includegraphics{\FigOne}}	       
	\sbox2{\includegraphics{\FigTwo}}       
	\sbox3{\includegraphics{\FigThree}}     
	\newcommand{\ColumnWidth}[1]
		{\dimexpr \LineRatio \linewidth * \AspectRatio{#1} / (\AspectRatio{1} + \AspectRatio{2} + \AspectRatio{3}) \relax
		}
	\newcommand{\ColumnGap}{\hspace {\dimexpr \linewidth /2 - \LineRatio\linewidth /2 }}

	\begin{tabular}{
		@{\ColumnGap}
		m{\ColumnWidth{1}}
		@{}
		m{\ColumnWidth{2}}
		@{}
		m{\ColumnWidth{3}}
		@{\ColumnGap}
		}
		\subfigimg[width=\linewidth,pos=ul,font=\fontfig{\subfigColor}]{$\;$(a)}{0.0}{\FigOne} &
		\subfigimg[width=\linewidth,pos=ul,font=\fontfig{\subfigColor}]{$\;$(b)}{0.0}{\FigTwo} &
		\subfigimg[width=\linewidth,pos=ul,font=\fontfig{\subfigColor}]{$\;$(c)}{0.0}{\FigThree}
	\end{tabular}
	
	
	
	\renewcommand{\FigOne}  {\PathFig \FlagAperture _ALPAO_repair.pdf}
	\renewcommand{\FigTwo}  {\PathFig \FlagAperture _SINC.pdf}      
	\renewcommand{\FigThree}{\PathFig \FlagAperture _Binary_mask.pdf}       
	
	\sbox1{\includegraphics{\FigOne}}	       
	\sbox2{\includegraphics{\FigTwo}}	   
	\sbox3{\includegraphics{\FigThree}}     
	\renewcommand{\ColumnWidth}[1]
		{\dimexpr \LineRatio \linewidth * \AspectRatio{#1} / (\AspectRatio{1} + \AspectRatio{2} + \AspectRatio{3}) \relax
		}
	\newcommand{\ColumnGapTwo}{\hspace {\dimexpr \linewidth /2 - \LineRatio\linewidth /2 }}
	
	\vspace{-5.5pt}
	
	\begin{tabular}{
		@{\ColumnGapTwo}
		m{\ColumnWidth{1}}
		@{}
		m{\ColumnWidth{2}}
		@{}
		m{\ColumnWidth{3}}
		@{\ColumnGapTwo}
		}
		\subfigimg[width=\linewidth,pos=ul,font=\fontfig{\subfigColor}]{$\;$(d)}{0.0}{\FigOne} &
		\subfigimg[width=\linewidth,pos=ul,font=\fontfig{\subfigColor}]{$\;$(e)}{0.0}{\FigTwo} &
		\subfigimg[width=\linewidth,pos=ul,font=\fontfig{\subfigColor}]{$\;$(f)}{0.0}{\FigThree}
	\end{tabular}
	
	\vspace{-5.5pt}

	\centering
	\subfigimg[width=\linewidth,pos=ul,font=\fontfig{black}]{}{0.0}{\PathFig bar.pdf}
	
	\caption{\label{fig:PSF_pha_r0_disk_95} Simulated PSF of the fitting residuals~$\PSFres$ for different ratios~$\rFried/\pitch$. Each panel shows the ratio value of $0.5$ (upper-left quadrant), $1$ (upper-right quadrant), $1.5$ (lower-left quadrant), and $2$ (lower-right quadrant). The grey squares represent the cut-off frequency~$\Abs{k_{i\in\Brace{1,2}}}<\Paren{2\pitch}^{-1}$. Panel~(a):~2D local piston. Panel~(b):~2D pyramid. Panel~(c):~2D axisymmetric Gaussian profile. Panel~(d):~2D axisymmetric ALPAO profile. Panel~(e):~2D~$\sinc$ function. Panel~(f):~Binary mask analytical model. -- Aperture: Disc of diameter~$\percent{95}\Dsim$.}
\end{figure}

\figfull{fig:PSF_pha_r0_disk_95} shows the different PSF of the fitting residuals for~$\rFried/\pitch\in\Brace{0.5,1,1.5,2}$. The dependence of the contrast due to the fitting error according to~$\rFried$ is clearly visible. Interestingly, it also appears that if~$\rFried$ directly impacts the contrast depth, it has a negligible impact on the shape of the transition area.

Additional situations are presented in \refapp{app:add_materials}. As mentioned for the PSD, the 2D maps of the simulated PSF of the fitting residuals obtained from the analytical PSD mode are highly similar whatever the aperture since they depend on it only negligibly via \eqs{eq:PSDpiston_rmv}{eq:PSDtt_rmv}. As for the PSD, \refsec{sec:res_PSD}, due to the limited sampling for the VLT-like case, the piston and pyramid profiles present some artefacts, as seen in \subfigspar{fig:PSF_PSD_appen}{4a,4b}. For the latter, the transition area was nonetheless properly retrieved.

\subsection{Predicted structure function}
\label{sec:res_SF}

\begin{figure}[t!]
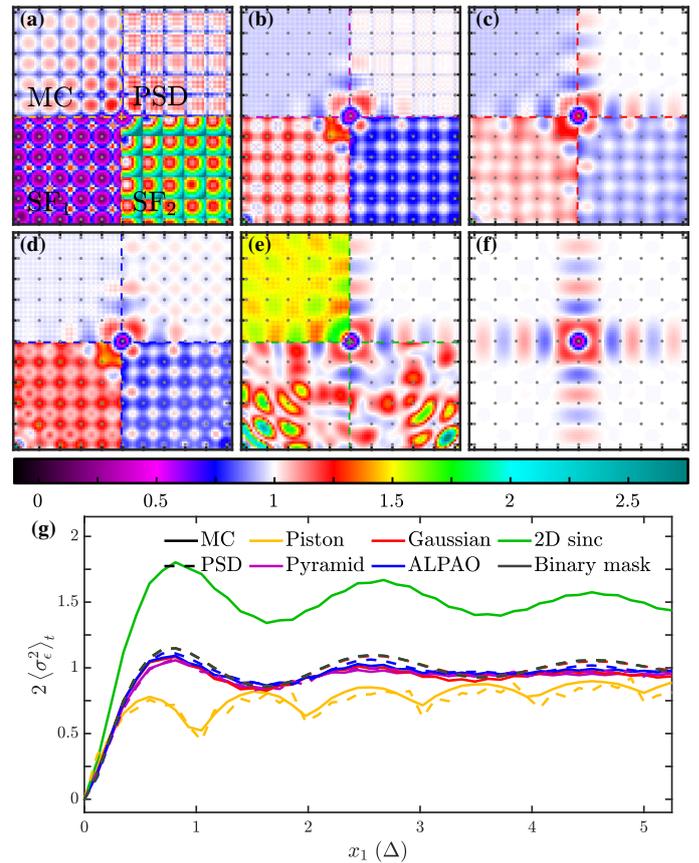
 
	\centering
	
	\newcommand{\PathFig}{figures_SF_}
	\newcommand{\FlagAperture}{disk_95}
	\newcommand{\subfigColor}{black}	
	
	\newcommand{\LineRatio}{1}
	
	\newcommand{\FigOne}{\PathFig \FlagAperture _Piston.pdf}
	\newcommand{\FigTwo}{\PathFig \FlagAperture _Pyramid.pdf}       
	\newcommand{\FigThree}{\PathFig \FlagAperture _Gaussian.pdf}
	
	\sbox1{\includegraphics{\FigOne}}	       
	\sbox2{\includegraphics{\FigTwo}}       
	\sbox3{\includegraphics{\FigThree}}     
	\newcommand{\ColumnWidth}[1]
		{\dimexpr \LineRatio \linewidth * \AspectRatio{#1} / (\AspectRatio{1} + \AspectRatio{2} + \AspectRatio{3}) \relax
		}
	\newcommand{\ColumnGap}{\hspace {\dimexpr \linewidth /2 - \LineRatio\linewidth /2 }}

	\begin{tabular}{
		@{\ColumnGap}
		m{\ColumnWidth{1}}
		@{}
		m{\ColumnWidth{2}}
		@{}
		m{\ColumnWidth{3}}
		@{\ColumnGap}
		}
		\subfigimg[width=\linewidth,pos=ul,font=\fontfig{\subfigColor}]{$\;$(a)}{0.0}{\FigOne} &
		\subfigimg[width=\linewidth,pos=ul,font=\fontfig{\subfigColor}]{$\;$(b)}{0.0}{\FigTwo} &
		\subfigimg[width=\linewidth,pos=ul,font=\fontfig{\subfigColor}]{$\;$(c)}{0.0}{\FigThree}
	\end{tabular}
	
	

	\renewcommand{\FigOne}  {\PathFig \FlagAperture _ALPAO_repair.pdf}
	\renewcommand{\FigTwo}  {\PathFig \FlagAperture _SINC.pdf}      
	\renewcommand{\FigThree}{\PathFig \FlagAperture _Binary_mask.pdf}       
	
	\sbox1{\includegraphics{\FigOne}}	       
	\sbox2{\includegraphics{\FigTwo}}	   
	\sbox3{\includegraphics{\FigThree}}     
	\renewcommand{\ColumnWidth}[1]
		{\dimexpr \LineRatio \linewidth * \AspectRatio{#1} / (\AspectRatio{1} + \AspectRatio{2} + \AspectRatio{3}) \relax
		}
	\newcommand{\ColumnGapTwo}{\hspace {\dimexpr \linewidth /2 - \LineRatio\linewidth /2 }}
	
	\vspace{-5.5pt}
	
	\begin{tabular}{
		@{\ColumnGapTwo}
		m{\ColumnWidth{1}}
		@{}
		m{\ColumnWidth{2}}
		@{}
		m{\ColumnWidth{3}}
		@{\ColumnGapTwo}
		}
		\subfigimg[width=\linewidth,pos=ul,font=\fontfig{\subfigColor}]{$\;$(d)}{0.0}{\FigOne} &
		\subfigimg[width=\linewidth,pos=ul,font=\fontfig{\subfigColor}]{$\;$(e)}{0.0}{\FigTwo} &
		\subfigimg[width=\linewidth,pos=ul,font=\fontfig{\subfigColor}]{$\;$(f)}{0.0}{\FigThree}
	\end{tabular}
	
	\vspace{-5.5pt}

	\centering
	\subfigimg[width=\linewidth,pos=ul,font=\fontfig{black}]{}{0.0}{\PathFig bar.pdf}

	\centering      
	\subfigimg[width=0.95\linewidth,pos=ul,font=\fontfig{black}]{(g)}{0.0}{\PathFig \FlagAperture _profile.pdf}
	
	\caption{\label{fig:SF_disk_95} Simulated structure function of the fitting residuals~$\SFAO$ normalised by~$2\AvgT{\sigAO^{2}}$ for the different profiles. Each panel shows the structure function simulated via Monte Carlo simulations (MC, upper-left corner), via the analytical PSD (upper-right corner), and via the analytical structure function (SF$_{1}$ and SF$_{2}$, lower-left and -right corners, respectively). Two 2D maps produced via the analytical structure function  are displayed: $\SFAO\Paren{\Vx,\V{0}}$ (SF$_{1}$) and $\SFAO\Paren{\Vx+\Vpitch/2,\Vpitch/2}$ (SF$_{2}$). The dots emphasise the actuator positions. Panel~(a):~2D local piston. Panel~(b):~2D pyramid. Panel~(c):~2D axisymmetric Gaussian profile. Panel~(d):~2D axisymmetric ALPAO profile. Panel~(e):~2D~$\sinc$ function. Panel~(f):~Binary mask analytical model. Panel~(g):~$x_{1}$-profiles of Panels~(a-f) of the Monte Carlo simulations (solid lines) and the PSD analytical models (dashed lines). -- Aperture: Disc of diameter~$\percent{95}\Dsim$.}
\end{figure}

The last physical quantity that can be studied is the structure function of the fitting residuals, $\SFAO$. As discussed in \refapp{app:SFf}, the structure function of the Monte Carlo simulations was obtained via the frequency domain, padding the pupil by a factor of~$\nPad = 3$. We note here that using this method to get the structure function implies its stationarity. The structure function induced by the analytical PSD was also computed in the Fourier domain via \eq{eq:SF_FT}. Finally, the structure function can also be analytically computed via \eq{eq:SF_analytic}. In \fig{fig:SF_disk_95}, we display two maps: $\SFAO\Paren{\Vx,\V{0}}$ and $\SFAO\Paren{\Vx+\Vpitch/2,\Vpitch/2}$. Under the hypothesis of stationarity, the maps should be equal. For clarity, the 2D maps were normalised by twice the expected variance averaged on the pupil,~$2\AvgT{\sigAO^{2}}$, which was obtained via the analytical PSD (see \tab{tab:var_disk}, last column). Indeed, one can expect that after AO correction, points that are far apart, $\Norm{\Vx}\gg\pitch$, are independent, implying via \eq{eq:SF_var} that
\begin{equation}
	\SF{\pha}\Paren{\Vx} \underset{\Norm{\Vx}\gg\pitch}{\simeq} 2\AvgT{\phaAO^{2}\Paren{\Vx,t}}
	\,,
\end{equation}
whose value should vary around the mean value across the pupil. Additional situations are presented in \refapp{app:add_materials}, \fig{fig:SF_appen}.

Once again, the predictions of the analytical PSD model are in very good agreement with the Monte Carlo simulations (two upper quadrants in \subfigs{fig:SF_disk_95}{a-e}) both in terms of amplitude and shape of the structure function, except for the 2D~$\sinc$ influence function. The small discrepancy in amplitude amongst the Monte Carlo simulations for the different profiles matches the ratio of the expected variance found in \refsec{sec:res_PSD} (see \tab{tab:var_disk}, two last columns).

As a general pattern, the structure functions present oscillations around twice the expected variance along the orthogonal directions of the actuator grid. The number and the amplitude of the oscillations depend on the profile, as seen in \subfig{fig:SF_disk_95}{g}. Once again, the Gaussian profile gives a result that is very close to the analytical prediction for the 2D~$\sinc$ influence function. The structure function obtained with the 2D~$\sinc$ case overlaps the binary mask model. The pyramid profile leads to less oscillations. The ALPAO profile lies between the two previous cases.

When it comes to the analytical model for the structure function (two lower quadrants in \subfigs{fig:SF_disk_95}{a-e}), the comparison is less favourable. Oscillations around twice the expected variance are still present, but their amplitudes are higher, and they are not attenuated, thus spreading all over the pupil. In addition, it clearly appears that $\SFAO\Paren{\Vx,\V{0}}\neq\SFAO\Paren{\Vx+\Vpitch/2,\Vpitch/2}$ ($\text{SF}_{1}\neq\text{SF}_{2}$). The situation gets worse for the 2D~$\sinc$ influence function, as the analytical structure functions do not present any oscillations but large-scale structured patterns.

These divergences from the analytical PSD model prediction support the fact that for all the influence function profiles, the structure function of the fitting residuals is not stationary. To further investigate this feature, we display in \fig{fig:map_SF_disk_95} the inhomogeneity in the pupil of the analytical structure function~$\SFAO$ of the fitting residuals for two different relative distances: $\SFAO\Paren{\Vx+\Vpitch/2,\Vx}$ and $\SFAO\paren{\Vx+\Paren{\pitch/2,0}\T,\Vx}$. A stationary structure function would produce constant maps: $\SFAO\Paren{\Vpitch/2}$ and $\SFAO\paren{\Paren{\pitch/2,0}\T},$ respectively, as given by the binary mask model in \subfig{fig:map_SF_disk_95}{f}. Additional situations are presented in \refapp{app:add_materials}, \fig{fig:map_SF_appen}. As previously stated, for the sake of readability, the 2D maps were normalised by~$2\AvgT{\sigAO^{2}}$.

\begin{figure}[t!] 
	\centering
	
	\newcommand{\PathFig}{figures_SF_map_}
	\newcommand{\FlagAperture}{disk_95}
	\newcommand{\subfigColor}{white}	
	
	\newcommand{\LineRatio}{1}
	
	\newcommand{\FigOne}{\PathFig \FlagAperture _Piston.pdf}
	\newcommand{\FigTwo}{\PathFig \FlagAperture _Pyramid.pdf}       
	\newcommand{\FigThree}{\PathFig \FlagAperture _Gaussian.pdf}
	
	\sbox1{\includegraphics{\FigOne}}	       
	\sbox2{\includegraphics{\FigTwo}}       
	\sbox3{\includegraphics{\FigThree}}     
	\newcommand{\ColumnWidth}[1]
		{\dimexpr \LineRatio \linewidth * \AspectRatio{#1} / (\AspectRatio{1} + \AspectRatio{2} + \AspectRatio{3}) \relax
		}
	\newcommand{\ColumnGap}{\hspace {\dimexpr \linewidth /2 - \LineRatio\linewidth /2 }}

	\begin{tabular}{
		@{\ColumnGap}
		m{\ColumnWidth{1}}
		@{}
		m{\ColumnWidth{2}}
		@{}
		m{\ColumnWidth{3}}
		@{\ColumnGap}
		}
		\subfigimg[width=\linewidth,pos=ul,font=\fontfig{\subfigColor}]{$\;$(a)}{0.0}{\FigOne} &
		\subfigimg[width=\linewidth,pos=ul,font=\fontfig{\subfigColor}]{$\;$(b)}{0.0}{\FigTwo} &
		\subfigimg[width=\linewidth,pos=ul,font=\fontfig{\subfigColor}]{$\;$(c)}{0.0}{\FigThree}
	\end{tabular}
	
	

	\renewcommand{\FigOne}  {\PathFig \FlagAperture _ALPAO_repair.pdf}
	\renewcommand{\FigTwo}  {\PathFig \FlagAperture _SINC.pdf}      
	\renewcommand{\FigThree}{\PathFig \FlagAperture _Binary_mask.pdf}       
	
	\sbox1{\includegraphics{\FigOne}}	       
	\sbox2{\includegraphics{\FigTwo}}	   
	\sbox3{\includegraphics{\FigThree}}     
	\renewcommand{\ColumnWidth}[1]
		{\dimexpr \LineRatio \linewidth * \AspectRatio{#1} / (\AspectRatio{1} + \AspectRatio{2} + \AspectRatio{3}) \relax
		}
	\newcommand{\ColumnGapTwo}{\hspace {\dimexpr \linewidth /2 - \LineRatio\linewidth /2 }}
	
	\vspace{-5.5pt}
	
	\begin{tabular}{
		@{\ColumnGapTwo}
		m{\ColumnWidth{1}}
		@{}
		m{\ColumnWidth{2}}
		@{}
		m{\ColumnWidth{3}}
		@{\ColumnGapTwo}
		}
		\subfigimg[width=\linewidth,pos=ul,font=\fontfig{\subfigColor}]{$\;$(d)}{0.0}{\FigOne} &
		\subfigimg[width=\linewidth,pos=ul,font=\fontfig{\subfigColor}]{$\;$(e)}{0.0}{\FigTwo} &
		\subfigimg[width=\linewidth,pos=ul,font=\fontfig{\subfigColor}]{$\;$(f)}{0.0}{\FigThree}
	\end{tabular}
	
	\vspace{-5.5pt}

	\centering
	\subfigimg[width=\linewidth,pos=ul,font=\fontfig{black}]{}{0.0}{\PathFig bar.pdf}
	
	\caption{\label{fig:map_SF_disk_95} Inhomogeneity in the pupil of the analytical structure function of the fitting residuals~$\SFAO$ for two different relative distances normalised by~$2\AvgT{\sigAO^{2}}$. Within each panel, on the left is $\SFAO\Paren{\Vx+\Vpitch/2,\Vx}$ (SF$_{1}$), and on right is $\SFAO\paren{\Vx+\Paren{\pitch/2,0}\T,\Vx}$ (SF$_{2}$). The dots emphasise the actuator positions. Panel~(a):~2D local piston. Panel~(b):~2D pyramid. Panel~(c):~2D axisymmetric Gaussian profile. Panel~(d):~2D axisymmetric ALPAO profile. Panel~(e):~2D~$\sinc$ function. Panel~(f):~Binary mask analytical model. -- Aperture: Disc of diameter~$\percent{95} \Dsim$.}
\end{figure}

Once again, it appears that the characteristic of stationarity is not met. The structure function does not only depend on the relative distance but also on the position in the pupil, and this fact applies to all the profiles. As discussed in \refsec{sec:SF_analytical}, the analytical structure function given by \eq{eq:SF_analytic} cannot be assumed to be stationary, thus jeopardising the analytical development of the PSD model in \refsec{sec:PSD_analytical}.

The resulting problem of this lack of stationarity is how to explain the similarities between the Monte Carlo simulations and the analytical model for the PSD in the previous sections. Except for the 2D~$\sinc$ influence function, the Monte Carlo simulations and the PSD analytical model indeed produce similar results. Looking closer at \fig{fig:map_SF_disk_95}, it appears that for all but the 2D~$\sinc$ shape, the structure functions are 2D periodic with a period~$\pitch$ on each axis of the actuator grid. As discussed in \refapp{app:SF_periodic}, this situation is encountered when the support of the influence function is finite. For the Gaussian and ALPAO influence functions, the profiles drop quickly to negligible values, leading to the assumption that they are indeed spatially limited. Thus, if the size of the pupil features (\eg shape, phase aberrations, size) are bigger than the actuator pitch~$\Paren{\gg\pitch}$ at the optical scale, the exponential of the structure function can be locally replaced by its average value in \eq{eq:FTlongPSF_SF}
\begin{equation}
	\label{eq:FTlongPSF_approx}
	\FTPSF\Paren{\Vx} \simeq \int\pupil\Paren{\Vxp}\conj{\pupil}\Paren{\Vxp+\Vx}\avg*{\E{-\frac{1}{2}\SFAO\Paren{\Vu,\Vx+\Vu}}}_{\Vu}\Dxp
	\,.
\end{equation}
If, in addition, the inhomogeneities of the structure function are small, the following expression  comes
\begin{align}
	\FTPSF\Paren{\Vx} {}\simeq{} & \int\pupil\Paren{\Vxp}\conj{\pupil}\Paren{\Vxp+\Vx}\E{-\frac{1}{2}\avg*{\SFAO\Paren{\Vu,\Vx+\Vu}}_{\Vu}}\Dxp
	\\
	{}\simeq{} & \E{-\frac{1}{2}\SFAO\Paren{\Vx}}\int\pupil\Paren{\Vxp}\conj{\pupil}\Paren{\Vxp+\Vx}\Dxp
	\,,
\end{align}
which is the result of \eq{eq:FTlongPSF_SF_stat} under the stationary hypothesis and where
\begin{align}
	\SFAO\Paren{\Vx} {}={} & \avg*{\SFAO\Paren{\Vxp,\Vx+\Vxp}}_{\Vxp}
	\\
	{}={} & \frac{\int\pupil\Paren{\Vxp}\pupil\Paren{\Vxp+\Vx}\SFAO\Paren{\Vxp,\Vx+\Vxp}\Dxp}{\int\pupil\Paren{\Vxp}\pupil\Paren{\Vxp+\Vx}\Dxp}
	\,,
\end{align}
as already proposed by \cite{Conan:94_PhD} and further studied by \cite{Veran:97_PSF_AO_telemetry}.

As a conclusion, under the previously mentioned hypotheses, the structure function can be considered stationary compared with the other features of the system, and the PSD analytical model provides a good approximation to predict the impact of the influence function profiles on the AO-corrected PSF. On the contrary, when these conditions are not met, such as for the 2D~$\sinc$ shape, the AO-corrected structure function cannot be approximated to be stationary, and the PSD analytical approach fails to match the Monte Carlo simulations.

We also mention that the hypothesis behind \eq{eq:FTlongPSF_approx} on the pupil spatial features relative to the actuator pitch is directly linked to the actuator density in the pupil. The XAO systems are high-performance systems with many actuators leading to small $\pitch/D$ ratios. They are thus very favourable cases to assume the stationarity of the structure function across the pupil compared to its spatial features. This is supported by the structure functions of the VLT-like simulations presented in \subfigspar{fig:SF_appen}{4a-4d} and \subfigspar{fig:map_SF_appen}{4a-4d} for spatially limited influence functions. The structure functions are still 2D periodic with a period~$\pitch$  smaller by a factor of two due to the increased number of actuators. This periodicity is broken only locally, in the vicinity of the central obscuration and the spider vanes, on a scale on the order of the actuator pitch~$\pitch$. We again highlight that the very poor performances in terms of periodicity and stationarity of the 2D~$\sinc$ shape, which is not spatially limited.

Finally, as mentioned in the introduction, we emphasise that these XAO cases are also the ones in which an accurate model of the fitting error is needed. Indeed, the other error terms become comparable when reaching the fundamental limits of the coronagraph instruments.

\subsection{Confrontation with a real case}
\label{sec:res_ALPAO}

\begin{figure}[t!]
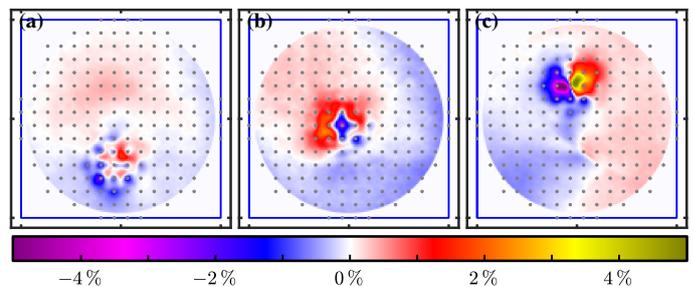
 
	\centering
	
	\newcommand{\LineRatio}{1}
	
	\newcommand{\PathFig}{figures_ALPAO_residuals_}
	\newcommand{\FlagAperture}{disk_95}
	\newcommand{\subfigColor}{black}		
	
	\newcommand{\FigOne}{\PathFig \FlagAperture _85.pdf}
	\newcommand{\FigTwo}{\PathFig \FlagAperture _88.pdf}    
	\newcommand{\FigThree}{\PathFig \FlagAperture _91.pdf}
	
	\sbox1{\includegraphics{\FigOne}}	       
	\sbox2{\includegraphics{\FigTwo}}	       
	\sbox3{\includegraphics{\FigThree}}     
	\newcommand{\ColumnWidth}[1]
		{\dimexpr \LineRatio \linewidth * \AspectRatio{#1} / (\AspectRatio{1} + \AspectRatio{2} + \AspectRatio{3}) \relax
		}
	\newcommand{\ColumnGap}{\hspace {\dimexpr \linewidth /2 - \LineRatio\linewidth /2 }}

	\begin{tabular}{
		@{\ColumnGap}
		m{\ColumnWidth{1}}
		@{}
		m{\ColumnWidth{2}}
		@{}
		m{\ColumnWidth{3}}
		@{}
		m{\ColumnWidth{4}}
		@{\ColumnGap}
		}
		\subfigimg[width=\linewidth,pos=ul,font=\fontfig{\subfigColor}]{$\;$(a)}{0.0}{\FigOne} &
		\subfigimg[width=\linewidth,pos=ul,font=\fontfig{\subfigColor}]{$\;$(b)}{0.0}{\FigTwo} &
		\subfigimg[width=\linewidth,pos=ul,font=\fontfig{\subfigColor}]{$\;$(c)}{0.0}{\FigThree}
	\end{tabular}
	
	\vspace{-5.5pt}

	\centering
	\subfigimg[width=\linewidth,pos=ul,font=\fontfig{black}]{}{0.0}{\PathFig bar.pdf}
	
	\caption{\label{fig:ALPAO_residuals} Influence function model residuals for different actuators on the circular aperture of diameter~$\percent{95}\Dsim$. The blue frames represent the simulation domain~$\Dsim$. The dots emphasise the actuator positions. The scale is the same as in \fig{fig:IF_profile} where the axisymmetric model is normalised between zero and one. In Panel~(a),~$a=85$; in  Panel~(b),~$a=88$, and in Panel~(c)~$a=91$.}
\end{figure}

In this section, we assess if the analytical model, assuming identical influence functions dispatched on a regular grid, can be used to predict the performances of a real system. The residuals of the ALPAO influence function model are presented in \fig{fig:ALPAO_residuals} for three different actuators. The residuals are defined as the difference between the true measured profile minus the 2D axisymmetric model. We note that this model is globally fitted on all the actuators jointly using a spline model, as described by \cite{Berdeu:22_SPIE_EvWaCo_AO_bench}. Consequently, this is the model that at best globally explains the influence function of all the actuators.

It is natural to expect that there will be some discrepancies between a real and an ideal model. For example, some actuators present an influence function that is larger, such as in \subfig{fig:ALPAO_residuals}{b}, or thinner than the fitted average profile. Others do not exactly lie on the regular grid, such as in \subfig{fig:ALPAO_residuals}{c}. Finally, even if the width and the position of the true influence function match the ideal model, the profile is not exactly axisymmetric, such as in \subfig{fig:ALPAO_residuals}{a} where it clearly appears that there is some cross-talk with the neighbouring actuators, which slightly pull back the deformable membrane.

\begin{figure}[t!]
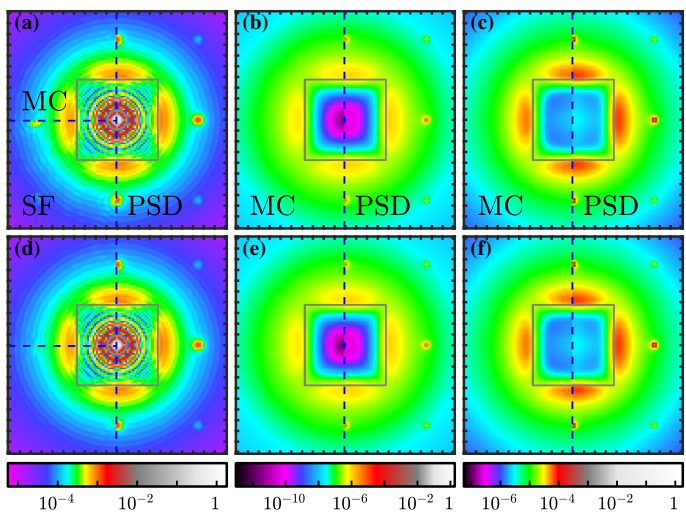
 
	\centering
	
	\newcommand{\LineRatio}{1}
	
	\newcommand{\PathFig}{figures_ALPAO_compare_}
	\newcommand{\FlagAperture}{disk_95}
	\newcommand{\subfigColor}{black}		
	
	\newcommand{\FigOne}{\PathFig \FlagAperture _PSF.pdf}
	\newcommand{\FigTwo}{\PathFig \FlagAperture _PSD.pdf}   
	\newcommand{\FigThree}{\PathFig \FlagAperture _PSF_pha.pdf}
	
	\sbox1{\includegraphics{\FigOne}}	       
	\sbox2{\includegraphics{\FigTwo}}       
	\sbox3{\includegraphics{\FigThree}}     
	\newcommand{\ColumnWidth}[1]
		{\dimexpr \LineRatio \linewidth * \AspectRatio{#1} / (\AspectRatio{1} + \AspectRatio{2} + \AspectRatio{3}) \relax
		}
	\newcommand{\ColumnGap}{\hspace {\dimexpr \linewidth /2 - \LineRatio\linewidth /2 }}

	\begin{tabular}{
		@{\ColumnGap}
		m{\ColumnWidth{1}}
		@{}
		m{\ColumnWidth{2}}
		@{}
		m{\ColumnWidth{3}}
		@{\ColumnGap}
		}
		\subfigimg[width=\linewidth,pos=ul,font=\fontfig{\subfigColor}]{$\;$(a)}{0.0}{\FigOne} &
		\subfigimg[width=\linewidth,pos=ul,font=\fontfig{\subfigColor}]{$\;$(b)}{0.0}{\FigTwo} &
		\subfigimg[width=\linewidth,pos=ul,font=\fontfig{\subfigColor}]{$\;$(c)}{0.0}{\FigThree}
	\end{tabular}
	
	\renewcommand{\FlagAperture}{disk_95_TT}
	
	\sbox1{\includegraphics{\FigOne}}	       
	\sbox2{\includegraphics{\FigTwo}}       
	\sbox3{\includegraphics{\FigThree}}     
	\renewcommand{\ColumnWidth}[1]
		{\dimexpr \LineRatio \linewidth * \AspectRatio{#1} / (\AspectRatio{1} + \AspectRatio{2} + \AspectRatio{3}) \relax
		}
	\renewcommand{\ColumnGap}{\hspace {\dimexpr \linewidth /2 - \LineRatio\linewidth /2 }}
	
	\vspace{-5.5pt}
	
	\begin{tabular}{
		@{\ColumnGap}
		m{\ColumnWidth{1}}
		@{}
		m{\ColumnWidth{2}}
		@{}
		m{\ColumnWidth{3}}
		@{\ColumnGap}
		}
		\subfigimg[width=\linewidth,pos=ul,font=\fontfig{\subfigColor}]{$\;$(d)}{0.0}{\FigOne} &
		\subfigimg[width=\linewidth,pos=ul,font=\fontfig{\subfigColor}]{$\;$(e)}{0.0}{\FigTwo} &
		\subfigimg[width=\linewidth,pos=ul,font=\fontfig{\subfigColor}]{$\;$(f)}{0.0}{\FigThree}
	\end{tabular}

	\renewcommand{\FlagAperture}{bar}
	
	\sbox1{\includegraphics{\FigOne}}	       
	\sbox2{\includegraphics{\FigTwo}}       
	\sbox3{\includegraphics{\FigThree}}     
	\renewcommand{\ColumnWidth}[1]
		{\dimexpr \LineRatio \linewidth * \AspectRatio{#1} / (\AspectRatio{1} + \AspectRatio{2} + \AspectRatio{3}) \relax
		}
	\renewcommand{\ColumnGap}{\hspace {\dimexpr \linewidth /2 - \LineRatio\linewidth /2 }}
	
	\vspace{-5.5pt}
	
	\begin{tabular}{
		@{\ColumnGap}
		m{\ColumnWidth{1}}
		@{}
		m{\ColumnWidth{2}}
		@{}
		m{\ColumnWidth{3}}
		@{\ColumnGap}
		}
		\subfigimg[width=\linewidth,pos=ul,font=\fontfig{\subfigColor}]{}{0.0}{\FigOne} &
		\subfigimg[width=\linewidth,pos=ul,font=\fontfig{\subfigColor}]{}{0.0}{\FigTwo} &
		\subfigimg[width=\linewidth,pos=ul,font=\fontfig{\subfigColor}]{}{0.0}{\FigThree}
	\end{tabular}
	
	\caption{\label{fig:ALPAO_compare} Comparison of the results obtained with the true influence functions (Monte Carlo simulations) and with the 2D axisymmetric profiles (analytical model).
	Panel~(a):~Simulated long-exposure PSF~$\PSF$.
	Panel~(b):~Simulated PSD of the fitting residuals~$\PSDres$ normalised by~$\rFried^{-5/3}$.
	Panel~(c):~Simulated PSF of the fitting residuals~$\PSFres$.
	The Monte Carlo simulations (MC) are performed with the true influence functions. The analytical structure function (SF) and analytical PSD are estimated with the axisymmetric profile.
	The grey squares represent the cut-off frequency~$\Abs{k_{i\in\Brace{1,2}}}<\Paren{2\pitch}^{-1}$.
	Panels~(d-f): Similar to Panels~(a-c) but with the tip-tilt removed.
	}
\end{figure}

\figfull{fig:ALPAO_compare} gathers the results of the simulations on the physical quantities studied in the previous sections, including the long-exposure PSF~$\PSF$, the PSD of the fitting residuals~$\PSDres$, and the PSF of the fitting residuals~$\PSFres$. The Monte Carlo simulations were run using the true influence functions, while the predictions of the analytical model (structure function and PSD) were obtained with the 2D axisymmetric analytical model.

We remark here that the prediction of the structure function analytical model  could have been obtained using the true influence functions in \eqs{eq:SF_analytic}{eq:Daap}. When doing so (not shown in this paper), the prediction of the analytical model exactly matches the result of the Monte Carlo simulations, as in the previous sections, once again validating our analytical expression.

Looking at \subfigs{fig:ALPAO_compare}{a-c}, the conclusions are similar to those of the previous sections: The analytical model is in very good agreement with the Monte Carlo predictions. The only noticeable difference is the absence of the diffracted spots in the turbulence halo. Except for this feature, the results validate that the hypothesis of identical influence functions replicated on a regular grid is a reasonable assumption and prove the robustness of the analytical approach when it comes to predicting the behaviour of real AO systems.

As discussed in the following paragraphs, these secondary spots are indeed not induced by a failure of this assumption, that is to say, a discrepancy between an ideal set of identical influence functions on a regular grid and a real set of influence functions whose profiles can slightly change in terms of width and position. Rather, the spots are due to the quality of the tipped-tilted flat that the DM can produce.

To test this hypothesis, the same simulations were run filtering out the tip-tilt from the incident wavefront. In the Monte Carlo simulations and the structure function analytical model, this was done by adding the tip-tilt mode to the influence function lists, which is equivalent to using  a fast steering mirror parallel to the DM to correct for these two modes. For the PSD analytical model, this is done via the method described in \refsec{sec:piston_TT}.

The results are given in \subfigs{fig:ALPAO_compare}{d-f}. The predictions of the structure function analytical model (see \subfig{fig:ALPAO_compare}{d}) became extremely similar to the predictions of the Monte Carlo simulations, showing once again the validity of the analytical model of \eq{eq:SF_analytic}. As predicted in \refsec{sec:piston_TT}, applying \eq{eq:PSDtt_rmv} in the analytical model only negligibly changes the core of the PSD and has no impact on the secondary spots, whose intensities are unchanged.

\begin{figure}[t!]
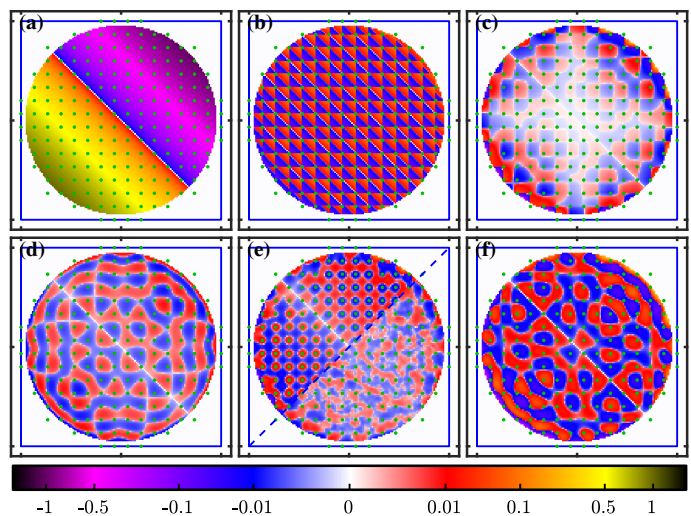
 
	\centering
	
	\newcommand{\LineRatio}{1}
	
	\newcommand{\PathFig}{figures_DM_flat_}
	\newcommand{\subfigColor}{black}		
	
	\newcommand{\FigOne}{\PathFig TT.pdf}
	\newcommand{\FigTwo}{\PathFig Piston.pdf}       
	\newcommand{\FigThree}{\PathFig Pyramid.pdf}
	
	\sbox1{\includegraphics{\FigOne}}	       
	\sbox2{\includegraphics{\FigTwo}}       
	\sbox3{\includegraphics{\FigThree}}     
	\newcommand{\ColumnWidth}[1]
		{\dimexpr \LineRatio \linewidth * \AspectRatio{#1} / (\AspectRatio{1} + \AspectRatio{2} + \AspectRatio{3}) \relax
		}
	\newcommand{\ColumnGap}{\hspace {\dimexpr \linewidth /2 - \LineRatio\linewidth /2 }}
	
	\begin{tabular}{
		@{\ColumnGap}
		m{\ColumnWidth{1}}
		@{}
		m{\ColumnWidth{2}}
		@{}
		m{\ColumnWidth{3}}
		@{\ColumnGap}
		}
		\subfigimg[width=\linewidth,pos=ul,font=\fontfig{\subfigColor}]{$\;$(a)}{0.0}{\FigOne} &
		\subfigimg[width=\linewidth,pos=ul,font=\fontfig{\subfigColor}]{$\;$(b)}{0.0}{\FigTwo} &
		\subfigimg[width=\linewidth,pos=ul,font=\fontfig{\subfigColor}]{$\;$(c)}{0.0}{\FigThree}
	\end{tabular}
	
	\renewcommand{\FigOne}{\PathFig Gaussian.pdf}
	\renewcommand{\FigTwo}{\PathFig ALPAO.pdf}      
	\renewcommand{\FigThree}{\PathFig SINC.pdf}
	
	\sbox1{\includegraphics{\FigOne}}	       
	\sbox2{\includegraphics{\FigTwo}}       
	\sbox3{\includegraphics{\FigThree}}     
		
	\vspace{-5.5pt}
	
	\begin{tabular}{
		@{\ColumnGap}
		m{\ColumnWidth{1}}
		@{}
		m{\ColumnWidth{2}}
		@{}
		m{\ColumnWidth{3}}
		@{\ColumnGap}
		}
		\subfigimg[width=\linewidth,pos=ul,font=\fontfig{\subfigColor}]{$\;$(d)}{0.0}{\FigOne} &
		\subfigimg[width=\linewidth,pos=ul,font=\fontfig{\subfigColor}]{$\;$(e)}{0.0}{\FigTwo} &
		\subfigimg[width=\linewidth,pos=ul,font=\fontfig{\subfigColor}]{$\;$(f)}{0.0}{\FigThree}
	\end{tabular}
	
	\vspace{-5.5pt}
	
	\centering
	\subfigimg[width=\linewidth,pos=ul,font=\fontfig{black}]{}{0.0}{\PathFig bar.pdf}
	
	\caption{\label{fig:DM_flat} Best tip-tilt correction residuals on a normalised tip-tilt. Panel~(a):~Tip-tilt normalised between minus one and one.  Panel~(b):~2D local piston. Panel~(c):~2D pyramid. Panel~(d):~2D axisymmetric Gaussian profile. Panel~(e):~Upper left: 2D axisymmetric ALPAO profile. Lower right: True ALPAO influence functions. Panel~(f):~2D~$\sinc$ function. The blue frames represent the simulation domain~$\Dsim$. The dots emphasise the actuator positions. -- Aperture: Disc of diameter~$\percent{95}\Dsim$.}
\end{figure}

To further support this hypothesis of the tipped-tilted flat quality of the DM, we analysed the residuals of a tip-tilt correction according to the different influence functions studied in this paper. The results are shown in \fig{fig:DM_flat}.

All the theoretical profiles presented structured patterns. But only the piston, the pyramid, and the axisymmetric ALPAO profiles presented patterns periodically reproduced on the actuator grid. The low order modes of the incident wavefront scatter on these regular patterns, producing the diffracted spots. Removing the tip-tilt with a fast steering mirror thus strongly dims the energy in the diffracted spots but not all of it, as the other low order modes are not corrected.

A closer look at \subfig{fig:DM_flat}{e} explains why the true influence functions from the DM192 do not produce diffraction spots in \subfig{fig:ALPAO_compare}{a}: The periodic structure mentioned for the 2D axisymmetric model is not present in the tip-tilt residuals. Thus, the incident light is not scattered.
This absence can be explained by the interactuator cross-talk mentioned previously and seen in \subfig{fig:ALPAO_residuals}{a}. The cross-talk is on the order of a percent of the actuator stroke, which is also the order of magnitude of the regular pattern obtained with the 2D axisymmetric model. We can consequently conclude that with the true influence functions, this pattern is pulled back by the neighbouring actuators, producing a better tipped-tilted flat than expected with the 2D axisymmetric model. Thus, a solution to obtain more accurate simulations would be to fit a more realistic 2D model of the influence functions to account for the cross-talk between the actuators, but this is beyond the scope of this paper.

\section{Discussion and conclusion}

In this work, we proposed an analytical model that accounts for the shape of the influence functions to model the fitting error. This model was tested on standard influence function profiles (local piston, pyramid, double Gaussian) as well as on a real set of influence functions (DM192 from ALPAO). This model is in very good agreement with the Monte Carlo simulations that validate the proposed approach.

In its most general framework, our analytical model gives access to the structure function after perfect AO correction. The main result is that contrary to the commonly made assumption, the structure function after AO correction is not stationary. Thanks to the analytical formula, our model enables the determination of the degree of the inhomogeneities that depends on the influence function profile. This non-stationarity has some implications, the main being that an AO-corrected wavefront cannot be defined by a stationary PSD. To be perfectly rigorous, this means that the general techniques to generate random wavefront from the PSD in simulations or to fit the PSF from the optical transfer function cannot be used. Similarly, some experimental developments in XAO use phase plates to emulate the residual wavefront after an initial stage of AO correction to feed a secondary AO loop. Due to the non-stationarity of the structure function, generating such a phase plate is then theoretically impossible or will just result in an approximation. This conclusion should nonetheless be tempered by the fact that phase plates are already an approximation of real turbulence (\eg atmosphere with few layers, frozen flow hypothesis, problem to properly emulate the inner and outer scales of the structure function) and that their main objectives are to validate and assess the performances of an OA system in controlled and reproducible conditions.

Under further hypotheses, the structure function can be approximated to be stationary. This is the case when the influence function is spatially limited and if the pupil is well behaved, that is to say, its spatial features spatially evolve on a longer scale than the actuator pitch. In this scenario, an analytical PSD model can be obtained. Our analysis shows negligible discrepancies with the Monte Carlo simulations. The proposed analytical model consequently provides a tool to quickly assess the impact of the influence function profiles in the instrument performances for different turbulence $\rFried$ parameters: PSD curves, expected variance on the pupil and fitting error, Strehl ratio, and contrast profile curves.

The 2D~$\sinc$ function, equivalent to the binary mask usually used in analytical models, does not meet the assumptions required for stationarity. Our study consequently shows that this ideal case is not representative of a real system, even when the system is as perfect and noiseless as it can be.

Contrary to standard analytical models, our analytical approach can predict the secondary spots due to the diffraction of the low spatial orders of the incident light on the WFC. These spots can be dimmed by the use of an additional tip-tilt modal corrector, as correctly predicted by our analytical structure function model, which is also in agreement with the Monte Carlo simulations. Nonetheless, our analytical PSD model, which relies on the structure function stationarity, cannot be correctly used in this context. Independently filtering the tip-tilt modes has no impact on these secondary spots. Despite this flaw, our analytical PSD model still correctly predicts the transition and the depth of the AO-corrected area, which are the prime interests of a model of the fitting residuals.

As already mentioned, the fitting error is the intrinsic optimal limit of an AO system. The pertinence of the model presented in this paper consequently lies in the possibility of extending it to other error terms modelled with similar accuracy and in the same formalism. Analytical models of additional terms (\eg WFS noise, aliasing, servo-lag) based on PSD modelling already exist, such as PAOLA developed by \cite{Jolissaint:06_AO_analytical}, and our analytical model of the fitting error is thus a natural extension of those models.

Finally, the limit of the stationarity hypothesis and its implications need to be further investigated. Indeed, it is expected that divergences from the stationarity occur when the pupil presents features on the order of the magnitude of the actuator pitch, for example when the WFC has dead actuators that can be masked or not. If the central obstruction of the secondary mirror or its spiders can also be a source of concern, this problem can be further enhanced for segmented primary mirrors. When static phase errors are introduced in the pupil to model static aberrations~\citep{Dohlen:16_SPHERE_performance_vs_prediction}, small patterns are also introduced, breaking this spatial assumption.

Indeed, if this non-stationarity seems to have negligible effects for standard PSF simulations, it may not be the case for high-contrast instruments where each diffracting element that is not correctly modelled can have a significant impact in the contrast curves. The contrast curves that are presented in this paper were obtained from the PSD and imply that the structure function is stationary. Preliminary results presented in this paper show that such XAO systems produce  a very favourable actuator density across the pupil, mitigating the effects on the structure function stationarity. Nonetheless, further analyses are needed to check the validity of the analytical PSD model in the framework of high-contrast imaging, for example, by combining our approach with the model proposed by \cite{Herscovici:17_coronagraph_PSF_analytical,Herscovici:19_NCPA_correction_coronagraph_analytic} that is based on the stationarity hypothesis.

\begin{acknowledgements}
	These activities have been performed in the framework of the EvWaCo project at the NARIT Center for Optics and Photonics. The authors are thankful to Alexis Jarry and Julien Charton from the ALPAO company who provided the measurements of the influence functions of the DM192 deformable mirror and for the fruitful discussions. The authors acknowledge the support from Chulalongkorn University’s CUniverse (CUAASC) grant.
\end{acknowledgements}

%
%

\bibliographystyle{aa}
\bibliography{bib_IF}

\flushcolsend

\begin{appendix}
\section{Notations}
\label{app:notations}
The following notations are used in this paper:
\begin{itemize}
	\item $\Vx = \Paren{x_{1}, x_{2}}\in\Reals^{2}$ is the 2D spatial coordinates, and $\Vk = \Paren{k_{1}, k_{2}}\in\Reals^{2}$ is its 2D frequency counterpart.
	\item $\Vpix = \Paren{\pix_{1}, \pix_{2}}\in\Reals^{2}$ is the 2D discrete spatial coordinates of the simulations, and $\Vpixk\in\Reals^{2}$ is its 2D discrete frequency counterpart.
	\item $\Vx\T\V{y} = x_{1}y_{1}+y_{2}k_{2}$ is the inner product of the two vectors~$\V{x}$ and~$\V{y}$.
	\item $\V{x}\EWprod\V{y} = \Paren{x_{1}y_{1},x_{2}y_{2}}\T$ is the element-wise product of the two vectors~$\V{x}$ and~$\V{y}$.
	\item $\FT{\pha}$ is the 2D Fourier transform of~$\pha$,
	\begin{align} 
		\label{eq:FT}
		\FT{\pha}\Paren{\Vk}
		& {}\triangleq{} \FTfull{\pha}{\Vk}
		{}={} \int{\pha\Paren{\Vx}\E{-2\I\pi\Vx\T\Vk}\Dx}
		\\ 
		\label{eq:FTinv}
		\pha\Paren{\Vx}
		& {}\triangleq{} \FTfullinv{\FT{\pha}}{\Vx}
		{}={} \int{\FT{\pha}\Paren{\Vk}\E{2\I\pi\Vx\T\Vk}\Dk}
		\,.
	\end{align}     
	\item $\DFT{\pha}\Paren{\Vpixk}$ is the 2D discrete Fourier transform of~$\pha\Paren{\Vpix}$. We note here that the DFT must be correctly normalised so that the energy is conserved (\ie so that the Parseval theorem holds).
	\item $\AvgT{\pha}$ is the temporal expectation of~$\pha\Paren{t}$.
	\item $\VarD{\pha}{t}$ is the temporal variance of~$\pha\Paren{t}$.
	\item $\PSD{\pha}\Paren{\Vk} = \AvgT{\Abs{\FT{\pha}\Paren{\Vk}}^{2}}$ is the power spectrum density of~$\pha\Paren{\Vx}$.
	\item $\SF{\pha}\Paren{\Vx,\Vxp} = \AvgT{\Paren{\pha\Paren{\Vx,t} - \pha\Paren{\Vxp,t}}^{2}}$ is the structure function of~$\pha\Paren{\Vx}$.
	\item $\SFd{\pha}\Paren{\Vpix}$ is the discrete structure function of~$\pha\Paren{\Vpix}$ computed in the direct space (see \refsec{app:SFd}).
	\item $\SFf{\pha}\Paren{\Vpix}$ is the discrete structure function of~$\pha\Paren{\Vpix}$ computed in the frequency domain (see \refsec{app:SFf}).
	\item $\conj{\pha}\Paren{\Vx}$ is the complex conjugate of~$\pha\Paren{\Vx}$.
	\item $\delta_{i,j}$ is the Kronecker delta.
	\item $\delta\Paren{\Vx}$ is the Dirac distribution.
	\item $\real{z}$ is the real part of the complex number~$z\in\Complexes$.
	\item $\Dsim$ is the diameter of the simulated domain.
	\item $\nactu$ is the number of actuators across~$\Dsim$.
	\item $\ntotactu$ is the total weighted number of actuators in the simulated domain.
	\item $\wa$ is the weight of the actuator~$\Va$ in the simulated domain.
	\item $\nwpair$ is the weighted number of actuator pairs in the simulated domain for the spacing~$\pair \in \Integers^{2}$.
\end{itemize}

\section{Optimal projector of the incident wavefront on the set of influence functions}
\label{app:proj_opti}

As derived by~\cite{Hudgin:77}, the optimal command~$\Vc\Paren{t}$ is the one minimising the residual variance (the so-called fitting error) of the fitting residuals,
\begin{equation}
	\Vc\Paren{t} = \argmin_{\Vc} \; \int\Abs{\phaAO\Paren{\Vx,t}}^{2}\Dx
	\,,
\end{equation}
where the fitting residuals~$\phaAO\Paren{\Vx,t}$ are defined in \eq{eq:pha_AO}. Expanding this equation via \eq{eq:phac} yields
\begin{equation}
	\label{eq:min_lin}
	\Vc\Paren{t} \triangleq \argmin_{\Vc} \; \Vc\T \M{A} \Vc - 2 \V{b}\T\Paren{t}\Vc
	\,,
\end{equation}
with 
\begin{equation}
	\Brack{\V{b}\Paren{t}}_{\ap} = \int\IFap\Paren{\Vx}\pha\Paren{\Vx,t}\Dx
	\,,
\end{equation}
and
\begin{equation}
	\Brack{\M{A}}_{a,\ap} = \int\IFa\Paren{\Vx}\IFap\Paren{\Vx}\Dx
	\,.
\end{equation}
The solution of \eq{eq:min_lin} is then classically given by
\begin{equation}
	\label{eq:proj_a}
	\Vc\Paren{t} = \M{A}\Inv\V{b}\Paren{t}
	\Rightarrow
	\forall a\in\setactu, \ca\Paren{t} = \int \proja \Paren{\Vx}\pha\Paren{\Vx,t}\Dx 
	\,,
\end{equation}
with
\begin{equation}
	\label{eq:proj_opti}
	\proja\Paren{\Vx} = \sum_{\ap\in\setactu} \Brack{\M{A}\Inv}_{a,\ap}\IFap\Paren{\Vx}
	\,.
\end{equation}
We note here that, by construction, $\proja\Paren{\Vx}$ is piston-free.

\section{Derivation of the analytical structure function of the fitting residuals}
\label{app:SF_analytical}

By definition, the structure function of the fitting residuals is
\begin{align}
	\SFAO\Paren{\Vx,\Vxp} & {}={} \AvgT{\Paren{\phaAO\Paren{\Vx,t} - \phaAO\Paren{\Vxp,t}}^{2}}
	\\
	& {}={} \AvgT{\Paren{
		\Paren{\pha\Paren{\Vx,t}-\pha\Paren{\Vxp,t}}
		-
		\Paren{\phac\Paren{\Vx,t}-\phac\Paren{\Vxp,t}}
		}^{2}}
	\\
	\label{eq:SF_AO_dev}
	& 
	\begin{aligned}
	{}={} & \SFpha\Paren{\Vx,\Vxp} + \SFc\Paren{\Vx,\Vxp} 
	\\
	& - 2\AvgT{
		\Paren{\pha\Paren{\Vx,t}-\pha\Paren{\Vxp,t}}
		\Paren{\phac\Paren{\Vx,t}-\phac\Paren{\Vxp,t}}
		}
	\,.
	\end{aligned}
\end{align}
We now focus on the cross term and the quadratic term of \eq{eq:SF_AO_dev}.

\subsection{Cross term}
From \eqs{eq:phac}{eq:proj_a}, it comes
\begin{equation}
	\label{eq:phac_dev}
	\begin{aligned}
	\phac\Paren{\Vx,t}-&\phac\Paren{\Vxp,t} = 
	\sum_{a\in\setactu} \Paren{\IFa\Paren{\Vx}-\IFa\Paren{\Vxp}}\int\proja\Paren{\Vu}\pha\Paren{\Vu,t}\Du
	\,.
	\end{aligned}
\end{equation}
Expanding the cross term in \eq{eq:SF_AO_dev} by using \eq{eq:phac_dev} yields
\begin{align}
	\chi\Paren{\Vx,\Vxp} {}\triangleq{} &
		\AvgT{
			\Paren{\pha\Paren{\Vx,t}-\pha\Paren{\Vxp,t}}
			\Paren{\phac\Paren{\Vx,t}-\phac\Paren{\Vxp,t}}
		}
	\\
	{}={} &
		\sum_{a \in \setactu} \Paren{\IFa\Paren{\Vx}-\IFa\Paren{\Vxp}}\PDa\Paren{\Vx,\Vxp}
	\,,
\end{align}
with
\begin{equation}
	\PDa\Paren{\Vx,\Vxp} \triangleq \int\proja\Paren{\Vu}\AvgT{\pha\Paren{\Vu,t}\Paren{\pha\Paren{\Vx,t}-\pha\Paren{\Vxp,t}}}\Du
	\,.
\end{equation}
Noticing that
\begin{equation} 
	\label{eq:cross_SF}
	\AvgT{\pha\Paren{\Vu,t}\pha\Paren{\Vup,t}} = \frac{1}{2}\Paren{
	\AvgT{\pha^{2}\Paren{\Vu,t}}
	+
	\AvgT{\pha^{2}\Paren{\Vup,t}}
	-\SFpha\Paren{\Vu,\Vup}
	}
	\,,
\end{equation}
and remembering that~$\proja$ is piston-free by construction, it comes
\begin{equation}
	\PDa\Paren{\Vx,\Vxp} = \frac{1}{2} \int\proja\Paren{\Vu}\Paren{\SFpha\Paren{\Vu,\Vxp}-\SFpha\Paren{\Vu,\Vx}}\Du
	\,.
\end{equation}
In total,
\begin{equation} 
	\label{eq:SF_cross_term}
	\begin{aligned}
		\chi\Paren{\Vx,\Vxp} = \frac{1}{2}
		\sum_{a\in\setactu} & \Paren{\IFa\Paren{\Vx}-\IFa\Paren{\Vxp}}
		\\
		& \times 
		\int\proja\Paren{\Vu}\Paren{\SFpha\Paren{\Vu,\Vxp}-\SFpha\Paren{\Vu,\Vx}}\Du
	\,.
	\end{aligned}
\end{equation}

\subsection{Quadratic term}

Expanding the quadratic term in \eq{eq:SF_AO_dev} by using \eq{eq:phac_dev} yields
\begin{equation} 
	\label{eq:SF_quad_term}
	\begin{aligned}
	\SFc\Paren{\Vx,\Vxp} = -\frac{1}{2}\sum_{a \in \setactu}\sum_{\ap \in \setactu} & 
	\Daap
	\Paren{\IFa\Paren{\Vx}-\IFa\Paren{\Vxp}}
	\\
	&\times
	\Paren{\IFap\Paren{\Vx}-\IFap\Paren{\Vxp}} 
	\,,
	\end{aligned}
\end{equation}
with
\begin{equation}
	\Daap = \iint -2\AvgT{\pha\Paren{\Vu,t}\pha\Paren{\Vup,t}}\proja\Paren{\Vu}\projap\Paren{\Vup}\Du\Dup
	\,.
\end{equation}
Using \eq{eq:cross_SF} and remembering that~$\proja$ and~$\projap$ are piston-free by construction, it comes
\begin{equation}
	\Daap = \iint \SFpha\Paren{\Vu,\Vup}\proja\Paren{\Vu}\projap\Paren{\Vup}\Du\Dup
	\,.
\end{equation}
In total, by combining \eqs{eq:SF_cross_term}{eq:SF_quad_term} in \eq{eq:SF_AO_dev}, we obtain the results claimed in \eqs{eq:SF_analytic}{eq:Daap}.

\section{Derivation of the analytical PSD of the fitting residuals}
\label{app:PSD_analytical}

By definition (see \eq{eq:phac}), $\phac\Paren{\Vx}$ belongs to the vector space~$\Vspacenot$ spanned by the set of the influence functions~$\IFa\Paren{\Vx}$ of the WFC,
\begin{equation}
	\phac\Paren{\Vx} \in \Vspace{\Brace{\IFa\Paren{\Vx} = \IFref\Paren{\Vx-\Pa}}_{\Va}}
	\,.
\end{equation}
As proved in \refapp{app:ortho_IF}, it is possible to design a generating function~$\IFperp\Paren{\Vx}$ from~$\IFref\Paren{\Vx}$ such that its associated basis spans the same vector space
\begin{equation}
	\Vspace{\Brace{\IFa\Paren{\Vx} = \IFref\Paren{\Vx-\Pa}}_{\Va}}
	=
	\Vspace{\Brace{\IFperpa\Paren{\Vx} = \IFperp\Paren{\Vx-\Pa}}_{\Va}}
	\,,
\end{equation}
and that the basis is orthonormal
\begin{equation} 
	\label{eq:ortho}
	\int{\IFperpa\Paren{\Vx}\IFperpap\Paren{\Vx}\Dx} = \int{\IFperp\Paren{\Vx - \Pa}\IFperp\Paren{\Vx - \Pap}\Dx} = \delta_{\Va, \Vap}
	\,.
\end{equation}
In this basis, according to \refapp{app:proj_opti}, 
\begin{equation}
	\Brack{\M{A}}_{a,\ap} = \Brack{\M{A}\Inv}_{a,\ap} = \delta_{\Va, \Vap}
	\,,
\end{equation}
and the projector~$\proja$ is the direct corresponding influence function
\begin{equation} 
	\label{eq:proj_perp}
	\proja\Paren{\Vx} = \IFperpa\Paren{\Vx}
	\,.
\end{equation}
The derivation of~$\IFperp\Paren{\Vx}$ from a given~$\IFref\Paren{\Vx}$ is discussed in~\refapp{app:ortho_IF}.

We now focus on the Fourier transform of the structure function. The structure function of~$\pha$ is defined as
\begin{equation}
	\SF{\pha}\Paren{\Vx,\Vxp} = \AvgT{\Paren{\pha\Paren{\Vx,t} - \pha\Paren{\Vxp,t}}^{2}}
	\,,
\end{equation}
and assuming that it is spatially stationary
\begin{equation} 
	\label{eq:SF_var}
	\SF{\pha}\Paren{\Vx} = 2\AvgT{\pha^{2}\Paren{\V{0},t}} - 2\AvgT{\pha\Paren{\Vx,t}\pha\Paren{\V{0},t}}
	\,.
\end{equation}
The first term is the expected constant variance over the domain given by the integral of the PSD, while the second term can be obtained via the Wiener–Khinchin theorem,
\begin{equation} 
	\label{eq:WK_theorem}
	\AvgT{\pha\Paren{\Vx,t}\pha\Paren{\Vxp,t}} = \FTfullinv{\PSD{\pha}\Paren{\Vk}}{\Vxp-\Vx}
	\,,
\end{equation}
leading to
\begin{equation} 
	\label{eq:SF_FT}
	\SF{\pha}\Paren{\Vx} = 2\int{\PSD{\pha}\Paren{\Vk}\Dk} - 2\FTfullinv{\PSD{\pha}\Paren{\Vk}}{\Vx}
	\,.
\end{equation}
The first term of the equation is a constant, giving a Dirac distribution centred on~$\V{0}$, thus
\begin{equation} 
	\label{eq:FTSF_PSD}
	\forall\Vk\neq\V{0}, -\frac{1}{2}\FTSFpha\Paren{\Vk} = \PSD{\pha}\Paren{\Vk}
	\,.
\end{equation}
On the other hand, taking the double Fourier transform on~$\Vx$ and~$\Vxp$ of the stationary structure function yields
\begin{align}
	\FTSFpha\Paren{\Vk,\Vkp} {}={} &
		\iint \SFpha\Paren{\Vx,\Vxp}\E{-2\I\pi\Vx\T\Vk}\E{-2\I\pi\Vxp\T\Vkp} \Dx\Dxp
	\\
	{}={} & \iint \SFpha\Paren{\Vx-\Vxp}\E{-2\I\pi\Vx\T\Vk}\E{-2\I\pi\Vxp\T\Vkp} \Dx\Dxp
	\\
	{}={} & \iint \SFpha\Paren{\V{\rho}}\E{-2\I\pi\Paren{\V{\rho}+\Vxp}\T\Vk}\E{-2\I\pi\Vxp\T\Vkp} \D{\V{\rho}}\Dxp
	\\
	\label{eq:SF_doubleFT}
	{}={} & \FTSFpha\Paren{\Vk}\delta\Paren{\Vk+\Vkp}
	\,.
\end{align}
Similarly, when taking the double Fourier transform of \eq{eq:SF_analytic}, for~$\Vk\neq\V{0}$ and~$\Vkp\neq\V{0}$, only the cross terms according to~$\Vx$ and~$\Vxp$ remain (the others give Dirac distributions in either~$\Vk = \V{0}$ or~$\Vkp = \V{0}$)
\begin{equation}
	\begin{aligned}
		\FTSFAO\Paren{\Vk,\Vkp} {}={} & \FTSFpha\Paren{\Vk,\Vkp}
		\\
		&
		- \sum_{a\in\setactu} \FTIFperpa\Paren{\Vk}
		\FTfull{\int\IFperpa\Paren{\Vu}\SFpha\Paren{\Vu,\Vxp}\Du}{\Vkp}
		\\
		&
		- \sum_{a\in\setactu} \FTIFperpa\Paren{\Vkp}
		\FTfull{\int\IFperpa\Paren{\Vu}\SFpha\Paren{\Vu,\Vx}\Du}{\Vk}
		\\
		& + \frac{1}{2}\sum_{a \in \setactu}\sum_{\ap \in \setactu}
		\Daap
		\Paren{
		\FTIFperpa\Paren{\Vk}\FTIFperpap\Paren{\Vkp}+
		\FTIFperpap\Paren{\Vk}\FTIFperpa\Paren{\Vkp}
		}
	\,.
	\end{aligned}
\end{equation}
Noticing that~$\Daap = \Dapa$ and computing the Fourier transform in the frequency domain via \eq{eq:D_conv} produces
\begin{equation}
	\begin{aligned}
		\FTSFAO\Paren{\Vk,\Vkp} {}={} & \FTSFpha\Paren{\Vk,\Vkp}
		\\
		&
		- \sum_{a\in\setactu} \Paren{
		\FTIFperpa\Paren{\Vk}
		\FTIFperpa\Paren{\Vkp}\FTSFpha\Paren{\Vkp}
		+
		\FTIFperpa\Paren{\Vkp}
		\FTIFperpa\Paren{\Vk}\FTSFpha\Paren{\Vk}
		}
		\\
		& + \sum_{a \in \setactu}\sum_{\ap \in \setactu}
		\Daap
		\FTIFperpa\Paren{\Vk}\FTIFperpap\Paren{\Vkp}
	\,.
	\end{aligned}
\end{equation}
Then,  
(1) applying \eq{eq:SF_doubleFT} for~$\Vkp = -\Vk$,
(2) noticing that~$\SFpha$ is even ($\FTSFpha\Paren{\Vk} = \FTSFpha\Paren{-\Vk}$, see \eqs{eq:SF_def}{eq:SF_stat}),
(3) noticing that~$\IFperpa\Paren{\Vx} \in \Reals \Rightarrow \FTIFperpa\Paren{-\Vk} = \conj{\FTIFperpa}\Paren{\Vk}$,
(4) noticing that $\IFperpa\Paren{\Vx} = \IFperp\Paren{\Vx - \Pa} \Leftrightarrow \FTIFperpa\Paren{\Vk} = \E{-2\I\pi\Pa\T\Vk}\FTIFperp\Paren{\Vk}$, and
(5) noting~$\PSDperp\Paren{\Vk} = \Abs{\FTIFperp\Paren{\Vk}}^{2}$,
lead to the following expression
\begin{equation} 
	\label{eq:SFAO_FT}
	\begin{aligned}
		\FTSFAO\Paren{\Vk} {}={} &
		\FTSFpha\Paren{\Vk}\times\Paren{1-2\ntotactu\PSDperp\Paren{\Vk}}
		\\
		& + \sum_{a \in \setactu}\sum_{\ap \in \setactu}
		\Daap
		\PSDperp\Paren{\Vk}\E{-2\I\pi\Paren{\Pa-\Pap}\T\Vk}
	\,.
	\end{aligned}
\end{equation}
We restate here that~$\ntotactu$ is the total number of actuators. In addition, performing the convolution in the frequency domain again produces
\begin{align}
	\Daap {}={} & \iint \SFpha\Paren{\Vu,\Vup}\IFperpa\Paren{\Vu}\IFperpap\Paren{\Vup}\Du\Dup
	\\
	{}={} & \iint \FTSFpha\Paren{\Vk}\FTIFperpa\Paren{\Vk}\IFperpap\Paren{\Vup}\E{2\I\Vup\T\Vk}\;\Dk\Dup
	\\
	{}={} & \int \FTSFpha\Paren{\Vk}\FTIFperpa\Paren{\Vk}\FTIFperpap\Paren{-\Vk}\Dk
	\\
	\label{eq:FTDaap}
	{}={} & \int \FTSFpha\Paren{\Vk}\PSDperp\Paren{\Vk}\E{-2\I\pi\Paren{\Pa-\Pap}\T\Vk}\;\Dk
	\,.
\end{align}
Finally, by applying \eq{eq:FTSF_PSD} in \eqs{eq:SFAO_FT}{eq:FTDaap}, we finally come the sought-after expression, \eq{eq:PSDres}
\begin{equation}
	\begin{aligned}
	\PSDres\Paren{\Vk} {}={} &
	\Paren{1 - 2 \ntotactu\PSDperp\Paren{\Vk}}\PSD{\pha}\Paren{\Vk} 
	+
	\sum_{a\in\setactu} \sum_{\ap\in\setactu} \PSDperp\Paren{\Vk}\E{-2\I\pi\Paren{\Pa-\Pap}\T\Vk}
	\\
	& \quad \times \Paren{\int{\PSD{\pha}\Paren{\Vkp}\PSDperp\Paren{\Vkp}\E{-2\I\pi\Paren{\Pa-\Pap}\T\Vkp}\Dkp}}
	\\
	 {}={} &
	\Paren{1 - 2 \ntotactu\PSDperp\Paren{\Vk}}\PSD{\pha}\Paren{\Vk} 
	+
	\sum_{a\in\setactu} \sum_{\ap\in\setactu} \PSDperp\Paren{\Vk}\E{-2\I\pi\Paren{\Pa-\Pap}\T\Vk}
	\\
	& \quad \times \FTfullinv{\PSD{\pha}\PSDperp}{\Pap-\Pa}
	\,.
	\end{aligned}
\end{equation}

We note here that it is possible to apply \eq{eq:FTSF_PSD} in \eq{eq:FTDaap} $\forall \Vk$ because~$\IFperp$ is piston-free and, thus its average value is null, implying that~$\PSDperp\Paren{\Vk=\V{0}}=0$. For the same reason, \eq{eq:PSDres} is valid for~$\Vk=\V{0}$ since the WFC does not apply any mode correction and~$\PSDres\Paren{\V{0}} = \PSD{\pha}\Paren{\V{0}}$.

\section{Orthogonalisation of the reference influence function}
\label{app:ortho_IF}

\begin{figure}[t!] 
	\centering
	
	\newcommand{\LineRatio}{1}
	
	\newcommand{\FigOne}  {figures_IF_ortho_IF_0.pdf}
	\newcommand{\FigTwo}  {figures_IF_ortho_IF_it.pdf}      
	\newcommand{\FigThree}{figures_IF_ortho_IF_it_shift.pdf}
	\newcommand{\FigFour} {figures_IF_ortho_bar.pdf}
	\newcommand{\subfigColor}{black}	
	
	\sbox1{\includegraphics{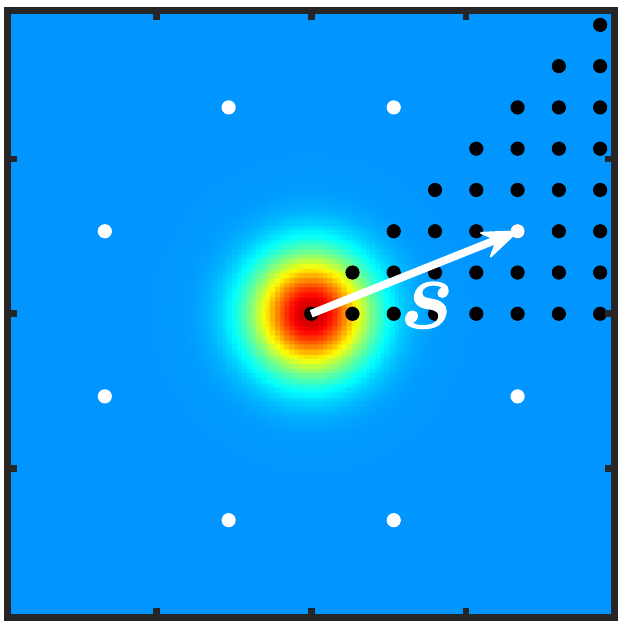}}	       
	\sbox2{\includegraphics{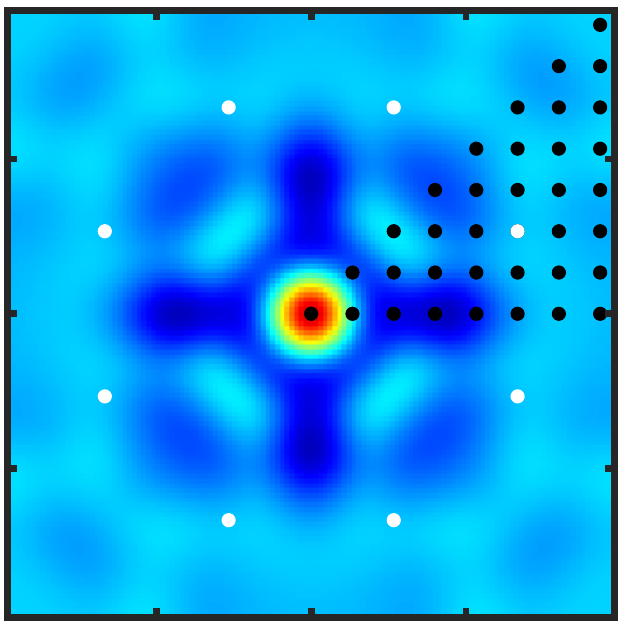}}	   
	\sbox3{\includegraphics{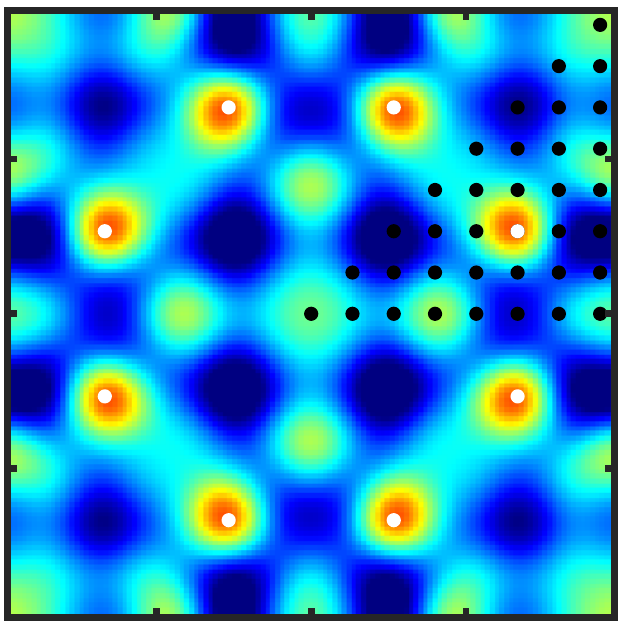}}     
	\sbox4{\includegraphics{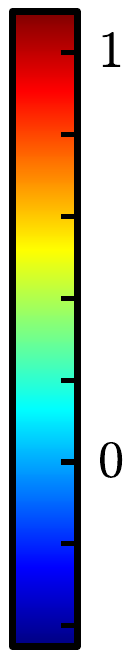}}	  
	\newcommand{\ColumnWidth}[1]
		{\dimexpr \LineRatio \linewidth * \AspectRatio{#1} / (\AspectRatio{1} + \AspectRatio{2} + \AspectRatio{3} + \AspectRatio{4}) \relax
		}
	\newcommand{\ColumnGap}{\hspace {\dimexpr \linewidth /5 - \LineRatio\linewidth /5 }}

	\begin{tabular}{
		@{\ColumnGap}
		m{\ColumnWidth{1}}
		@{\ColumnGap}
		m{\ColumnWidth{2}}
		@{\ColumnGap}
		m{\ColumnWidth{3}}
		@{\ColumnGap}
		m{\ColumnWidth{4}}
		@{\ColumnGap}
		}
		\subfigimg[width=\linewidth,pos=ul,font=\fontfig{\subfigColor}]{$\,$(a)}{0.0}{\FigOne} &
		\subfigimg[width=\linewidth,pos=ul,font=\fontfig{\subfigColor}]{$\,$(b)}{0.0}{\FigTwo} &
		\subfigimg[width=\linewidth,pos=ul,font=\fontfig{\subfigColor}]{$\,$(c)}{0.0}{\FigThree} &
		\subfigimg[width=\linewidth,pos=ul,font=\fontfig{\subfigColor}]{}{0.0}{\FigFour}
	\end{tabular}
	\caption{\label{fig:IF_ortho} Visualisation of an iteration~$i$ of the influence function orthonormalisation procedure. Black dots: Position of the actuators on which the influence function must be orthonormalised. White dots: The eight positions corresponding to the shift~$\shift$ of the iteration~$i$. Panel~(a):~Initial influence function~$\IFref$. Panel~(b):~Orthonormalised influence function $\IFiti$ at the iteration~$i$. Panel~(c):~Replicated shifted pattern $\IFits$ at the iteration~$i$.}
\end{figure}

We describe in this appendix the procedure to orthonormalise a given reference influence function~$\IFref$ to obtain~$\IFperp$. We restate here that the actuators lie on a regular 2D Cartesian grid of pitch~$\pitch_{1} = \pitch_{2} = \pitch$ and whose axes are assumed to be along~$x_{1}$ and~$x_{2}$. It is reasonable to consider that none of the axes of the WFC are peculiar and that, as a consequence, the reference influence function~$\IFref$ satisfies the following set of symmetries
\begin{equation}
	\IFref\paren{x_{1},x_{2}}
	=
	\IFref\paren{-x_{1},x_{2}}
	=
	\IFref\paren{x_{1},-x_{2}}
	=
	\IFref\paren{x_{2},x_{1}}
	\,.
\end{equation}
Using the symmetries of the causes leading to the symmetries of the effects, we built an orthonormalised shape~$\IFperp$ that follows the same symmetries. As a consequence, orthonormalising the shape on the~$\nactu^{\Tag{ortho}}$ actuator positions emphasised by the black dots in \fig{fig:IF_ortho} is sufficient.

The orthonormalisation procedure was inspired by the Gram-Schmidt process. It implies iterative orthonormalisation performed step by step (i.e. actuator by actuator) and initialised with~$\IFit{0} = \IFref$.

We note~$\IFiti$ the orthonormalised shape at the step~$i$ of the algorithm (see \subfig{fig:IF_ortho}{b}). We let~$\Vj=\Paren{\jOne,\jTwo}\T\in\PosIntegers^{2}$ be the index of the position on the 2D Cartesian grid on which~$\IFiti$ must be orthonormalised. As shown in \subfig{fig:IF_ortho}{a}, this corresponds to a shift~$\shift = \Paren{\pitch\jOne, \pitch\jTwo}\T=\pitch\Vj$. To preserve the symmetry, the new shape~$\IFitip$ must be a linear combination of~$\IFiti$ and its shifted replicates on the eight symmetric positions corresponding to~$\shift$, as emphasised by the white dots in \fig{fig:IF_ortho}
\begin{equation}
	\begin{aligned}
		\mathcal{S} =
				\pitch & \left\lbrace
		\Paren{+\jOne, +\jTwo}\T,
		\Paren{+\jOne, -\jTwo}\T,
		\Paren{-\jOne, +\jTwo}\T,
		\Paren{-\jOne, -\jTwo}\T,
		\right.
		\\
		& 
		\left.
		\Paren{+\jTwo, +\jOne}\T,
		\Paren{+\jTwo, -\jOne}\T,
		\Paren{-\jTwo, +\jOne}\T,
		\Paren{-\jTwo, -\jOne}\T
		\right\rbrace
		\,.
	\end{aligned}
\end{equation}
We note~$\IFits$ is the combination of the shifted replicates presented in \subfig{fig:IF_ortho}{c}
\begin{equation} 
	\label{eq:IF_shift}
	\IFits\Paren{\Vx} \triangleq \sum_{\bar{\shift}\in\mathcal{S}}\IFiti\Paren{\Vx-\bar{\shift}}
	\,.
\end{equation}
Thus,
\begin{equation} 
	\label{eq:IFitip}
	\IFitip = \IFiti + \alpha\IFits
	\,.
\end{equation}
The value of~$\alpha$ is determined by imposing the orthogonality for the shift~$\shift$
\begin{align}
	0 & {}={} \int\IFitip\Paren{\Vx}\IFitip\Paren{\Vx-\shift}\Dx
	\\
	& \begin{aligned}
	{}={} & 
		\int\IFiti\Paren{\Vx}\IFiti\Paren{\Vx-\shift}\Dx
		+ \alpha\int\IFiti\Paren{\Vx}\IFits\Paren{\Vx-\shift}\Dx
		\\
		& + \alpha\int\IFits\Paren{\Vx}\IFiti\Paren{\Vx-\shift}\Dx 
		+ \alpha^2\int\IFits\Paren{\Vx}\IFits\Paren{\Vx-\shift}\Dx 
	\end{aligned}
	\\
	\label{eq:eq_alpha}
	& {}={} C+2B\alpha+A\alpha^2
	\,,
\end{align}
with
\begin{equation} 
	\label{eq:eq_coef}
	\begin{cases}
		A = \int\IFits\Paren{\Vx}\IFits\Paren{\Vx-\shift}\Dx 
		\\
		B = \int\IFiti\Paren{\Vx}\IFits\Paren{\Vx-\shift}\Dx = \int\IFits\Paren{\Vx}\IFiti\Paren{\Vx-\shift}\Dx
		\\
		C = \int\IFiti\Paren{\Vx}\IFiti\Paren{\Vx-\shift}\Dx
		\\
		D = B^2 - A \times C
	\end{cases}
	\,,
\end{equation}
where the double equality on~$B$ is obtained using the symmetries of~$\IFiti$ and~$\IFits$.

If~$D\geq0$, the second order equation \eq{eq:eq_alpha} has two solutions,~$\alpha \in \Brace{\frac{-B\pm\sqrt{D}}{A}}$. To prevent any divergence, the solution with the smallest absolute value is kept
\begin{equation} 
	\label{eq:sol_alpha}
	\alpha = \argmin_{\alpha_{\pm}\in\Brace{\frac{-B\pm\sqrt{D}}{A}}} \; \Abs{\alpha_{\pm}}
	\,,
\end{equation}
and~$\IFitip$ is updated using \eq{eq:IFitip}.

If~$D<0$, there is no real solution, and $\IFitip$ is left equal to~$\IFiti$. This means that the current position~$\shift$ cannot be orthogonalised. If this situation occurs for each run on the whole actuator set, it means that~$\IFref$ cannot be orthonormalised. For all the profiles presented in this paper, after a few loops on the whole set of actuators, this situation was never encountered, meaning that all the possible shifts~$\shift$ were successfully orthogonalised.

These steps were repeated on all the~$\nactu^{\Tag{ortho}}$ actuators of the orthonormalisation set until convergence. The stopping criteria was when the normalised root mean square error~$\RMS$ of the evolution of~$\IFiti$, defined as
\begin{equation} 
	\label{eq:RMS}
	\RMS\Paren{\IFiti,\IFit{i+\nactu^{\Tag{ortho}}}} \triangleq 
	\sqrt{\frac{\int \Paren{\IFit{i+\nactu^{\Tag{ortho}}}\Paren{\Vx}-\IFiti\Paren{\Vx}}^{2}}{\int \Paren{\IFiti\Paren{\Vx}}^{2} \Dx}}
	\,,
\end{equation}
was below a given threshold~$\epsilon_{\Tag{th}}=10^{-8}$.

We note here that the shifting operations in \eqs{eq:IF_shift}{eq:eq_coef} can be numerically performed either in the frequency domain or by updating a list of mixing coefficients on the basis~$\Brace{\IFa}_{a\in\setactu}$. The results presented in this paper were obtained by performing the shifts in the frequency domain.

This procedure is summarised in the pseudo-code of Algorithm~\ref{alg:IF_ortho}. At the end of this procedure,~$\IFperp$ was normalised to one and is piston-free by construction ($\IFref$ is piston-free).

\begin{algorithm}
\caption{\label{alg:IF_ortho} Orthonormalisation algorithm~$\IFref \rightarrow \IFperp$}
\begin{algorithmic}[1]
\small

\State $\IFperp \gets \IFref$
	\commentalgo{Initialisation of the orthonormalised shape}

\State $\IFiti \gets \IFref$
	\commentalgo{Initialisation of the iteration}

\State $\epsilon \gets 1$
	\commentalgo{Initialisation of the error}
	
\State $\nactu^{\Tag{rad}} \gets \Floor{\frac{\nactu-1}{2}}$
	\commentalgo{Radial number of actuators for the orthonormalisation}

\While{$\epsilon \geq \epsilon_{\Tag{th}}$}
	\commentalgo{Until convergence, orthogonalisation on the whole set}
     
	\For{$\jOne$ from $1$ to $\nactu^{\Tag{rad}}$}
		\commentalgo{Loop on the~$x_{1}$ axis}

		\For{$\jTwo$ from $0$ to $\jOne$}
			\commentalgo{Loop on the~$x_{2}$ axis}
			
			\State $\shift \gets \Paren{\pitch\jOne, \pitch\jTwo}$
				\commentalgo{Corresponding shift of the iteration}
			
			\State $\IFits\Paren{\Vx} \gets \sum_{\bar{\shift}\in\mathcal{S}}\IFiti\Paren{\Vx-\bar{\shift}}$
				\commentalgo{\eq{eq:IF_shift}}
												
			\State $A \gets \int\IFits\Paren{\Vx}\IFits\Paren{\Vx-\shift}\Dx$
				\commentalgo{\eq{eq:eq_coef}}
			\State $B \gets \int\IFiti\Paren{\Vx}\IFits\Paren{\Vx-\shift}\Dx$
				\commentalgo{\eq{eq:eq_coef}}
			\State $C \gets \int\IFiti\Paren{\Vx}\IFiti\Paren{\Vx-\shift}\Dx$
				\commentalgo{\eq{eq:eq_coef}}
			\State $D \gets B^2 - A \times C$       
				\commentalgo{\eq{eq:eq_coef}}   
			\State $\alpha \gets \argmin_{\alpha_{\pm}\in\Brace{\frac{-B\pm\sqrt{D}}{A}}} \; \Abs{\alpha_{\pm}}$  
				\commentalgo{\eq{eq:sol_alpha}}
				
		\If{$\alpha$ is real}
		\commentalgo{Orthonormalisation if possible}

				\State $\IFiti \gets \IFiti + \alpha\IFits$
					\commentalgo{Update of the iteration, \eq{eq:IFitip}}
				\State $\IFiti \gets \IFiti/\int \Paren{\IFiti\Paren{\Vx}}^{2} \Dx$
					\commentalgo{normalisation}

		\EndIf	  
		\EndFor		 
	\EndFor

    \State $\epsilon \gets \RMS\Paren{\IFperp,\IFiti}$
	\commentalgo{Update of the RMS error, \eq{eq:RMS}}
		
    \State $\IFperp \gets \IFiti$
	\commentalgo{Update of the orthonormalised shape}
			
\EndWhile

\State \textbf{return} $\IFperp$
\end{algorithmic}
\end{algorithm}

By construction
\begin{equation}
	\Vspace{\Brace{\IFperpa\Paren{\Vx} = \IFperp\Paren{\Vx-\Pa}}_{a\in\setactu}}
	\subset
	\Vspace{\Brace{\IFa\Paren{\Vx} = \IFref\Paren{\Vx-\Pa}}_{a\in\setactu}}
	\,,
\end{equation}
but one needs to check that after the orthonormalisation procedure, the basis spans the same vector space. That is to say,
\begin{equation} 
	\label{eq:vect_space}
	\Vspace{\Brace{\IFa\Paren{\Vx} = \IFref\Paren{\Vx-\Pa}}_{a\in\setactu}}
	\subset
	\Vspace{\Brace{\IFperpa\Paren{\Vx} = \IFperp\Paren{\Vx-\Pa}}_{a\in\setactu}}
	\,.
\end{equation}
To do so, the reference influence function~$\IFref$ was projected onto this new basis, from \eq{eq:proj_perp},
\begin{equation}
	\IFrefproj\Paren{\Vx} = \sum_{a\in\setactu} \IFperpa\Paren{\Vx} \int\IFperpa\Paren{\Vu}\IFref\Paren{\Vu}\Du
	\,.     
\end{equation}

\tab{tab:RMS} gathers the RMS error~$\RMS\Paren{\IFref, \IFrefproj}$ for the different profiles. The error is negligible, ensuring that~$\IFref=\IFrefproj$. Thus, \eq{eq:vect_space} was satisfied, as the basis generated with~$\IFref$ and the basis generated with~$\IFperp$ span the same vector space. We also note that these RMS values are consistent with the orthogonalisation threshold~$\epsilon_{\Tag{th}}=10^{-8}$.

\begin{table}[t!] 
	\caption{\label{tab:RMS} Projection of ~$\IFref$ on the basis defined by~$\IFperp$}
	\centering
	\begin{tabular}{cccccc}
	\hline
	\hline
	$\IFref$   & Piston	      & Pyramid	     & Gaussian	    & ALPAO	       & 2D~$\sinc$     \\
	\hline
	$\frac{\RMS}{10^{-8}}$ & $3.4$ & $2.6$ & $2.2$ & $2.8$ & $2.0$ \\
	\hline
	\end{tabular}
	\tablefoot{Root mean square errors~$\RMS$ of the residuals between~$\IFref$ and its projection on the basis defined by~$\IFperp$ for the different shapes.}
\end{table}

\section{Periodicity of the analytical structure function}
\label{app:SF_periodic}

In this appendix, we discuss the fact that if the influence function~$\IFref$ is spatially limited on a support~$\Supref$, the inhomogeneities in the non-stationary structure function after AO correction  are $\pitch$ 2D periodic. That is to say,
\begin{equation}
	\SFAO\paren{\Vx+\Vpitchper,\Vxp+\Vpitchper} = \SFAO\paren{\Vx,\Vxp}
	\,,
\end{equation}
with~$\Vpitchper\in\Brace{\Paren{\pitch,0}\T,\Paren{0,\pitch}\T}$.

With~$\Supa$ as the support of~$\IFa$, all the influence functions become a translated version of~$\IFref$ by~$\Pa$. $\Supa$ is a also translated version of the domain~$\Supref$ by~$\Pa$.

For positions~$\Vx$ that are `far' (in the sense of the influence function support) from the pupil edges, only the local actuators~$a \in \Brace{a \st \Vx \in \Supa}$ can participate in the wavefront correction. As the influence functions of all the actuators are identical and on a regular grid of pitch~$\pitch$ on each axis and as the Kolmogorov structure function is stationary, the problem becomes 2D periodic of period~$\pitch$ on each axis. As the symmetry elements of the causes must be found in their effects, the inhomogeneities of the structure function $\SFAO$ are 2D periodic with a period~$\pitch$ on each axis of the actuator grid. Of course, when close to the pupil edges, the problem is not isotropic anymore, but this concerns only a small portion of the full pupil.

For an influence function shape that is not spatially limited, this reasoning does not hold. According to the position in the pupil, the geometrical repartition of the actuators participating in the local correction changes. For example, there are potentially more actuators on one side than the other, not the same number of actuators, and so on. The problem is not 2D periodic, and the $\SFAO$ may not be 2D periodic either.

We remark here that having a finite support is a sufficient but not a necessary condition. Indeed, for the 2D Gaussian profile, the orthogonalised influence function is spatially extended (see \subfig{fig:IF_profile}{c}), but the inhomogeneities of the structure function are periodic (see \subfig{fig:map_SF_disk_95}{c}).

\section{Numerical estimate of the structure function}
\label{app:SF}

Different solutions exist to compute the empirical structure function of a set of numerical wavefronts. This computation can be done either in the direct space using \eq{eq:SF_def} or via the frequency domain using \eq{eq:SF_FT}.

\subsection{In the direct space}
\label{app:SFd}

Similarly to~\refsec{sec:Domain_periodisation}, the expected value according to~$\Vpixp$ can be computed in the direct space by pairing the position~$\Vpix$ in the discrete pupil. From \eq{eq:SF_def}, it comes
\begin{align}
	\SFd{\pha}\Paren{\Vpix} {}={} & \frac{\sum_{\Vpixp}\pupil\Paren{\Vpixp}\pupil\Paren{\Vpix + \Vpixp}\AvgT{\Paren{\pha\Paren{\Vpixp,t} - \pha\Paren{\Vpix + \Vpixp,t}}^{2}}}{\sum_{\Vpixp}\pupil\Paren{\Vpixp}\pupil\Paren{\Vpix + \Vpixp}}
	\\
	& \hspace{-33pt}= \frac{\AvgT{\iDFT{\real{\conj{\DFT{\pupil}}\DFT{\pupil\pha^{2}}} - \Abs{\DFT{\pupil\pha}}^{2}}\Paren{\Vpix}}}{0.5\times\iDFT{\Abs{\DFT{\pupil}}^{2}}\Paren{\Vpix}}
	\,.
\end{align}
In simulation, the temporal expected value is the average of the generated frames. We note here that when using DFTs, the pupil must be padded at least twice to avoid aliasing.

This method is only used in \refsec{sec:res_Screen}, as it gives good results when the phase screen is strongly not periodic. This is the case for a turbulent wavefront without AO correction. In this situation, the edges lead to artefacts if the structure function is computed as described below via the frequency domain.

\subsection{In the frequency domain}
\label{app:SFf}

The structure function can also be obtained using \eq{eq:SF_FT} via the empirical estimate of the PSD, weighted by the pupil area to conserve the energy on the simulated domain
\begin{equation}
	\PSD{\pha}\Paren{\Vpixk} = \frac{\AvgT{\Abs{\DFT{\pupil\Paren{\Vpix}\pha\Paren{\Vpix}}\Paren{\Vpixk}}^{2}}}{\sum_{\Vpix}\pupil\Paren{\Vpix}}
	\,.
\end{equation}
Applying \eq{eq:SF_FT} then yields
\begin{equation}
	\SFf{\pha}\Paren{\Vpix} = 2\sum_{\Vpixk}\PSD{\pha}\Paren{\Vpixk} - 2\times\iDFT{\PSD{\pha}\Paren{\Vpixk}}\Paren{\Vpix}
	\,.
\end{equation}
Using DFTs, the pupil must be padded at least twice to avoid aliasing. In this work, the computation was done as $\nPad=3$.

This method gives good results when the phase screen is roughly periodic but with a sharp pupil edge in the domain, such as after AO correction. This method is the one mainly used in this work.

\section{Simulations: Additional materials}
\label{app:add_materials}

\begin{figure*}[t!] 
	\centering
	
    \newcommand{\FirstCol}{8pt}
	\newcommand{\PathFig}{figures_PSF_}
	\newcommand{\FlagApertureOne}{square}
	\newcommand{\FlagApertureTwo}{disk}
	\newcommand{\FlagApertureThree}{disk_TT}
	\newcommand{\FlagApertureFour}{SPHERE_97p5}
	\newcommand{\subfigColor}{white}
	
	\newcommand{\LineRatio}{1}
		
	\sbox1{\includegraphics{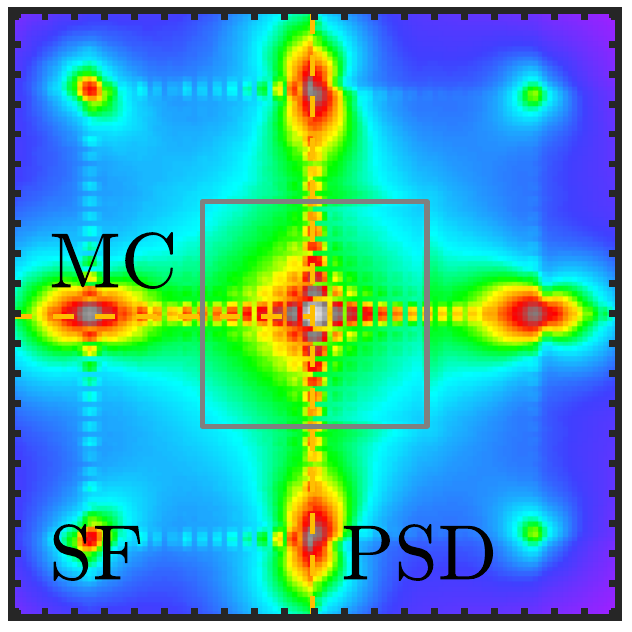}}		 
	\sbox2{\includegraphics{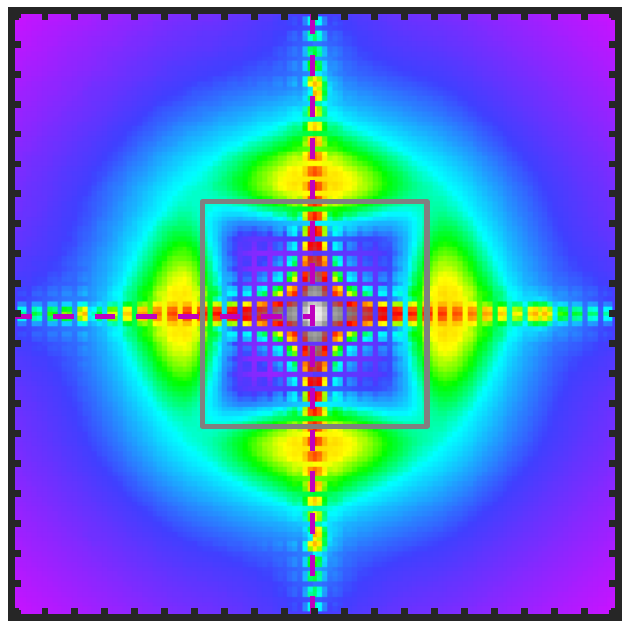}}	 
	\sbox3{\includegraphics{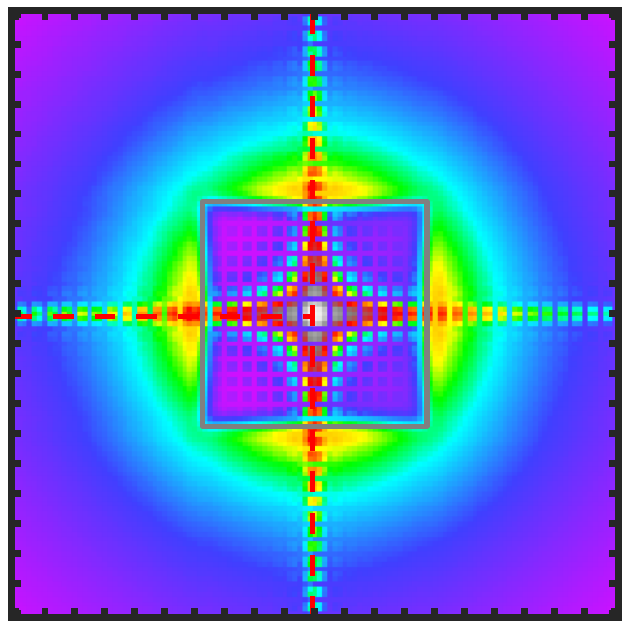}}	 
	\sbox4{\includegraphics{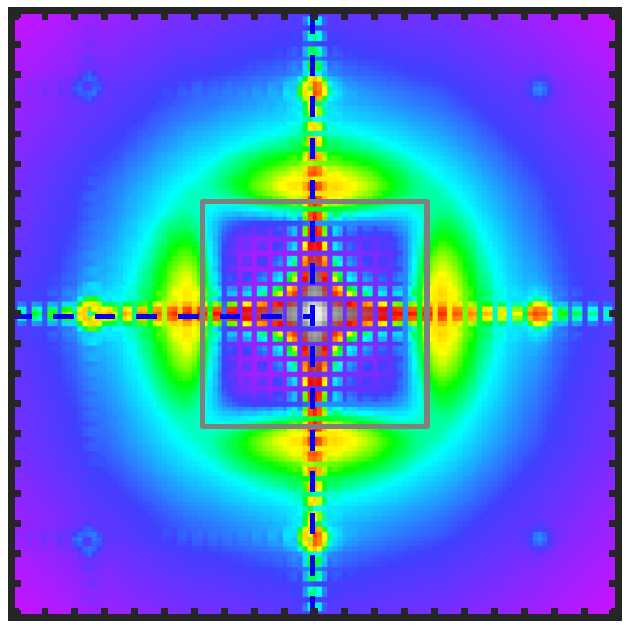}}	 
	\sbox5{\includegraphics{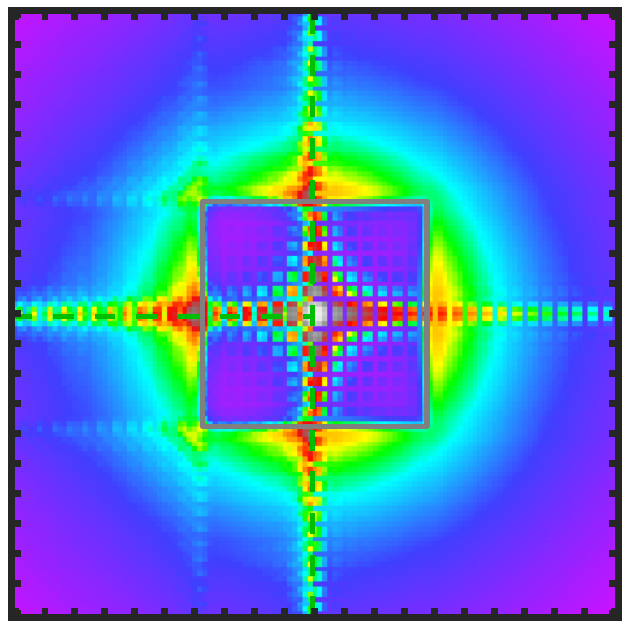}}   
	\sbox6{\includegraphics{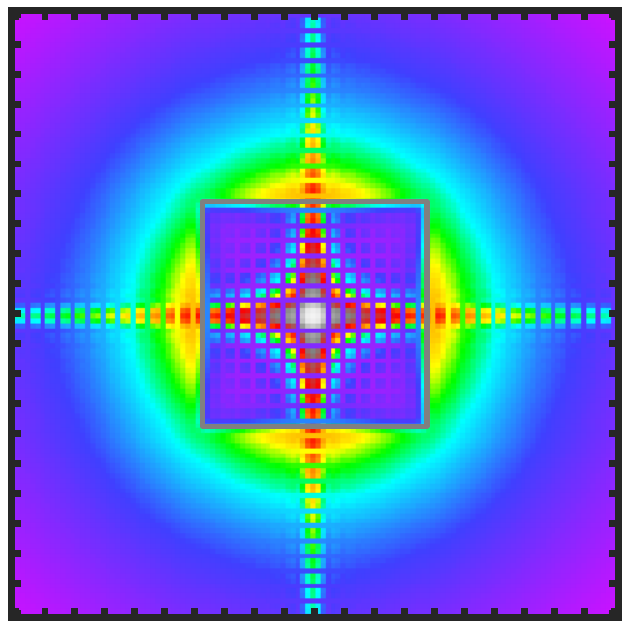}}	 
	
	\newcommand{\ColumnWidth}[1]
		{\the \dimexpr (\linewidth-\FirstCol) * \LineRatio / 6 \relax}

	\begin{tabular}{
		@{}
		C{\FirstCol}
		@{}
		C{\ColumnWidth{1}}
		@{}
		C{\ColumnWidth{2}}
		@{}
		C{\ColumnWidth{3}}
		@{}
		C{\ColumnWidth{4}}
		@{}
		C{\ColumnWidth{5}}
		@{}
		C{\ColumnWidth{6}}
		@{}
		}
		&
		\small{Piston}
		&
		\small{Pyramid}
		&
		\small{Gaussian}
		&
		\small{ALPAO}
		&
		\small{2D~$\sinc$}
		&
		\small{Binary mask}
		\\[-1pt]
		\rotatebox[origin=l]{90}{\small Square of side~$\Dsim$}
		&
		\subfigimg[width=\linewidth,pos=ul,font=\fontfig{\subfigColor}]{$\;$(1a)}{0.0}{\PathFig \FlagApertureOne _Piston.pdf} &
		\subfigimg[width=\linewidth,pos=ul,font=\fontfig{\subfigColor}]{$\;$(1b)}{0.0}{\PathFig \FlagApertureOne _Pyramid.pdf} &
		\subfigimg[width=\linewidth,pos=ul,font=\fontfig{\subfigColor}]{$\;$(1c)}{0.0}{\PathFig \FlagApertureOne _Gaussian.pdf} &
		\subfigimg[width=\linewidth,pos=ul,font=\fontfig{\subfigColor}]{$\;$(1d)}{0.0}{\PathFig \FlagApertureOne _ALPAO_repair.pdf} &
		\subfigimg[width=\linewidth,pos=ul,font=\fontfig{\subfigColor}]{$\;$(1e)}{0.0}{\PathFig \FlagApertureOne _SINC.pdf} &
		\subfigimg[width=\linewidth,pos=ul,font=\fontfig{\subfigColor}]{$\;$(1f)}{0.0}{\PathFig \FlagApertureOne _Binary_mask.pdf}
		\\[-4pt]
		\rotatebox[origin=l]{90}{\small Disc of side~$\Dsim$}
		&
		\subfigimg[width=\linewidth,pos=ul,font=\fontfig{\subfigColor}]{$\;$(2a)}{0.0}{\PathFig \FlagApertureTwo _Piston.pdf} &
		\subfigimg[width=\linewidth,pos=ul,font=\fontfig{\subfigColor}]{$\;$(2b)}{0.0}{\PathFig \FlagApertureTwo _Pyramid.pdf} &
		\subfigimg[width=\linewidth,pos=ul,font=\fontfig{\subfigColor}]{$\;$(2c)}{0.0}{\PathFig \FlagApertureTwo _Gaussian.pdf} &
		\subfigimg[width=\linewidth,pos=ul,font=\fontfig{\subfigColor}]{$\;$(2d)}{0.0}{\PathFig \FlagApertureTwo _ALPAO_repair.pdf} &
		\subfigimg[width=\linewidth,pos=ul,font=\fontfig{\subfigColor}]{$\;$(2e)}{0.0}{\PathFig \FlagApertureTwo _SINC.pdf} &
		\subfigimg[width=\linewidth,pos=ul,font=\fontfig{\subfigColor}]{$\;$(2f)}{0.0}{\PathFig \FlagApertureTwo _Binary_mask.pdf}
		\\[-4pt]
		\rotatebox[origin=l]{90}{\small (2) + tip-tilt removed}
		&
		\subfigimg[width=\linewidth,pos=ul,font=\fontfig{\subfigColor}]{$\;$(3a)}{0.0}{\PathFig \FlagApertureThree _Piston.pdf} &
		\subfigimg[width=\linewidth,pos=ul,font=\fontfig{\subfigColor}]{$\;$(3b)}{0.0}{\PathFig \FlagApertureThree _Pyramid.pdf} &
		\subfigimg[width=\linewidth,pos=ul,font=\fontfig{\subfigColor}]{$\;$(3c)}{0.0}{\PathFig \FlagApertureThree _Gaussian.pdf} &
		\subfigimg[width=\linewidth,pos=ul,font=\fontfig{\subfigColor}]{$\;$(3d)}{0.0}{\PathFig \FlagApertureThree _ALPAO_repair.pdf} &
		\subfigimg[width=\linewidth,pos=ul,font=\fontfig{\subfigColor}]{$\;$(3e)}{0.0}{\PathFig \FlagApertureThree _SINC.pdf} &
		\subfigimg[width=\linewidth,pos=ul,font=\fontfig{\subfigColor}]{$\;$(3f)}{0.0}{\PathFig \FlagApertureThree _Binary_mask.pdf}
		\\[-4pt]
		\rotatebox[origin=l]{90}{\small VLT-like ($\times 10$)}
		&
		\subfigimg[width=\linewidth,pos=ul,font=\fontfig{black}]{$\;$(4a)}{0.0}{\PathFig \FlagApertureFour _Piston.pdf} &
		\subfigimg[width=\linewidth,pos=ul,font=\fontfig{black}]{$\;$(4b)}{0.0}{\PathFig \FlagApertureFour _Pyramid.pdf} &
		\subfigimg[width=\linewidth,pos=ul,font=\fontfig{black}]{$\;$(4c)}{0.0}{\PathFig \FlagApertureFour _Gaussian.pdf} &
		\subfigimg[width=\linewidth,pos=ul,font=\fontfig{black}]{$\;$(4d)}{0.0}{\PathFig \FlagApertureFour _ALPAO_repair.pdf} &
		\subfigimg[width=\linewidth,pos=ul,font=\fontfig{black}]{$\;$(4e)}{0.0}{\PathFig \FlagApertureFour _SINC.pdf} &
		\subfigimg[width=\linewidth,pos=ul,font=\fontfig{black}]{$\;$(4f)}{0.0}{\PathFig \FlagApertureFour _Binary_mask.pdf}
	\end{tabular}
	
	\caption{\label{fig:PSF_appen} Simulated long-exposure PSF~$\PSF$. See caption and colour bar of \fig{fig:PSF_disk_95}. -- (1) Aperture: Square of side~$\Dsim$. (2) Aperture: Disc of diameter~$\Dsim$. (3) Aperture: Disc of diameter~$\Dsim$, tip-tilt removed. (4) Aperture: VLT-like pupil of diameter~$\percent{97.5}\Dsim$. To match the colour bar, the PSF were multiplied by ten.}
\end{figure*}

\begin{figure*}[t!] 
	\centering
	
    \newcommand{\FirstCol}{8pt}
	\newcommand{\PathFig}{figures_PSD_}
	\newcommand{\FlagApertureOne}{square}
	\newcommand{\FlagApertureTwo}{disk}
	\newcommand{\FlagApertureThree}{disk_TT}
	\newcommand{\FlagApertureFour}{SPHERE_97p5}
	\newcommand{\subfigColor}{black}
	
	\newcommand{\LineRatio}{1}
		
	\sbox1{\includegraphics{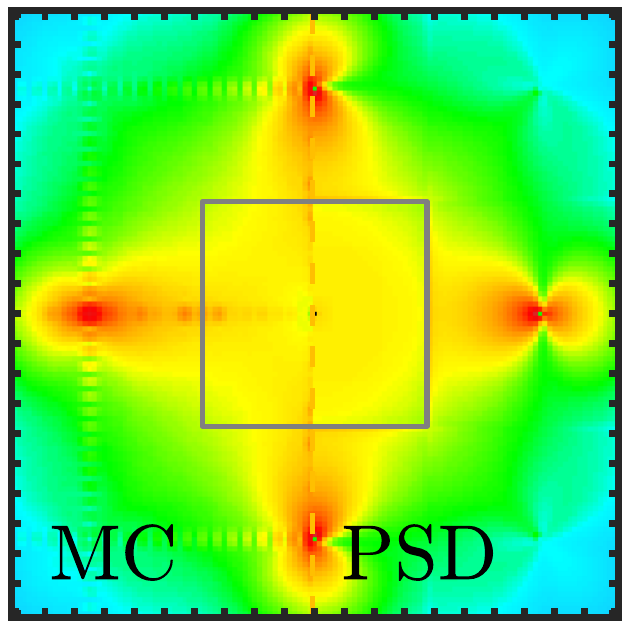}}		 
	\sbox2{\includegraphics{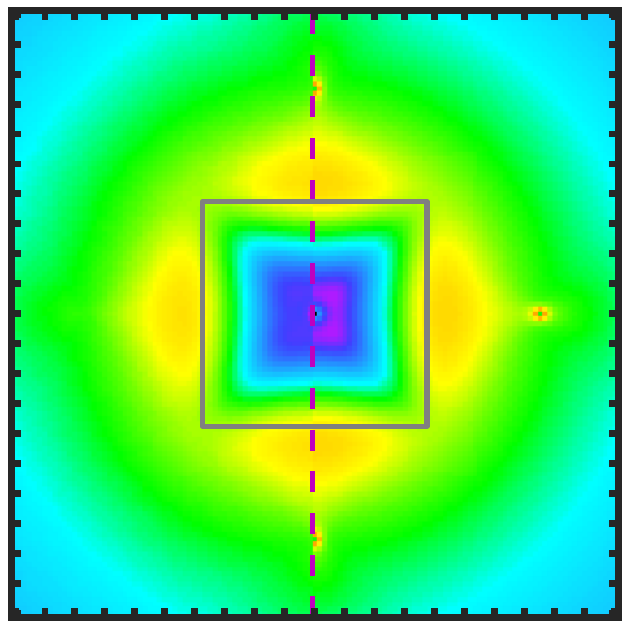}}	 
	\sbox3{\includegraphics{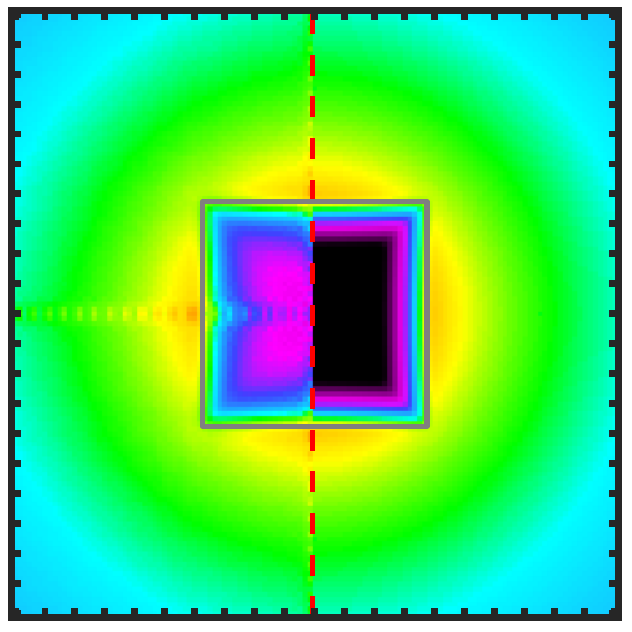}}	 
	\sbox4{\includegraphics{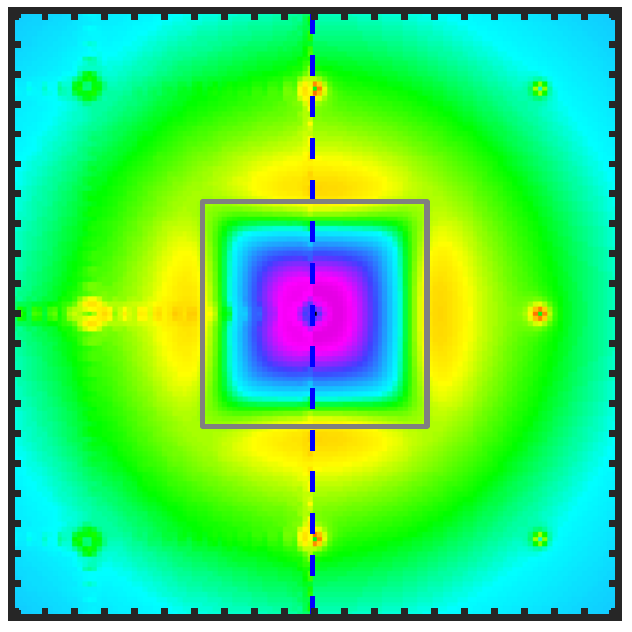}}	 
	\sbox5{\includegraphics{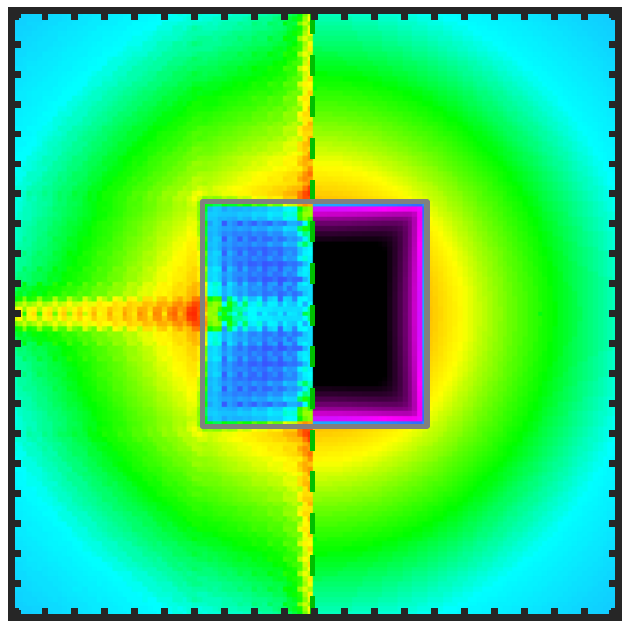}}   
	\sbox6{\includegraphics{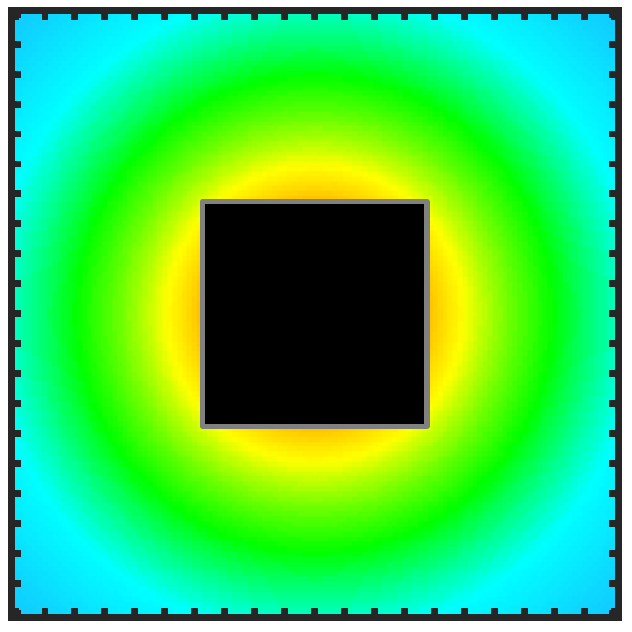}}	 
	
	\newcommand{\ColumnWidth}[1]
		{\the \dimexpr (\linewidth-\FirstCol) * \LineRatio / 6 \relax}

	\begin{tabular}{
		@{}
		C{\FirstCol}
		@{}
		C{\ColumnWidth{1}}
		@{}
		C{\ColumnWidth{2}}
		@{}
		C{\ColumnWidth{3}}
		@{}
		C{\ColumnWidth{4}}
		@{}
		C{\ColumnWidth{5}}
		@{}
		C{\ColumnWidth{6}}
		@{}
		}
		&
		\small{Piston}
		&
		\small{Pyramid}
		&
		\small{Gaussian}
		&
		\small{ALPAO}
		&
		\small{2D~$\sinc$}
		&
		\small{Binary mask}
		\\[-1pt]
		\rotatebox[origin=l]{90}{\small Square of side~$\Dsim$}
		&
		\subfigimg[width=\linewidth,pos=ul,font=\fontfig{\subfigColor}]{$\;$(1a)}{0.0}{\PathFig \FlagApertureOne _Piston.pdf} &
		\subfigimg[width=\linewidth,pos=ul,font=\fontfig{\subfigColor}]{$\;$(1b)}{0.0}{\PathFig \FlagApertureOne _Pyramid.pdf} &
		\subfigimg[width=\linewidth,pos=ul,font=\fontfig{\subfigColor}]{$\;$(1c)}{0.0}{\PathFig \FlagApertureOne _Gaussian.pdf} &
		\subfigimg[width=\linewidth,pos=ul,font=\fontfig{\subfigColor}]{$\;$(1d)}{0.0}{\PathFig \FlagApertureOne _ALPAO_repair.pdf} &
		\subfigimg[width=\linewidth,pos=ul,font=\fontfig{\subfigColor}]{$\;$(1e)}{0.0}{\PathFig \FlagApertureOne _SINC.pdf} &
		\subfigimg[width=\linewidth,pos=ul,font=\fontfig{\subfigColor}]{$\;$(1f)}{0.0}{\PathFig \FlagApertureOne _Binary_mask.pdf}
		\\[-4pt]
		\rotatebox[origin=l]{90}{\small Disc of side~$\Dsim$}
		&
		\subfigimg[width=\linewidth,pos=ul,font=\fontfig{\subfigColor}]{$\;$(2a)}{0.0}{\PathFig \FlagApertureTwo _Piston.pdf} &
		\subfigimg[width=\linewidth,pos=ul,font=\fontfig{\subfigColor}]{$\;$(2b)}{0.0}{\PathFig \FlagApertureTwo _Pyramid.pdf} &
		\subfigimg[width=\linewidth,pos=ul,font=\fontfig{\subfigColor}]{$\;$(2c)}{0.0}{\PathFig \FlagApertureTwo _Gaussian.pdf} &
		\subfigimg[width=\linewidth,pos=ul,font=\fontfig{\subfigColor}]{$\;$(2d)}{0.0}{\PathFig \FlagApertureTwo _ALPAO_repair.pdf} &
		\subfigimg[width=\linewidth,pos=ul,font=\fontfig{\subfigColor}]{$\;$(2e)}{0.0}{\PathFig \FlagApertureTwo _SINC.pdf} &
		\subfigimg[width=\linewidth,pos=ul,font=\fontfig{\subfigColor}]{$\;$(2f)}{0.0}{\PathFig \FlagApertureTwo _Binary_mask.pdf}
		\\[-4pt]
		\rotatebox[origin=l]{90}{\small (2) + tip-tilt removed}
		&
		\subfigimg[width=\linewidth,pos=ul,font=\fontfig{\subfigColor}]{$\;$(3a)}{0.0}{\PathFig \FlagApertureThree _Piston.pdf} &
		\subfigimg[width=\linewidth,pos=ul,font=\fontfig{\subfigColor}]{$\;$(3b)}{0.0}{\PathFig \FlagApertureThree _Pyramid.pdf} &
		\subfigimg[width=\linewidth,pos=ul,font=\fontfig{\subfigColor}]{$\;$(3c)}{0.0}{\PathFig \FlagApertureThree _Gaussian.pdf} &
		\subfigimg[width=\linewidth,pos=ul,font=\fontfig{\subfigColor}]{$\;$(3d)}{0.0}{\PathFig \FlagApertureThree _ALPAO_repair.pdf} &
		\subfigimg[width=\linewidth,pos=ul,font=\fontfig{\subfigColor}]{$\;$(3e)}{0.0}{\PathFig \FlagApertureThree _SINC.pdf} &
		\subfigimg[width=\linewidth,pos=ul,font=\fontfig{\subfigColor}]{$\;$(3f)}{0.0}{\PathFig \FlagApertureThree _Binary_mask.pdf}
		\\[-4pt]
		\rotatebox[origin=l]{90}{\small VLT-like}
		&
		\subfigimg[width=\linewidth,pos=ul,font=\fontfig{\subfigColor}]{$\;$(4a)}{0.0}{\PathFig \FlagApertureFour _Piston.pdf} &
		\subfigimg[width=\linewidth,pos=ul,font=\fontfig{\subfigColor}]{$\;$(4b)}{0.0}{\PathFig \FlagApertureFour _Pyramid.pdf} &
		\subfigimg[width=\linewidth,pos=ul,font=\fontfig{\subfigColor}]{$\;$(4c)}{0.0}{\PathFig \FlagApertureFour _Gaussian.pdf} &
		\subfigimg[width=\linewidth,pos=ul,font=\fontfig{\subfigColor}]{$\;$(4d)}{0.0}{\PathFig \FlagApertureFour _ALPAO_repair.pdf} &
		\subfigimg[width=\linewidth,pos=ul,font=\fontfig{\subfigColor}]{$\;$(4e)}{0.0}{\PathFig \FlagApertureFour _SINC.pdf} &
		\subfigimg[width=\linewidth,pos=ul,font=\fontfig{\subfigColor}]{$\;$(4f)}{0.0}{\PathFig \FlagApertureFour _Binary_mask.pdf}
	\end{tabular}
	
	\caption{\label{fig:PSD_appen} Simulated PSD of the fitting residuals~$\PSDres$ normalised by~$\rFried^{-5/3}$. See caption and colour bar of \fig{fig:PSD_disk_95}. -- (1) Aperture: Square of side~$\Dsim$. (2) Aperture: Disc of diameter~$\Dsim$. (3) Aperture: Disc of diameter~$\Dsim$, tip-tilt removed. (4) Aperture: VLT-like pupil of diameter~$\percent{97.5}\Dsim$.}
\end{figure*}

\begin{figure*}[t!] 
	\centering
	
    \newcommand{\FirstCol}{8pt}
	\newcommand{\PathFig}{figures_PSF_PSD_}
	\newcommand{\FlagApertureOne}{square}
	\newcommand{\FlagApertureTwo}{disk}
	\newcommand{\FlagApertureThree}{disk_TT}
	\newcommand{\FlagApertureFour}{SPHERE_97p5}
	\newcommand{\subfigColor}{black}
	
	\newcommand{\LineRatio}{1}
		
	\sbox1{\includegraphics{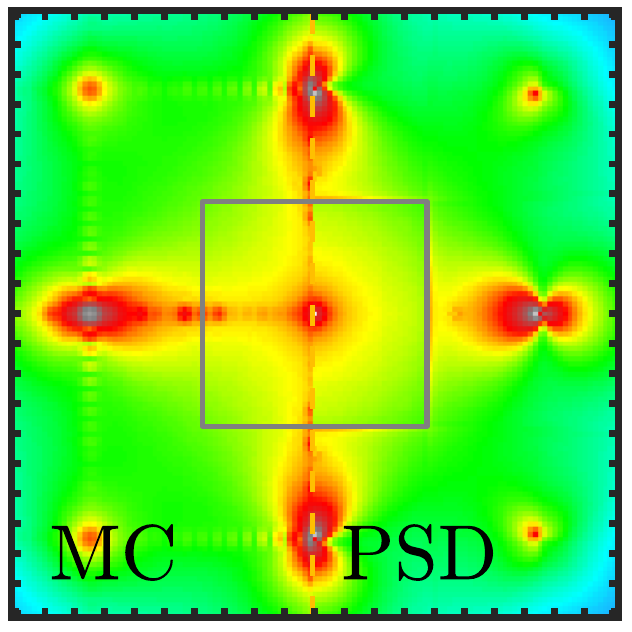}}		 
	\sbox2{\includegraphics{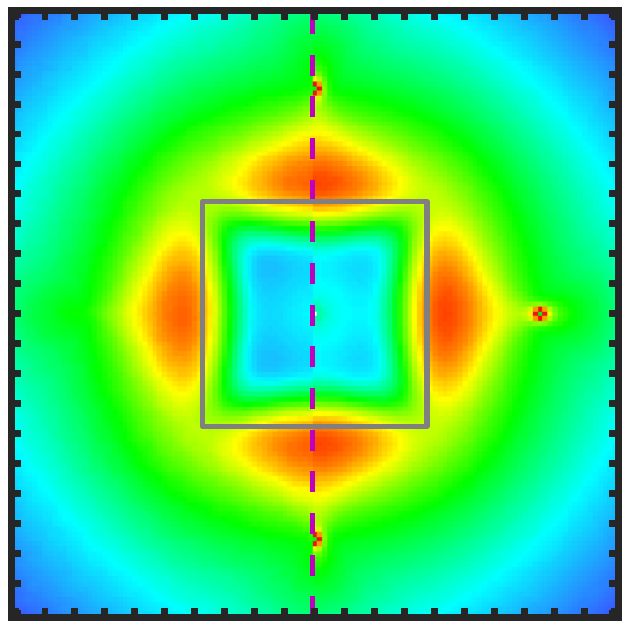}}	 
	\sbox3{\includegraphics{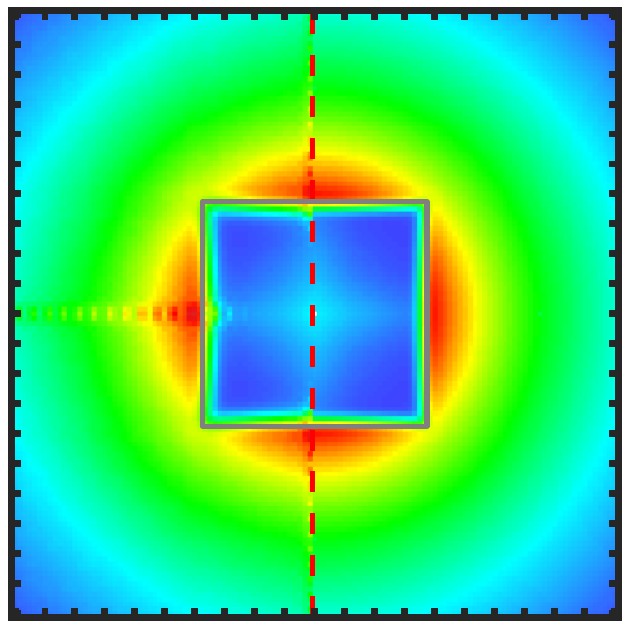}}	 
	\sbox4{\includegraphics{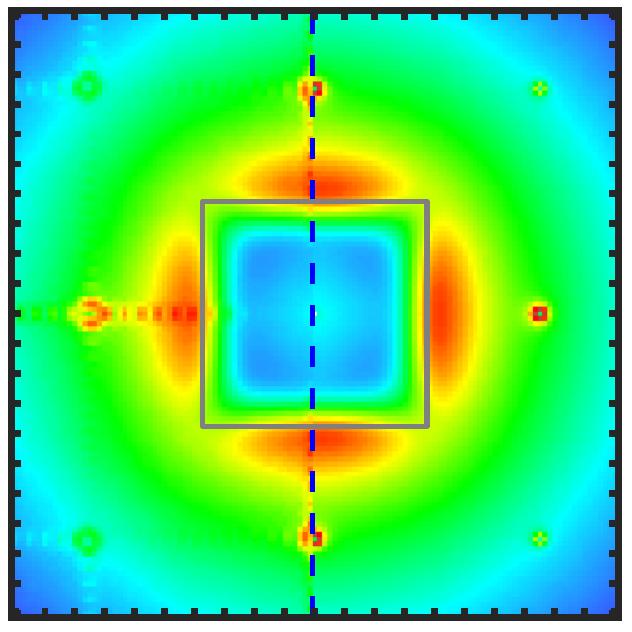}}	 
	\sbox5{\includegraphics{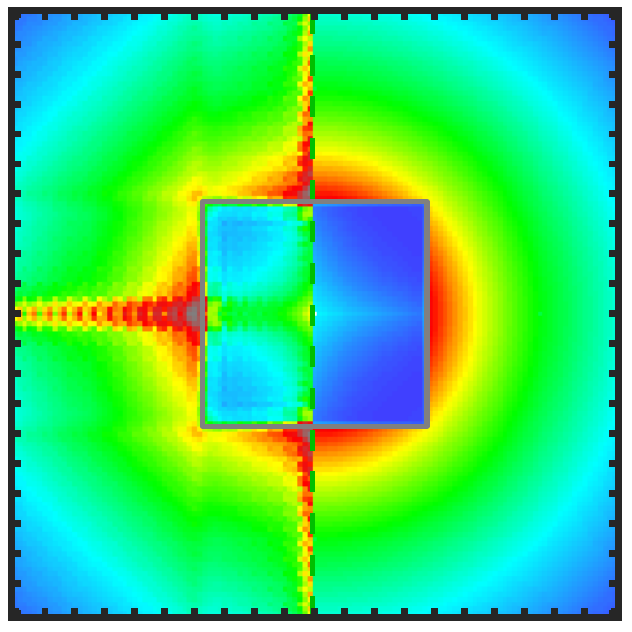}}   
	\sbox6{\includegraphics{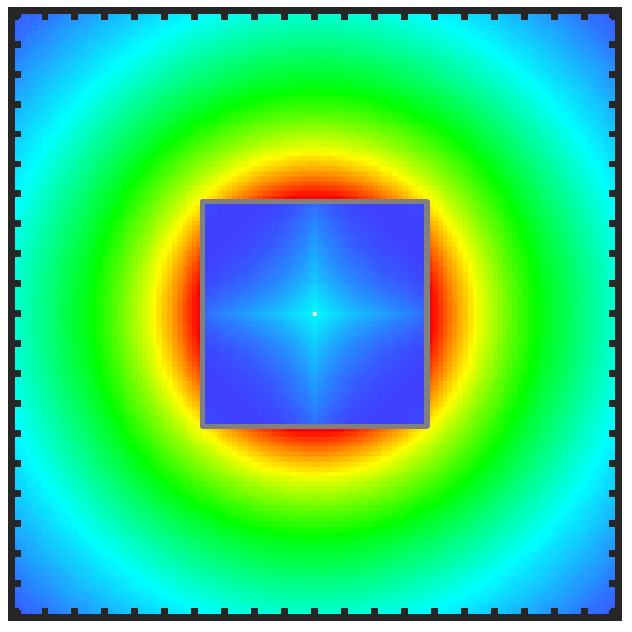}}	 
	
	\newcommand{\ColumnWidth}[1]
		{\the \dimexpr (\linewidth-\FirstCol) * \LineRatio / 6 \relax}

	\begin{tabular}{
		@{}
		C{\FirstCol}
		@{}
		C{\ColumnWidth{1}}
		@{}
		C{\ColumnWidth{2}}
		@{}
		C{\ColumnWidth{3}}
		@{}
		C{\ColumnWidth{4}}
		@{}
		C{\ColumnWidth{5}}
		@{}
		C{\ColumnWidth{6}}
		@{}
		}
		&
		\small{Piston}
		&
		\small{Pyramid}
		&
		\small{Gaussian}
		&
		\small{ALPAO}
		&
		\small{2D~$\sinc$}
		&
		\small{Binary mask}
		\\[-1pt]
		\rotatebox[origin=l]{90}{\small Square of side~$\Dsim$}
		&
		\subfigimg[width=\linewidth,pos=ul,font=\fontfig{\subfigColor}]{$\;$(1a)}{0.0}{\PathFig \FlagApertureOne _Piston.pdf} &
		\subfigimg[width=\linewidth,pos=ul,font=\fontfig{\subfigColor}]{$\;$(1b)}{0.0}{\PathFig \FlagApertureOne _Pyramid.pdf} &
		\subfigimg[width=\linewidth,pos=ul,font=\fontfig{\subfigColor}]{$\;$(1c)}{0.0}{\PathFig \FlagApertureOne _Gaussian.pdf} &
		\subfigimg[width=\linewidth,pos=ul,font=\fontfig{\subfigColor}]{$\;$(1d)}{0.0}{\PathFig \FlagApertureOne _ALPAO_repair.pdf} &
		\subfigimg[width=\linewidth,pos=ul,font=\fontfig{\subfigColor}]{$\;$(1e)}{0.0}{\PathFig \FlagApertureOne _SINC.pdf} &
		\subfigimg[width=\linewidth,pos=ul,font=\fontfig{\subfigColor}]{$\;$(1f)}{0.0}{\PathFig \FlagApertureOne _Binary_mask.pdf}
		\\[-4pt]
		\rotatebox[origin=l]{90}{\small Disc of side~$\Dsim$}
		&
		\subfigimg[width=\linewidth,pos=ul,font=\fontfig{\subfigColor}]{$\;$(2a)}{0.0}{\PathFig \FlagApertureTwo _Piston.pdf} &
		\subfigimg[width=\linewidth,pos=ul,font=\fontfig{\subfigColor}]{$\;$(2b)}{0.0}{\PathFig \FlagApertureTwo _Pyramid.pdf} &
		\subfigimg[width=\linewidth,pos=ul,font=\fontfig{\subfigColor}]{$\;$(2c)}{0.0}{\PathFig \FlagApertureTwo _Gaussian.pdf} &
		\subfigimg[width=\linewidth,pos=ul,font=\fontfig{\subfigColor}]{$\;$(2d)}{0.0}{\PathFig \FlagApertureTwo _ALPAO_repair.pdf} &
		\subfigimg[width=\linewidth,pos=ul,font=\fontfig{\subfigColor}]{$\;$(2e)}{0.0}{\PathFig \FlagApertureTwo _SINC.pdf} &
		\subfigimg[width=\linewidth,pos=ul,font=\fontfig{\subfigColor}]{$\;$(2f)}{0.0}{\PathFig \FlagApertureTwo _Binary_mask.pdf}
		\\[-4pt]
		\rotatebox[origin=l]{90}{\small (2) + tip-tilt removed}
		&
		\subfigimg[width=\linewidth,pos=ul,font=\fontfig{\subfigColor}]{$\;$(3a)}{0.0}{\PathFig \FlagApertureThree _Piston.pdf} &
		\subfigimg[width=\linewidth,pos=ul,font=\fontfig{\subfigColor}]{$\;$(3b)}{0.0}{\PathFig \FlagApertureThree _Pyramid.pdf} &
		\subfigimg[width=\linewidth,pos=ul,font=\fontfig{\subfigColor}]{$\;$(3c)}{0.0}{\PathFig \FlagApertureThree _Gaussian.pdf} &
		\subfigimg[width=\linewidth,pos=ul,font=\fontfig{\subfigColor}]{$\;$(3d)}{0.0}{\PathFig \FlagApertureThree _ALPAO_repair.pdf} &
		\subfigimg[width=\linewidth,pos=ul,font=\fontfig{\subfigColor}]{$\;$(3e)}{0.0}{\PathFig \FlagApertureThree _SINC.pdf} &
		\subfigimg[width=\linewidth,pos=ul,font=\fontfig{\subfigColor}]{$\;$(3f)}{0.0}{\PathFig \FlagApertureThree _Binary_mask.pdf}
		\\[-4pt]
		\rotatebox[origin=l]{90}{\small VLT-like ($\times 10$)}
		&
		\subfigimg[width=\linewidth,pos=ul,font=\fontfig{\subfigColor}]{$\;$(4a)}{0.0}{\PathFig \FlagApertureFour _Piston.pdf} &
		\subfigimg[width=\linewidth,pos=ul,font=\fontfig{\subfigColor}]{$\;$(4b)}{0.0}{\PathFig \FlagApertureFour _Pyramid.pdf} &
		\subfigimg[width=\linewidth,pos=ul,font=\fontfig{\subfigColor}]{$\;$(4c)}{0.0}{\PathFig \FlagApertureFour _Gaussian.pdf} &
		\subfigimg[width=\linewidth,pos=ul,font=\fontfig{\subfigColor}]{$\;$(4d)}{0.0}{\PathFig \FlagApertureFour _ALPAO_repair.pdf} &
		\subfigimg[width=\linewidth,pos=ul,font=\fontfig{\subfigColor}]{$\;$(4e)}{0.0}{\PathFig \FlagApertureFour _SINC.pdf} &
		\subfigimg[width=\linewidth,pos=ul,font=\fontfig{\subfigColor}]{$\;$(4f)}{0.0}{\PathFig \FlagApertureFour _Binary_mask.pdf}
	\end{tabular}
	
	\caption{\label{fig:PSF_PSD_appen} Simulated PSF of the fitting residuals~$\PSFres$. See caption and colour bar of \fig{fig:PSF_pha_disk_95}.  -- (1) Aperture: Square of side~$\Dsim$. (2) Aperture: Disc of diameter~$\Dsim$. (3) Aperture: Disc of diameter~$\Dsim$, tip-tilt removed. (4) Aperture: VLT-like pupil of diameter~$\percent{97.5}\Dsim$. To match the colour bar, the PSF were multiplied by ten.}
\end{figure*}

\begin{figure*}[t!] 
	\centering
	
    \newcommand{\FirstCol}{8pt}
	\newcommand{\PathFig}{figures_SF_}
	\newcommand{\FlagApertureOne}{square}
	\newcommand{\FlagApertureTwo}{disk}
	\newcommand{\FlagApertureThree}{disk_TT}
	\newcommand{\FlagApertureFour}{SPHERE_97p5}
	\newcommand{\subfigColor}{black}
	
	\newcommand{\LineRatio}{1}
		
	\sbox1{\includegraphics{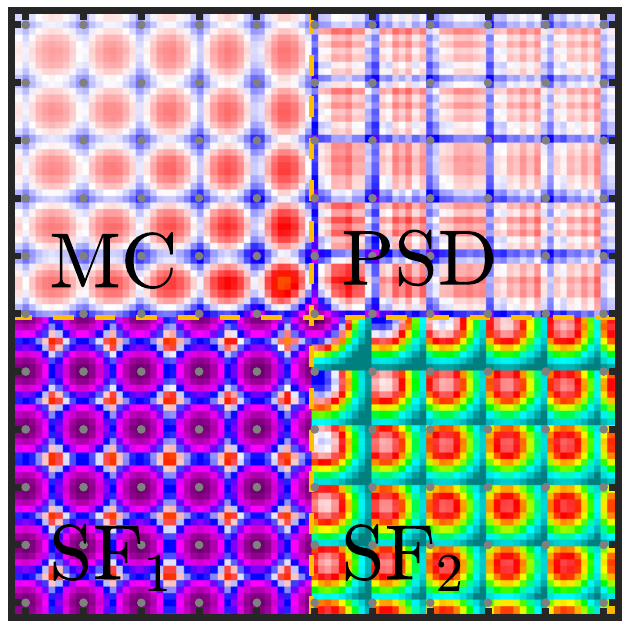}}		 
	\sbox2{\includegraphics{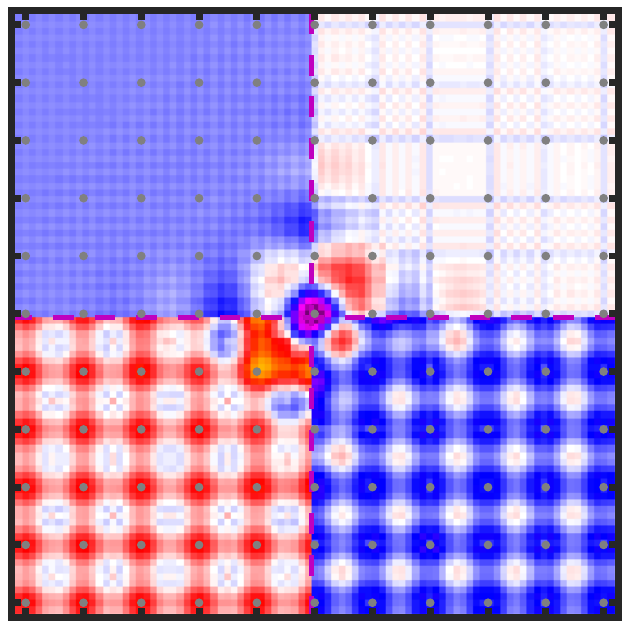}}	 
	\sbox3{\includegraphics{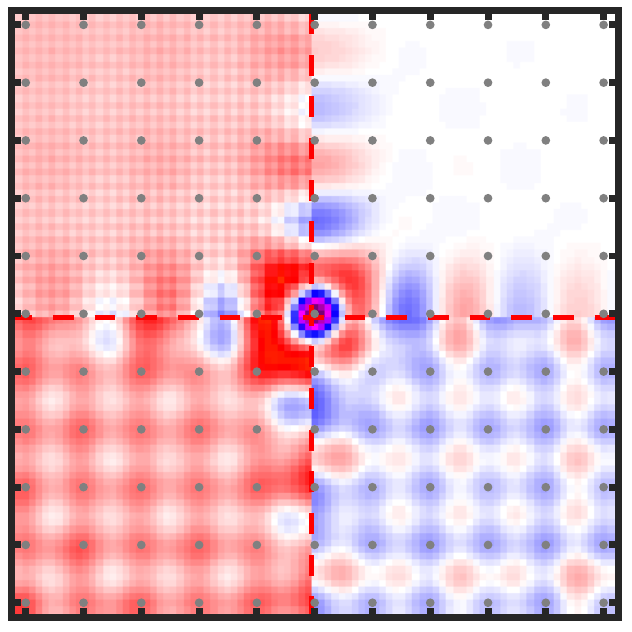}}	 
	\sbox4{\includegraphics{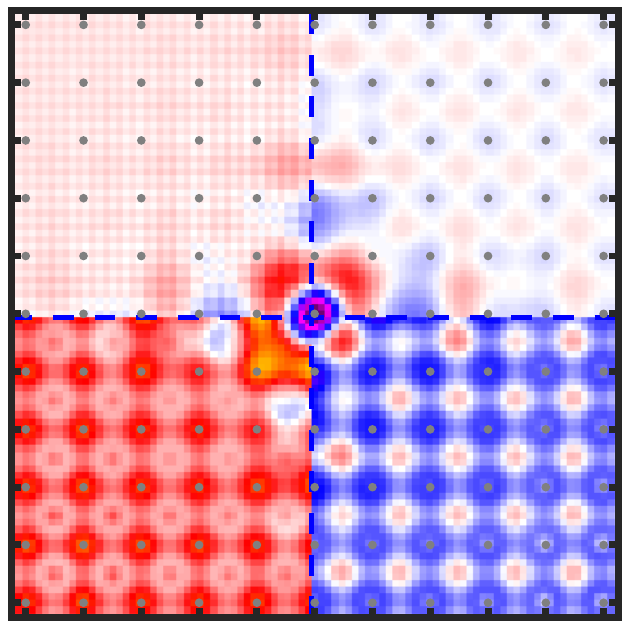}}	 
	\sbox5{\includegraphics{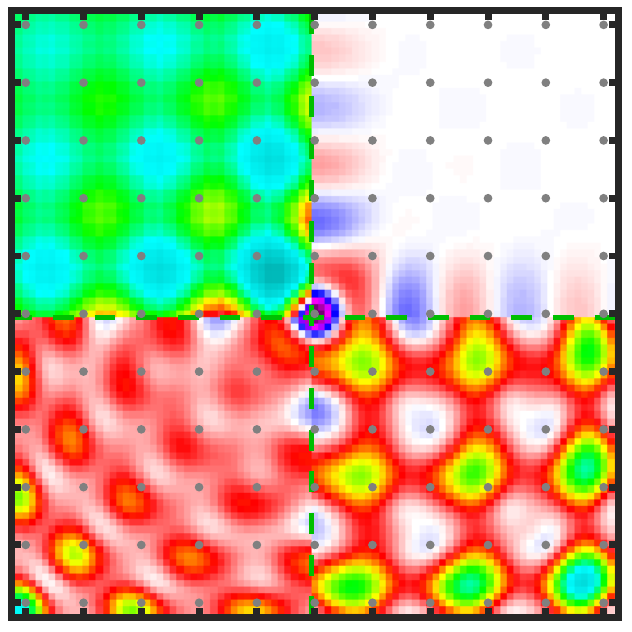}}   
	\sbox6{\includegraphics{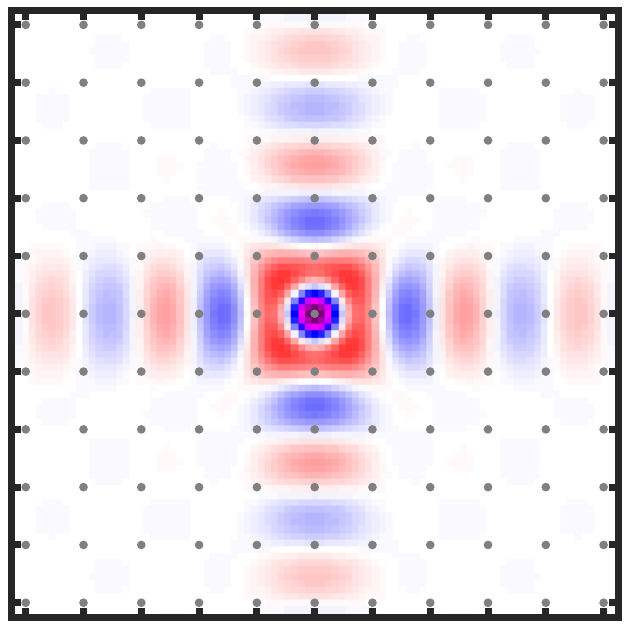}}	 
	
	\newcommand{\ColumnWidth}[1]
		{\the \dimexpr (\linewidth-\FirstCol) * \LineRatio / 6 \relax}

	\begin{tabular}{
		@{}
		C{\FirstCol}
		@{}
		C{\ColumnWidth{1}}
		@{}
		C{\ColumnWidth{2}}
		@{}
		C{\ColumnWidth{3}}
		@{}
		C{\ColumnWidth{4}}
		@{}
		C{\ColumnWidth{5}}
		@{}
		C{\ColumnWidth{6}}
		@{}
		}
		&
		\small{Piston}
		&
		\small{Pyramid}
		&
		\small{Gaussian}
		&
		\small{ALPAO}
		&
		\small{2D~$\sinc$}
		&
		\small{Binary mask}
		\\[-1pt]
		\rotatebox[origin=l]{90}{\small Square of side~$\Dsim$}
		&
		\subfigimg[width=\linewidth,pos=ul,font=\fontfig{\subfigColor}]{$\;$(1a)}{0.0}{\PathFig \FlagApertureOne _Piston.pdf} &
		\subfigimg[width=\linewidth,pos=ul,font=\fontfig{\subfigColor}]{$\;$(1b)}{0.0}{\PathFig \FlagApertureOne _Pyramid.pdf} &
		\subfigimg[width=\linewidth,pos=ul,font=\fontfig{\subfigColor}]{$\;$(1c)}{0.0}{\PathFig \FlagApertureOne _Gaussian.pdf} &
		\subfigimg[width=\linewidth,pos=ul,font=\fontfig{\subfigColor}]{$\;$(1d)}{0.0}{\PathFig \FlagApertureOne _ALPAO_repair.pdf} &
		\subfigimg[width=\linewidth,pos=ul,font=\fontfig{\subfigColor}]{$\;$(1e)}{0.0}{\PathFig \FlagApertureOne _SINC.pdf} &
		\subfigimg[width=\linewidth,pos=ul,font=\fontfig{\subfigColor}]{$\;$(1f)}{0.0}{\PathFig \FlagApertureOne _Binary_mask.pdf}
		\\[-4pt]
		\rotatebox[origin=l]{90}{\small Disc of side~$\Dsim$}
		&
		\subfigimg[width=\linewidth,pos=ul,font=\fontfig{\subfigColor}]{$\;$(2a)}{0.0}{\PathFig \FlagApertureTwo _Piston.pdf} &
		\subfigimg[width=\linewidth,pos=ul,font=\fontfig{\subfigColor}]{$\;$(2b)}{0.0}{\PathFig \FlagApertureTwo _Pyramid.pdf} &
		\subfigimg[width=\linewidth,pos=ul,font=\fontfig{\subfigColor}]{$\;$(2c)}{0.0}{\PathFig \FlagApertureTwo _Gaussian.pdf} &
		\subfigimg[width=\linewidth,pos=ul,font=\fontfig{\subfigColor}]{$\;$(2d)}{0.0}{\PathFig \FlagApertureTwo _ALPAO_repair.pdf} &
		\subfigimg[width=\linewidth,pos=ul,font=\fontfig{\subfigColor}]{$\;$(2e)}{0.0}{\PathFig \FlagApertureTwo _SINC.pdf} &
		\subfigimg[width=\linewidth,pos=ul,font=\fontfig{\subfigColor}]{$\;$(2f)}{0.0}{\PathFig \FlagApertureTwo _Binary_mask.pdf}
		\\[-4pt]
		\rotatebox[origin=l]{90}{\small (2) + tip-tilt removed}
		&
		\subfigimg[width=\linewidth,pos=ul,font=\fontfig{\subfigColor}]{$\;$(3a)}{0.0}{\PathFig \FlagApertureThree _Piston.pdf} &
		\subfigimg[width=\linewidth,pos=ul,font=\fontfig{\subfigColor}]{$\;$(3b)}{0.0}{\PathFig \FlagApertureThree _Pyramid.pdf} &
		\subfigimg[width=\linewidth,pos=ul,font=\fontfig{\subfigColor}]{$\;$(3c)}{0.0}{\PathFig \FlagApertureThree _Gaussian.pdf} &
		\subfigimg[width=\linewidth,pos=ul,font=\fontfig{\subfigColor}]{$\;$(3d)}{0.0}{\PathFig \FlagApertureThree _ALPAO_repair.pdf} &
		\subfigimg[width=\linewidth,pos=ul,font=\fontfig{\subfigColor}]{$\;$(3e)}{0.0}{\PathFig \FlagApertureThree _SINC.pdf} &
		\subfigimg[width=\linewidth,pos=ul,font=\fontfig{\subfigColor}]{$\;$(3f)}{0.0}{\PathFig \FlagApertureThree _Binary_mask.pdf}
		\\[-4pt]
		\rotatebox[origin=l]{90}{\small VLT-like}
		&
		\subfigimg[width=\linewidth,pos=ul,font=\fontfig{\subfigColor}]{$\;$(4a)}{0.0}{\PathFig \FlagApertureFour _Piston.pdf} &
		\subfigimg[width=\linewidth,pos=ul,font=\fontfig{\subfigColor}]{$\;$(4b)}{0.0}{\PathFig \FlagApertureFour _Pyramid.pdf} &
		\subfigimg[width=\linewidth,pos=ul,font=\fontfig{\subfigColor}]{$\;$(4c)}{0.0}{\PathFig \FlagApertureFour _Gaussian.pdf} &
		\subfigimg[width=\linewidth,pos=ul,font=\fontfig{\subfigColor}]{$\;$(4d)}{0.0}{\PathFig \FlagApertureFour _ALPAO_repair.pdf} &
		\subfigimg[width=\linewidth,pos=ul,font=\fontfig{\subfigColor}]{$\;$(4e)}{0.0}{\PathFig \FlagApertureFour _SINC.pdf} &
		\subfigimg[width=\linewidth,pos=ul,font=\fontfig{\subfigColor}]{$\;$(4f)}{0.0}{\PathFig \FlagApertureFour _Binary_mask.pdf}
	\end{tabular}
	
	\caption{\label{fig:SF_appen}  Simulated structure function of the fitting residuals~$\SFAO$ normalised by~$2\AvgT{\sigAO^{2}}$. See caption and colour bar of \fig{fig:SF_disk_95}. -- (1) Aperture: Square of side~$\Dsim$. (2) Aperture: Disc of diameter~$\Dsim$. (3) Aperture: Disc of diameter~$\Dsim$, tip-tilt removed. (4) Aperture: VLT-like pupil of diameter~$\percent{97.5}\Dsim$. To avoid the central obscuration, the two 2D maps of the analytical structure function are displayed for $\SFAO\Paren{\Vx+\Paren{3\pitch,0}\T,\Paren{3\pitch,0}\T}$ (SF$_{1}$) and $\SFAO\Paren{\Vx+\Vpitch/2+\Paren{3\pitch,0}\T,\Vpitch/2+\Paren{3\pitch,0}\T}$ (SF$_{2}$).}
\end{figure*}

\begin{figure*}[t!] 
	\centering
	
    \newcommand{\FirstCol}{8pt}
	\newcommand{\PathFig}{figures_SF_map_}
	\newcommand{\FlagApertureOne}{square}
	\newcommand{\FlagApertureTwo}{disk}
	\newcommand{\FlagApertureThree}{disk_TT}
	\newcommand{\FlagApertureFour}{SPHERE_97p5}
	\newcommand{\subfigColor}{black}
	
	\newcommand{\LineRatio}{1}
		
	\sbox1{\includegraphics{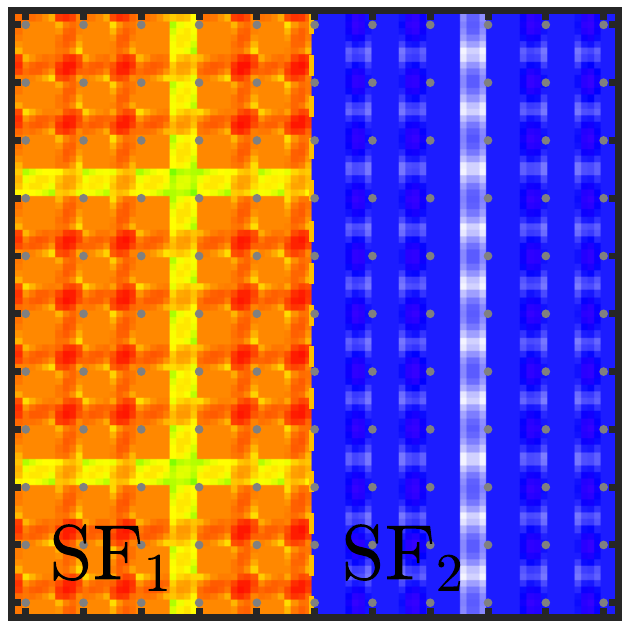}}		 
	\sbox2{\includegraphics{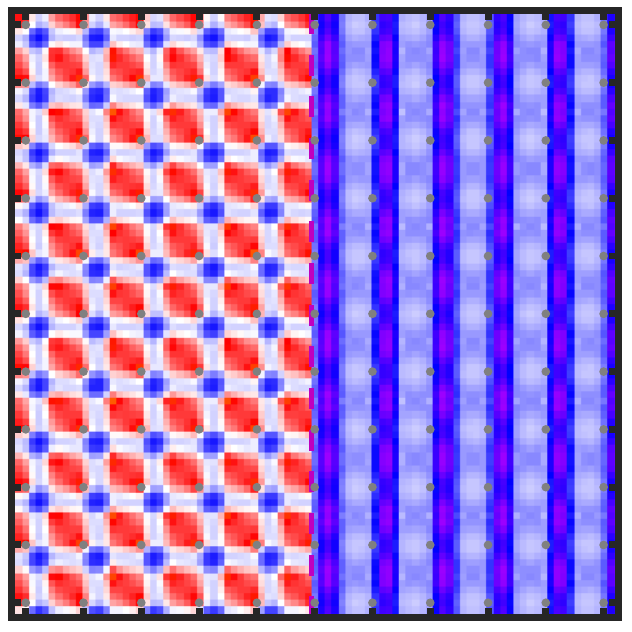}}	 
	\sbox3{\includegraphics{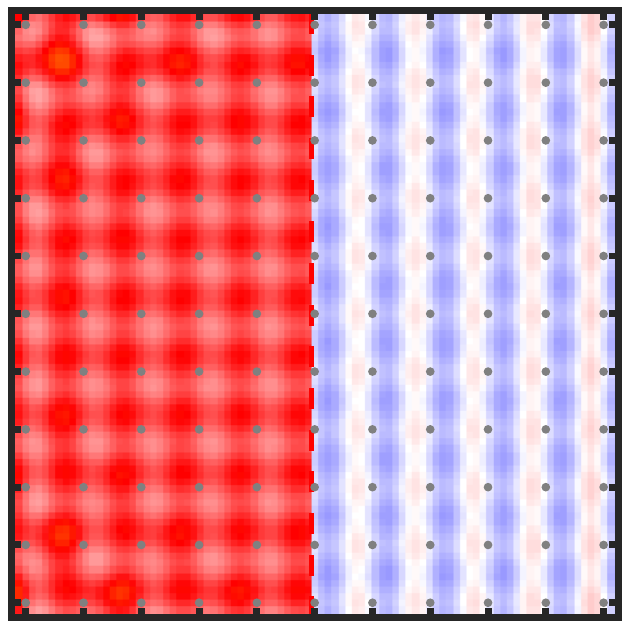}}	 
	\sbox4{\includegraphics{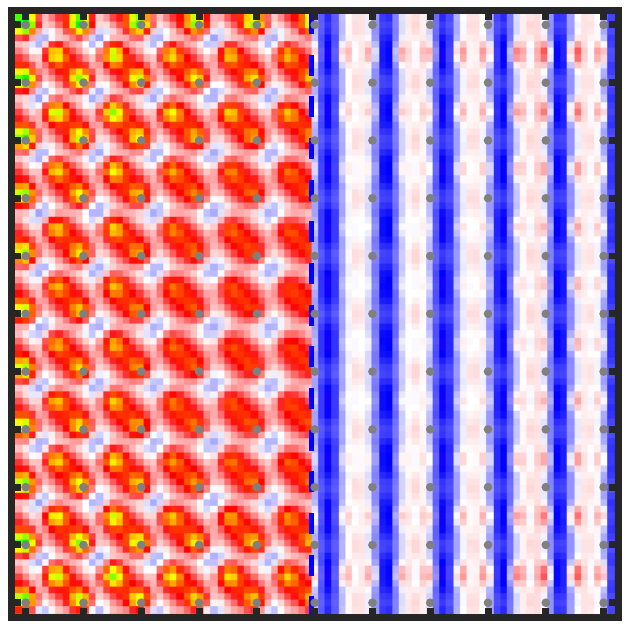}}	 
	\sbox5{\includegraphics{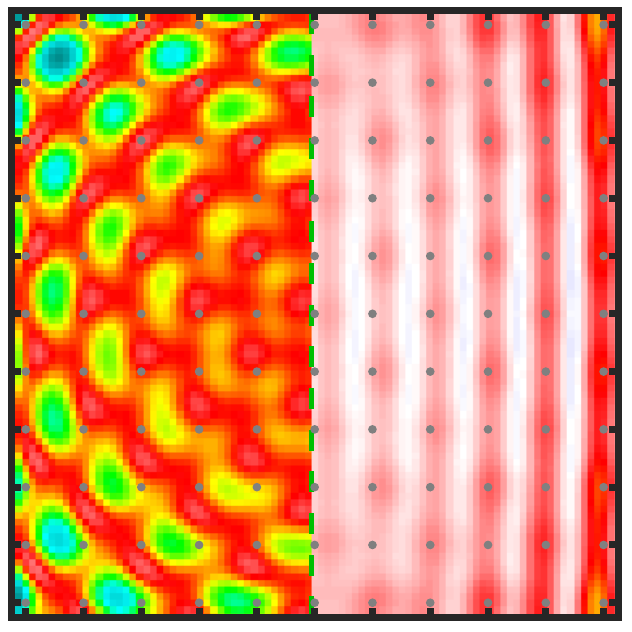}}   
	\sbox6{\includegraphics{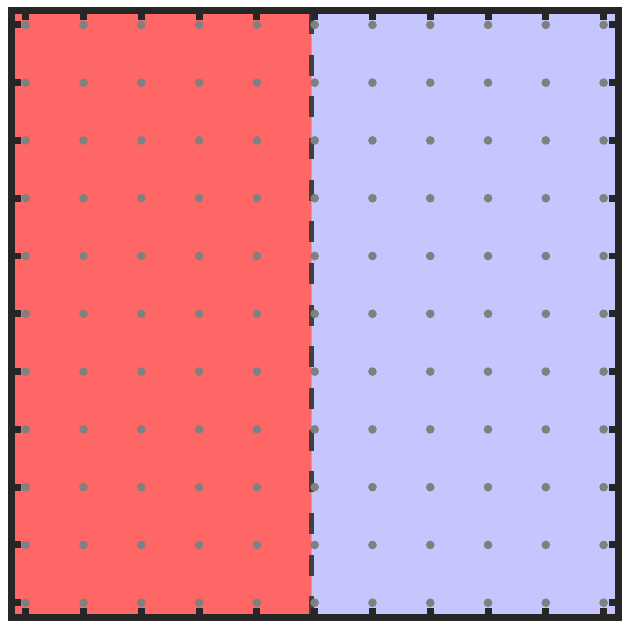}}	 
	
	\newcommand{\ColumnWidth}[1]
		{\the \dimexpr (\linewidth-\FirstCol) * \LineRatio / 6 \relax}

	\begin{tabular}{
		@{}
		C{\FirstCol}
		@{}
		C{\ColumnWidth{1}}
		@{}
		C{\ColumnWidth{2}}
		@{}
		C{\ColumnWidth{3}}
		@{}
		C{\ColumnWidth{4}}
		@{}
		C{\ColumnWidth{5}}
		@{}
		C{\ColumnWidth{6}}
		@{}
		}
		&
		\small{Piston}
		&
		\small{Pyramid}
		&
		\small{Gaussian}
		&
		\small{ALPAO}
		&
		\small{2D~$\sinc$}
		&
		\small{Binary mask}
		\\[-1pt]
		\rotatebox[origin=l]{90}{\small Square of side~$\Dsim$}
		&
		\subfigimg[width=\linewidth,pos=ul,font=\fontfig{\subfigColor}]{$\;$(1a)}{0.0}{\PathFig \FlagApertureOne _Piston.pdf} &
		\subfigimg[width=\linewidth,pos=ul,font=\fontfig{\subfigColor}]{$\;$(1b)}{0.0}{\PathFig \FlagApertureOne _Pyramid.pdf} &
		\subfigimg[width=\linewidth,pos=ul,font=\fontfig{\subfigColor}]{$\;$(1c)}{0.0}{\PathFig \FlagApertureOne _Gaussian.pdf} &
		\subfigimg[width=\linewidth,pos=ul,font=\fontfig{\subfigColor}]{$\;$(1d)}{0.0}{\PathFig \FlagApertureOne _ALPAO_repair.pdf} &
		\subfigimg[width=\linewidth,pos=ul,font=\fontfig{\subfigColor}]{$\;$(1e)}{0.0}{\PathFig \FlagApertureOne _SINC.pdf} &
		\subfigimg[width=\linewidth,pos=ul,font=\fontfig{\subfigColor}]{$\;$(1f)}{0.0}{\PathFig \FlagApertureOne _Binary_mask.pdf}
		\\[-4pt]
		\rotatebox[origin=l]{90}{\small Disc of side~$\Dsim$}
		&
		\subfigimg[width=\linewidth,pos=ul,font=\fontfig{\subfigColor}]{$\;$(2a)}{0.0}{\PathFig \FlagApertureTwo _Piston.pdf} &
		\subfigimg[width=\linewidth,pos=ul,font=\fontfig{\subfigColor}]{$\;$(2b)}{0.0}{\PathFig \FlagApertureTwo _Pyramid.pdf} &
		\subfigimg[width=\linewidth,pos=ul,font=\fontfig{\subfigColor}]{$\;$(2c)}{0.0}{\PathFig \FlagApertureTwo _Gaussian.pdf} &
		\subfigimg[width=\linewidth,pos=ul,font=\fontfig{\subfigColor}]{$\;$(2d)}{0.0}{\PathFig \FlagApertureTwo _ALPAO_repair.pdf} &
		\subfigimg[width=\linewidth,pos=ul,font=\fontfig{\subfigColor}]{$\;$(2e)}{0.0}{\PathFig \FlagApertureTwo _SINC.pdf} &
		\subfigimg[width=\linewidth,pos=ul,font=\fontfig{\subfigColor}]{$\;$(2f)}{0.0}{\PathFig \FlagApertureTwo _Binary_mask.pdf}
		\\[-4pt]
		\rotatebox[origin=l]{90}{\small (2) + tip-tilt removed}
		&
		\subfigimg[width=\linewidth,pos=ul,font=\fontfig{\subfigColor}]{$\;$(3a)}{0.0}{\PathFig \FlagApertureThree _Piston.pdf} &
		\subfigimg[width=\linewidth,pos=ul,font=\fontfig{\subfigColor}]{$\;$(3b)}{0.0}{\PathFig \FlagApertureThree _Pyramid.pdf} &
		\subfigimg[width=\linewidth,pos=ul,font=\fontfig{\subfigColor}]{$\;$(3c)}{0.0}{\PathFig \FlagApertureThree _Gaussian.pdf} &
		\subfigimg[width=\linewidth,pos=ul,font=\fontfig{\subfigColor}]{$\;$(3d)}{0.0}{\PathFig \FlagApertureThree _ALPAO_repair.pdf} &
		\subfigimg[width=\linewidth,pos=ul,font=\fontfig{\subfigColor}]{$\;$(3e)}{0.0}{\PathFig \FlagApertureThree _SINC.pdf} &
		\subfigimg[width=\linewidth,pos=ul,font=\fontfig{\subfigColor}]{$\;$(3f)}{0.0}{\PathFig \FlagApertureThree _Binary_mask.pdf}
		\\[-4pt]
		\rotatebox[origin=l]{90}{\small VLT-like}
		&
		\subfigimg[width=\linewidth,pos=ul,font=\fontfig{\subfigColor}]{$\;$(4a)}{0.0}{\PathFig \FlagApertureFour _Piston.pdf} &
		\subfigimg[width=\linewidth,pos=ul,font=\fontfig{\subfigColor}]{$\;$(4b)}{0.0}{\PathFig \FlagApertureFour _Pyramid.pdf} &
		\subfigimg[width=\linewidth,pos=ul,font=\fontfig{\subfigColor}]{$\;$(4c)}{0.0}{\PathFig \FlagApertureFour _Gaussian.pdf} &
		\subfigimg[width=\linewidth,pos=ul,font=\fontfig{\subfigColor}]{$\;$(4d)}{0.0}{\PathFig \FlagApertureFour _ALPAO_repair.pdf} &
		\subfigimg[width=\linewidth,pos=ul,font=\fontfig{\subfigColor}]{$\;$(4e)}{0.0}{\PathFig \FlagApertureFour _SINC.pdf} &
		\subfigimg[width=\linewidth,pos=ul,font=\fontfig{\subfigColor}]{$\;$(4f)}{0.0}{\PathFig \FlagApertureFour _Binary_mask.pdf}
	\end{tabular}
	
	\caption{\label{fig:map_SF_appen} Inhomogeneity in the pupil of the analytical structure function of the fitting residuals~$\SFAO$ for two different relative distances normalised by~$2\AvgT{\sigAO^{2}}$. See caption and colour bar of \fig{fig:map_SF_disk_95}. -- (1) Aperture: Square of side~$\Dsim$. (2) Aperture: Disc of diameter~$\Dsim$. (3) Aperture: Disc of diameter~$\Dsim$, tip-tilt removed. (4) Aperture: VLT-like pupil of diameter~$\percent{97.5}\Dsim$.}
\end{figure*}

In this appendix, we present additional simulations on the apertures presented in \subfigs{fig:aperture}{a,b,d}:
\begin{itemize}
	\item The long-exposure PSF~$\PSF$ in \fig{fig:PSF_appen}.
	\item The expected variance on the pupil of the fitting residuals~$\AvgT{\sigAO^{2}}$, the so-called fitting error, in \tab{tab:var_sup}.
	\item The PSD of the fitting residuals~$\PSDres$ in \fig{fig:PSD_appen}.
	\item The Strehl ratio~$\strehl$ in \tab{tab:strehl_sup}.
	\item The PSF of the fitting residuals~$\PSFres$ in \fig{fig:PSF_PSD_appen}.
	\item The structure function of the fitting residuals~$\SFAO$ normalised by~$2\AvgT{\sigAO^{2}}$ in \fig{fig:SF_appen}.
	\item The inhomogeneity in the pupil of the analytical structure function of the fitting residuals~$\SFAO$ for two different relative distances normalised by~$2\AvgT{\sigAO^{2}}$ in \fig{fig:map_SF_appen}.
\end{itemize}


\begin{table*}[t!] 
	\caption{\label{tab:var_sup} Fitting error~$\AvgT{\sigAO^{2}}$ on the pupil, normalised by~$\Paren{\pitch/\rFried}^{5/3}$.}
	\centering
	\begin{tabular}{cccccccccc}
	\hline
	\hline
	\multirow{2}{*}{$\IFref$} & \multirow{2}{*}{\cite{Tyson:15_principles_of_AO}} & \multicolumn{2}{c}{Square~$\Paren{\Dsim}$} & \multicolumn{2}{c}{Disc~$\Paren{\Dsim}$} & \multicolumn{2}{c}{Disc~$\Paren{\Dsim}$, no tip-tilt} & \multicolumn{2}{c}{VLT-like}
	\\ \cline{3-10} 
			 & & \multicolumn{1}{c}{MC} & \multicolumn{1}{c}{PSD} & \multicolumn{1}{c}{MC} & \multicolumn{1}{c}{PSD} & \multicolumn{1}{c}{MC} & \multicolumn{1}{c}{PSD} & \multicolumn{1}{c}{MC} & \multicolumn{1}{c}{PSD}
    \\
	\hline
	Piston & 1.26 & 1.26$\pm0.41$ & 1.23 & 1.35$\pm0.52$ & 1.23 & 0.85$\pm0.14$ & 1.23 & 1.16$\pm0.10$ & 16.15
	\\
	Pyramid & 0.28 & 0.26$\pm0.02$ & 0.30 & 0.79$\pm0.48$ & 0.29 & 0.32$\pm0.04$ & 0.30 & 0.29$\pm0.01$ & 0.47
	\\
	Gaussian & 0.23 & 0.24$\pm0.02$ & 0.23 & 0.24$\pm0.02$ & 0.23 & 0.23$\pm0.01$ & 0.23 & 0.21$\pm0.01$ & 0.20
	\\
	ALPAO & -- & 0.26$\pm0.02$ & 0.26 &  0.39$\pm0.13$ & 0.26 & 0.26$\pm0.02$ & 0.26 & 0.27$\pm0.01$ & 0.25
	\\
	2D~$\sinc$ & -- & 0.42$\pm0.18$ & 0.23 & 1.118$\pm0.81$ & 0.23 & 0.32$\pm0.07$ & 0.23 & 0.42$\pm0.06$ & 0.20
	\\
	Binary mask & -- & -- & 0.23 & -- & 0.23 & -- & 0.23 & -- & 0.20
	\\ \hline
	\end{tabular}
	\tablefoot{See caption of \tab{tab:var_disk}.}
\end{table*}


\begin{table*}[t!] 
	\caption{\label{tab:strehl_sup} Strehl ratio~$\strehl$.}
	\centering
	\begin{tabular}{ccccccccc}
	\hline
	\hline
	\multirow{2}{*}{$\IFref$} & \multicolumn{2}{c}{Square~$\Paren{\Dsim}$} & \multicolumn{2}{c}{Disc~$\Paren{\Dsim}$} & \multicolumn{2}{c}{Disc~$\Paren{\Dsim}$, no tip-tilt} & \multicolumn{2}{c}{VLT-like}
	\\ \cline{2-9} 
			 & \multicolumn{1}{c}{MC} & \multicolumn{1}{c}{PSD} & \multicolumn{1}{c}{MC} & \multicolumn{1}{c}{PSD} & \multicolumn{1}{c}{MC} & \multicolumn{1}{c}{PSD} & \multicolumn{1}{c}{MC} & \multicolumn{1}{c}{PSD}
    \\
    \hline 
	Piston & $\percent{32.4}$ & $\percent{29.2}$ & $\percent{31.1}$ & $\percent{29.2}$ & $\percent{44.5}$ & $\percent{29.2}$ & $\percent{70.2}$ & $\percent{0.08}$
	\\
	Pyramid & $\percent{76.9}$ & $\percent{74.2}$ & $\percent{64.7}$ & $\percent{75.1}$ & $\percent{73.6}$ & $\percent{74.2}$ & $\percent{91.7}$ & $\percent{86.2}$
	\\
	Gaussian & $\percent{78.6}$ & $\percent{79.7}$ & $\percent{78.4}$ & $\percent{79.7}$ & $\percent{79.4}$ & $\percent{79.7}$ & $\percent{93.5}$ & $\percent{93.9}$
	\\
	ALPAO & $\percent{76.9}$ & $\percent{77.4}$ & $\percent{71.2}$ & $\percent{77.4}$ & $\percent{76.9}$ & $\percent{77.4}$ & $\percent{91.8}$ & $\percent{92.3}$
	\\
	2D~$\sinc$ & $\percent{68.5}$ & $\percent{79.8}$ & $\percent{53.8}$ & $\percent{79.8}$ & $\percent{74.0}$ & $\percent{79.8}$ & $\percent{89.0}$ & $\percent{93.8}$
	\\ 
	Binary mask & -- & $\percent{79.8}$ & -- & $\percent{79.8}$ & -- & $\percent{79.8}$ & -- & $\percent{93.9}$
	\\
	\hline
	\end{tabular}
	\tablefoot{See caption of \tab{tab:strehl_disk}.}
\end{table*}


\end{appendix}

\end{document}